\documentclass[12pt,preprint]{aastex}

\slugcomment{Accepted for publication in the Astrophysical Journal
  Supplements }

\shorttitle{The Intrinsically X-ray Weak Quasar PHL 1811. II.}
\shortauthors{Leighly et al.}

\begin{document}


\title{The Intrinsically X-ray Weak Quasar PHL 1811. II. Optical and UV Spectra and
  Analysis\footnote{Based on observations made 
with the NASA/ESA Hubble Space Telescope, obtained at the Space
Telescope Science Institute, which is operated by the Association of
Universities for Research in Astronomy, Inc., under NASA contract NAS
5-26555. These observations are associated with proposal
\#9181.}\hphantom{l}{$^,$}\footnote{Based on observations obtained at
  Kitt Peak National Observatory, a division of the National Optical
Astronomy Observatories, which is operated by the Association of
Universities for Research in Astronomy, Inc. under cooperative
agreement with the National Science Foundation.}}


\author{Karen M.\ Leighly}
\affil{Homer L.\ Dodge Department of Physics and Astronomy, The
  University of   Oklahoma, 440 W.\ Brooks St., Norman, OK 73019}
\altaffiltext{1}{Visiting Professor, The Ohio State University,
  Department of Astronomy, 4055 McPherson Laboratory, 140 West 18th
  Avenue, Columbus, OH 43210-1173}
\email{leighly@nhn.ou.edu}
\author{Jules P. Halpern}
\affil{Department of Astronomy, Columbia University, 550 W.\ 120th
  St., New York, NY 10027-6601}
\author{Edward B.\ Jenkins}
\affil{Princeton University Observatory, Princeton, NJ 08544-1001}
\author{Darrin Casebeer}
\affil{Homer L.\ Dodge Department of Physics and Astronomy, The University of
  Oklahoma, 440 W.\ Brooks St., Norman, OK 73019}



\begin{abstract}

This is the second of two papers reporting observations and analysis
of the unusually bright ($m_b=14.4$), luminous ($M_B=-25.5$), nearby
($z=0.192$) narrow-line quasar PHL~1811.  The first paper reported
that PHL~1811 is intrinsically X-ray weak, and presented a spectral
energy distribution (SED).  Here we present {\it HST} STIS optical and
UV spectra, and ground-based optical spectra.  The optical and UV line
emission is very unusual.  There is no evidence for forbidden or
semiforbidden lines.  The near-UV spectrum is dominated by very strong
\ion{Fe}{2} and \ion{Fe}{3}, and unusual low-ionization lines such as
\ion{Na}{1}~D and \ion{Ca}{2}~H\&K are observed.  High-ionization
lines are very weak; \ion{C}{4} has an equivalent width of 6.6\AA\/, a
factor of $\sim 5$ smaller than measured from quasar composite
spectra. An unusual feature near 1200\AA\/ can be deblended in terms
of Ly$\alpha$, \ion{N}{5}, \ion{Si}{2} and \ion{C}{3}* using the
blueshifted \ion{C}{4} profile as a template.   Photoionization
modeling shows that the unusual line emission can be explained
qualitatively by the unusually soft SED.  Principally, a low gas
temperature results in inefficient emission of  collisionally-excited
lines, including the semiforbidden lines generally used as density
diagnostics.  The emission resembles that of high-density gas; in both
cases this is a consequence of inefficient cooling.   PHL~1811 is very
unusual, but we note that quasar surveys are generally biased against 
finding similar objects.  

\end{abstract}


\keywords{quasars: emission lines---quasars: individual (PHL 1811)}


\section{Introduction}

Strong, broad emission lines are an identifying feature of Active
Galactic Nuclei (AGN) optical and UV spectra.  The gas emitting these
lines is illuminated and photoionized by the continuum spectrum
emitted by the accretion disk and corona in the central engine.  The
emission line spectrum in AGN is broadly similar from object to object
over many decades of luminosity.  This makes sense because
photoionization and thermal equilibrium tends to produce strong
recombination lines, as well as strong resonance lines from easily
excited abundant ions.

Different elements and ions in the line-emitting gas have a range of
ionization potentials, so naively one would think that the variations
in the spectral energy distribution should influence the emission-line
equivalent widths and emission-line ratios.  The influence is muted by
the fact that many of the metal ions are excited predominately by collisions, so
that the ionization state of the gas depends on the ionization
parameter $U=\Phi/nc$, where $\Phi$ is the photon flux, $n$ is the gas
density, and $c$ is the speed of light.  This means that the dependence
on spectral energy distribution should be a secondary effect
\citep[e.g.,][]{kk88,clb05}. Second, the line emitting region is most
probably extended and comprises gas with a range of densities
\citep{baldwin95}.  Another complication is that the broad-line region
may not see the continuum that we see because some components of the
central engine may radiate more isotropically than others
\citep[e.g.,][]{netzer87}.

Despite these complications, evidence that differences in the spectral
energy distribution have an effect on the broad-line region emission
and kinematics has been found recently.  It is now 
accepted that the Baldwin effect \citep[the empirical anticorrelation
between the BLR emission-line equivalent widths and the continuum
luminosity;][]{baldwin77} is a result of the softening of the
spectral energy distribution in more luminous objects as a result of a
larger black hole mass 
\citep[see,  e.g.,][for a review]{clb05}; the observed anticorrelation
between the Baldwin effect slope and the ionization potential for each line is
the most compelling evidence \citep{dietrich02}.  The accretion rate, which
also may affect the spectral energy distribution,  may also influence
the emission-line equivalent widths \citep{bl04}, and the fact that
two properties may influence the equivalent width perhaps contributes
to the large scatter in the Baldwin effect correlation.  In addition,
the spectral energy distribution can affect the line fluxes and
equivalent widths in indirect ways.  \citet{leighly04} discussed the
fact that an X-ray deficient spectral energy distribution better
explained the broad, blueshifted high-ionization line ratios in two
extreme Narrow-line Seyfert 1 Galaxies (NLS1s).  Then, she showed that
the narrow, intermediate-ionization lines were better explained if the
continuum incident upon them  had been first transmitted through the
gas emitting the high-ionization lines.  In addition, they found that the
strong high-ionization line emission, in particular, the strong
\ion{O}{6} line, in the low-luminosity NLS1 RE~1034$+$39 is a
consequence of its hard spectral energy distribution
\citep{clb05}.

This is the second of two papers that report optical, UV and X-ray
observations and analysis of the nearby ($z=0.192$), luminous
($M_B=-25.5$) narrow-line quasar PHL~1811 \citep{leighly06}.  PHL~1811
was first cataloged as a blue object in the Palomar-Haro-Luyten plate survey
\citep{hl62}.  It then rediscovered 
in the optical followup of the VLA Faint Images of the Radio Sky at
Twenty Centimeters (FIRST) survey \citep{white97,bwh95}.  It is
extremely bright (B=14.4, R=14.1); it is the second brightest quasar
at $z>0.1$ after 3C~273.  Being so bright, it is a very good
background source for studies of the intergalactic and interstellar
medium; furthermore, a {\it FUSE} observation found it to have a rare
Lyman limit system which has been studied by
\citet{jenkins03,jenkins05}.  It was odd, however, that such a bright
quasar was not detected in the ROSAT All Sky Survey (RASS). In comparison with
other quasars of its luminosity, the expected RASS count rate is about
$0.5 \rm\, s^{-1}$; we placed an upper limit of $1.3 \times 10^{-2}
\rm \, counts\, s^{-1}$ \citep{lhhbi01}.  A pointed {\it BeppoSAX}
observation detected the object, but it was still anomalously
weak. Too few photons were obtained in the {\it BeppoSAX} observation
to unambiguously determine the cause of the X-ray weakness;
\citet{lhhbi01} speculated that either it is intrinsically X-ray weak,
or it is a nearby broad-absorption line quasar and the X-ray emission
is absorbed, or it is highly variable, and we observed it both times in
a low state. 

In the companion paper \citep[][hereafter Paper I]{leighly06}, we
reported the results of five X-ray observations of PHL~1811 using {\it
Chandra}, {\it XMM-Newton}, and {\it Swift}.  We also reported the
simultaneous UV photometry using the OM on {\it XMM-Newton} and the
UVOT on {\it Swift}, as well as optical photometry obtained at the MDM
observatory. 
These observations confirmed that PHL~1811 is an anomalously X-ray weak
quasar.  The X-ray spectrum is steep, with photon index between 2 and
2.6, and there is no evidence for absorption in excess of the Galactic
\ion{H}{1} column.  The X-ray flux varied by a factor of $\sim 5$ among the 5
observations, but the UV photometry, when compared with the {\it HST}
spectrum presented and analyzed in this paper, showed no detectable
variability.  The inferred $\alpha_{ox}$, defined as the
point-to-point slope between 2500\AA\/ and 2 keV, is measured to be
$-2.3 \pm 0.1$.  Typical quasars with PHL~1811's optical luminosity
have $\alpha_{ox}$ of $-1.6$ \citep{steffen06}.  Accounting for scatter,
this means that PHL~1811 is between 13 and 450 times fainter in X-rays
than other quasars with the same UV luminosity.  Along with the {\it
  ROSAT} All Sky Survey upper limit and the {\it BeppoSAX}
observation, we have now observed PHL~1811 in X-rays seven times and
it is always found to be  X-ray weak. While we can never disprove the
hypothesis that we coincidentally always observe it in a transient low
state, the odds of that being true are decreasing.  We concluded
in Paper I that PHL~1811 is the best example of an intrinsically X-ray
weak quasar.  We present an updated spectral energy distribution of
PHL~1811 in Paper~I and discuss possible reasons for its X-ray weakness.

In this paper, we present {\it HST} STIS UV and optical spectra, and
ground-based optical spectra obtained at the KPNO 2.1 meter telescope
(\S 2).  As shown in \S 3, the emission-line properties of PHL~1811
are very unusual.  There are no forbidden or semi-forbidden lines, and
the high-ionization lines are weak.  The spectra are dominated by
low-ionization lines, and \ion{Fe}{2} and \ion{Fe}{3} are strong.  In
\S 4 we investigate, using {\it Cloudy} models, the physics of gas
illuminated by an X-ray weak spectral energy distribution (SED).  It
turns out that gas illuminated by a soft SED has much different
properties than gas illuminated by a normal SED, and furthermore,
many of the peculiar features of the optical and UV spectra of
PHL~1811, including the lack of semiforbidden lines, can be explained
by this single factor.    We summarize the principal results of
the paper in \S 5. We also include a brief Appendix in which we
explore additional properties of the soft SED that are not directly
related to the observations.   Some of the results have been published
in \citet{lhj04}.  We assume a flat Universe with $H_0=70\rm\, km
\,s^{-1}\,Mpc^{-1}$ and $\Omega_{vac}=0.73$, unless otherwise
specified.

\section{Optical and UV Observations and Analysis}

The {\it HST} observations were reduced using the standard pipeline.
To reduce the effects of any possible fixed-pattern noise, four
exposures with each of the chosen MAMA gratings were made with the
target stepped along the slit.  The separate exposures for each
detector were each cross correlated with a preliminary average
spectrum to search for any systematic offsets; none were found.  The
average spectrum was then computed using the {\it IRAF} task {\tt
scombine} and resampled to a linear binning.  The {\it HST} STIS CCD
observation was split into two separate exposures in order to aid in
the rejection of cosmic rays.

We presented an optical spectrum taken at the MMT telescope in 1997 in
\citet{lhhbi01}.  We obtained several other spectra at KPNO that have
better resolution and signal-to-noise ratios that we present here
(Table 1). Two of the KPNO spectra were taken within 1.5 months of the
{\it HST} and {\it Chandra} observations. 

To check the wavelength calibration of the UV spectra, we fit
Galactic absorption lines (FUV: \ion{Si}{2}~$\lambda 1260$, $\lambda
1527$; \ion{C}{2}~$\lambda 1335$; \ion{C}{4}~$\lambda 1548,1551$; NUV:
\ion{Mg}{2}~$\lambda 2796,2804$; \ion{Fe}{2}~$\lambda 2344,2374, 2383,
2567, 2600$).
We find that the FUV spectrum is consistent with no
systematic shift.  However, all five lines in the NUV were offset from
their rest wavelengths by 1.0--1.7 \AA\/.  To rectify this, the
wavelength scale in the NUV was adjusted blueward by 1.35 \AA\/ (the
mean of the shifts of the seven NUV lines).  

There were no convenient absorption lines available to check
the wavelength calibration in the {\it HST} STIS CCD and ground-based
optical spectra.  Assuming no systematic shifts does not lead to any
anomalous or unexplainable results (see below) so we assume that the
wavelength calibration in these spectra is satisfactory.  

The {\it HST} spectra are all given in the vacuum wavelengths; we
convert the optical spectra to vacuum wavelengths uniformly for
consistency.  

We next estimate the redshift of the object.  There is no prominent
emission feature in the FUV spectrum that could confidently be
considered to be emitted at the rest wavelength. There is,
however, a weak absorption line that \citet{jenkins03} speculated
originates in Ly$\alpha$ in the rest frame of the object. This
absorption line gives an estimate of the redshift of $z=0.1903$.
Interestingly, however, in the medium resolution {\it HST} spectra,
there can be seen two lines near the Ly$\alpha$ restframe; one is 
consistent with a redshift of 0.1901, and another, smaller one that
does not appear in the low resolution data is consistent with a
redshift of 0.1920.  Alternatively, these may originate in galaxies
other than the quasar host that are in the same group.  

In the NUV spectrum, there is a narrow emission line clearly
originating in \ion{Fe}{2} UV191.  The three multiplet components are
closely spaced, lying within 3\AA\/ of one another; the $gf_{ij}$
weighted average wavelength is 1786.2\AA\/.  Fitting this feature in
the NUV spectrum gives an estimated redshift of $z=0.1900$.  In the
{\it HST} CCD spectrum, we may consider \ion{Mg}{2} to be emitted at
the rest wavelength; it gives an estimate of the redshift of $0.1920$.
The Balmer lines in the optical spectra also give estimates near this
value (0.1919 estimated from H$\alpha$ in the July spectrum; 0.1924
estimate from H$\beta$ in the October blue spectrum; 0.1923 from
H$\beta$ in the October red spectrum; 0.1921 from H$\alpha$ in the
October red spectrum).  Note there is no observed [\ion{O}{3}] or any
other narrow-line region emission line.  Thus, we see a difference in
the estimated redshift between the UV and optical spectra by
0.002. This corresponds to wavelength shifts of 2.9\AA\/ at observed
frame Ly$\alpha$, 4.3\AA\/ at \ion{Fe}{2} UV191, and 6.7\AA\/ at
\ion{Mg}{2}.  The estimated wavelength calibration uncertainty,
obtained from the STIS data handbook \citep{quijano03}, is 0.5--1.0
pixel in the MAMA and 0.2--0.5 pixel in the CCD.  This corresponds to
0.3--0.6\AA\/ in the FUV spectrum, 0.8--1.6\AA\/ in the NUV spectrum,
and 0.5--1.4\AA\/ in the CCD spectrum -- much smaller than the shift
required to explain the $\Delta z=0.002$ between the redshift
estimates.  Therefore, we assume that the Balmer and \ion{Mg}{2} lines
are most likely to reflect rest wavelength of the object, we take the
redshift to be z=0.192, and assume that the \ion{Fe}{2} UV191 is
blueshifted. 

We next construct a merged spectrum using the {\it HST} and
ground-based spectra. For the {\it HST} spectra, we start at the far
UV end and successively adjust the normalization and construct
average, resampled spectra in the overlap regions.  The absolute
photometry is estimated to be 4\% and 5\% for the MAMA and CCD
spectroscopy, according to the STIS data handbook.  We found good
agreement in the overlap regions throughout the FUV, NUV and optical
HST spectra. 

Turning to the ground-based spectra, we found that the October red and blue
KPNO spectra were consistent in slope and normalization.  However, the
July red spectrum had better SNR than the October red 
spectrum, and we wanted to use it.  It had a slight difference in
slope compared with the July spectrum.  We resample
the overlap region, which was most of both spectra, onto the same
wavelength binning, divided the two, fit a quadratic function to the
result and correct the slope difference in the July spectrum.  We
then merge the October blue spectrum with the corrected July red
spectrum.   

Next, we merge the {\it HST} and KPNO spectra.  The {\it HST}
spectrum is brighter by about  30\%, most likely because of slit
losses in the KPNO spectra, and there was also a small
difference in slope.  We resampled the overlapping regions onto the
same wavelength scale, fitted the ratio to a 9-node spline and
corrected the KPNO spectrum to match the slope of the {\it HST}
spectrum.  

Finally, we deredden the merged merged spectrum using $E(B-V)=0.046$,
estimated from the infrared cirrus \citep{sfd98} and the \citet{ccm89}
reddening law, and apply the redshift correction ($z=0.192$), and

The merged spectrum is shown in Fig.\ 1.  Except for the small
absorption line near rest-frame Ly$\alpha$, the other prominent
absorption lines are all from our Galaxy or from intervening galaxies;
they are all identified in \citet{jenkins03}, and the reader is
referred to that paper for the identifications.  Emission lines
expected from active galaxies are marked. We also mark emission lines
from a particular type of ion - those in which the lowest excited
level has  $\Delta S \ne 0$ compared with the ground state, but
rather that transition to the ground state is semiforbidden.  These
ions include Si$^{+1}$, Si$^{+2}$, C$^{+2}$, and Al$^{+1}$. The emission
lines that we mark are those that have the first excited level as the
lower level.  This type of line is analogous to auroral lines, except
that the metastable transitions to ground are semiforbidden rather
than forbidden.  The motivation for looking at these lines is
discussed in \S 3.3.2.

\section{Emission-line Properties}

In this section, we discuss the optical and UV emission-line
properties of PHL~1811.  We organize the discussion by moving
shortward in the spectrum: we first discuss the optical emission
lines, then the near UV, and finally the far UV lines.

To model the spectra, we use a combination of the IRAF task {\tt
Specfit}, and programs that we developed ourselves written in IDL.
{\tt Specfit} requires 1-sigma error bars on the spectra, which are
not available for the optical spectra.  The signal-to-noise ratio in
the continuum of the optical spectra is greater than 30, so fit
results are not dominated by statistics.  We estimate uniform 1-sigma
errors as the standard deviation in line-free segments of the data,
and interpolate between these estimates.

\subsection{Optical Emission-line Properties}

We use the merged spectrum to analyze the emission-line properties.
Fig.\ 2 displays the region of the spectrum near H$\beta$ and
H$\alpha$.  The strong emission lines from \ion{Fe}{2} that are common
in NLS1s are clearly seen.  In order to determine which lines besides
\ion{Fe}{2} are present in the spectrum, and to measure the properties
of the Balmer lines, we perform an \ion{Fe}{2} template subtraction,
as is commonly done \citep[e.g.,][]{bg92,leighly99b} using a template
obtained from the prototypical NLS1 I~Zw~1.  We use two different
templates. The first one is the \citet{bg92} template with a few
additional iron lines added in the blue by Dirk Grupe.  The second one
was recently developed by \citet{vjv04}.  Each template is broadened
by convolution with a Gaussian with $\rm FWHM=400\rm\,km\,s^{-1}$, scaled
appropriately, and subtracted from the spectra. Neither template
matches the \ion{Fe}{2} emission in the region of H$\beta$ very well
(Fig.\ 2).  We note that this might be expected for the V\'eron-Cetty,
Joly \& V\'eron template, since those authors remove forbidden
\ion{Fe}{2} emission, leaving only permitted \ion{Fe}{2} in their
template.

The \citet{bg92} template produces a better fit in the vicinity of
H$\beta$, so we use that spectrum for further analysis in that region.
\citet{vjv04} identified \ion{Fe}{2} usually hidden in the base of
H$\alpha$, and therefore their template yields an H$\alpha$ line with
a narrower, more symmetric base.  The H$\alpha$ and H$\beta$ line
profiles have extended wings and clearly can not be modeled using a
Gaussian.  We use a Lorentzian profile plus a local linear continuum to
fit them.  The results are given in Table 2.

We clearly observe H$\gamma$, and report the results of the Lorentzian
fit to this line in Table 2.  However, because the \ion{Fe}{2}
subtraction is really not very good in the region of this line, we do
not consider these results to be very reliable.

The optical spectra contain very few other lines besides hydrogen
Balmer lines and \ion{Fe}{2}.  \ion{He}{2}~$\lambda 4686$ is not
apparent, although it could be present if it is weak and masked by the
imperfect \ion{Fe}{2} subtraction.  There is no trace of the generally
strong narrow-line region line, [\ion{O}{3}]~$\lambda 5007$, or any
other narrow-line region line.  There is an excess near 6350 \AA\/
that can be seen after subtraction with the V\'eron-Cetty, Joly, \&
V\'eron template (Fig.\ 2).  Based on the line identifications for
I~Zw~1 made in V\'eron-Cetty, Joly, \& V\'eron (2004), this could be
\ion{Si}{2}~$\lambda\lambda 6347.10, 6371.40$.   

The most interesting lines that are seen in the optical part of the
spectrum are the very low ionization lines \ion{Na}{1}~D and
\ion{Ca}{2}~H\& K.  The \ion{Na}{1}~D~$\lambda\lambda 5889.89,
5895.89$ doublet is frequently blended with the \ion{He}{1}~$\lambda
5875.70$ \citep[e.g.,][]{thompson91}, but in this case, as seen in
Fig.\ 3, we are fairly confident that the entire feature is
\ion{Na}{1}~D and no \ion{He}{1} is present.  We modeled the feature
in two ways, using the Veron-template-subtracted spectrum.  First, we
modelled the doublet with two Lorentzians of equal flux, width and
fixed separation.  The results of this model comprises the entry for
\ion{Na}{1}D in Table 2.  Notably, we find that the mean wavelength of
the doublet, which would be the weighted mean of the feature if the
gas were very optically thick to the line, is equal to the lab value.
The second model includes an additional Lorentzian to model
\ion{He}{1} also.  The separations of the lines were fixed, and the
widths and fluxes of the \ion{Na}{1}D lines were constrained to be
equal (as before).  Since we do not see \ion{He}{1}, we constrain the
fit by assuming that the width is the same as that of H$\alpha$,
although we note that it is observed to be somewhat broader in Seyfert
galaxies \citep{crenshaw86}.  The improvement in the fit was not
significant ($\Delta\chi^2=1.2$ for 201 fitted points), and the
uncertainty on the flux of the \ion{He}{1} component is almost as
large as the flux (i.e, relative uncertainty is 90\%). In addition,
the mean wavelength for the \ion{Na}{1}D line is now redward of the
lab wavelength, although it is still consistent with that value.  We
conclude that \ion{He}{1} is not detected, and evaluate the upper
limit by varying the normalization until the $\chi^2$ is larger by
6.63 (99\% confidence for one parameter of interest).  In addition, we
search for \ion{He}{1}~$\lambda 7067$.  We find no evidence for it, and
we estimate the upper limit in the same way.  The results are given in
Table 2.

The weakness of \ion{He}{1}~$\lambda 5876$ is physically interesting.
\citet{crenshaw86} tabulate the fluxes of \ion{He}{1} and H$\alpha$
for 9 Seyfert galaxies; the mean \ion{He}{1}/H$\alpha$ ratio is 0.061.
\citet{thompson91} tabulate \ion{He}{1}/H$\alpha$ ratio for 6 quasars, and
the mean ratio is 0.025.  We measure an upper limit on the ratio of
0.0062 in PHL~1811.  Such a low ratio is difficult to explain using
photoionization models; we address this issue in \S 4.4. 

The strength of \ion{Na}{1}D is also interesting.  \citet{thompson91}
discussed the presence of this line in quasars.  The ionization
potential of Na$^+$ is only 5.14~eV, so rather special conditions are
necessary to produce this line.  Using {\it  Cloudy} models, he found
that a very high column density was necessary to shield the neutral
sodium from photoionization in order to produce sufficiently large
\ion{Na}{1}D/H$\alpha$.   

Turning to the \ion{Ca}{2}~H\& K doublet, we initially modelled these
lines as  Lorentzians, as the other lines
are. The line widths are then found to be only $770 \pm 83 \rm\, km\,
s^{-1}$, about half the width of the other lines.  Gaussian profiles
fit these lines somewhat better ($\Delta\chi^2=15.4$ for 152 points),
and the widths are still somewhat narrow ($980 \pm 80 \rm \,
km\,s^{-1}$).  Indeed, if we overlay their profiles on that of
H$\beta$ (Fig.\ 4), they appear to be narrower than that line.  The
\ion{Na}{1}~D lines are also somewhat narrower than H$\alpha$ ($1365
\pm 180\rm \, km\,s^{-1}$), although we model those with a Lorentzian
profile. 

Observations of the \ion{Ca}{2} infrared triplet in AGN were reported
by \citet{persson88}.  He found that these lines were detected in at 
least nine of the forty objects surveyed, and that the line strengths
were roughly correlated with the optical \ion{Fe}{2}.  Interestingly,
reports of observations of \ion{Ca}{2} H\&K are rather rare
\citep{fp89}, although we note the recent detection of these lines in
IRAS~07598$+$6508 by \citet{veroncetty06}.  \citet{fp89} and
\citet{joly89} present models that show that very large column
densities are required to produce the observed infrared triplet
emission.  There are two hypotheses for the weakness of the
\ion{Ca}{2}~H\&K lines in most AGNs.   First, the structure of the
Ca$^+$ ion is similar to 
the Fe$^+$ ion.  In each case, there are resonance transitions to a
high level, which decay back to the ground state, or to an
intermediate metastable state.   Some of the UV \ion{Fe}{2} lines and
the \ion{Ca}{2}~H\&K lines correspond to the resonance transitions to
the high level, while the optical \ion{Fe}{2} lines and the
\ion{Ca}{2} IR triplet correspond to the decay to the intermediate
metastable state.  Thus, just as scattering of the UV \ion{Fe}{2} lines
reprocess the iron emission into optical \ion{Fe}{2} lines
\citep{kk81}, scattering in a high-column-density gas converts the
\ion{Ca}{2}~H\&K lines into the infrared triplet.  Alternatively  or in
addition, absorption in the host galaxy may make \ion{Ca}{2}~H\&K in
emission difficult to detect. Regardless, the similarities
between Ca$^+$ and Fe$^+$ suggest that the study of \ion{Ca}{2} may be
a profitable way to understand the physical conditions of the
low-ionization line-emitting region in AGN, as Ca$^+$ is a much
simpler atom than Fe$^+$.    

The \ion{Ca}{2} lines are well resolved, so in principle we can try to
constrain the optical depth of the line-emitting region.  We fit the
lines with equal width and a fixed separation, but allow their fluxes
to be independent.  We find that the K component flux is then larger
than the H component flux, and the ratio between the two is measured
to be 1.28.  The $gf$ values are 1.36 and 0.66 for K and H,
respectively, thus predicting a ratio equal to 2 for optically thin
gas.  We note that there is some uncertainty in the measurement of the
line fluxes because the continuum is not smooth and is difficult to
identify; however, the line ratio can be seen by eye to closer to
one than to two.  \citet{laor97b} measure the ratio between the
similar \ion{Mg}{2} doublet components in I~Zw~1 to be 1.2.
\ion{Mg}{2} is isoelectronic with \ion{Ca}{2} and these lines are both
transitions from the lowest-level excited state to the ground state,
so the interpretation of the low ratio should be similar.  In both
cases it means that that the gas is sufficiently optically thick that
enough scatterings occur to redistribute the lines; the ratio would be
1 to 1 if completely optically thick.

The ratio of \ion{Fe}{2} to H$\beta$ is a parameter frequently used to
described the strength of the \ion{Fe}{2} emission.  As was done in
\citet{bg92} and \citet{leighly99b}, we integrate the
subtracted \ion{Fe}{2} model between 4344 and 4684 \AA\/.  Dividing
the flux measured from the model by the integrated and modeled
H$\beta$ fluxes yields an estimated ratio in the range of 1.22--1.35.
Comparing with the sample of NLS1s reported in \citet{leighly99b}, we
find that the estimated ratio for PHL~1811 is in the middle of the range
reported there, but somewhat larger than the median and mean (1.0 and
1.1, respectively).  Thus, PHL~1811 is a moderately strong \ion{Fe}{2}
emitter. 

In summary, we find that the optical spectrum from PHL~1811 contains
prominent Balmer and \ion{Fe}{2} lines, similar to many other
NLS1s. The spectrum is somewhat unusual in that there are no forbidden
lines; in particular the generally strong [\ion{O}{3}]~$\lambda 5007$
line is absent. 
NLS1s are identified by their weak forbidden lines, and a few other
objects appear to have no [\ion{O}{3}] also \citep[e.g.,
RX~J0134.2$-$4258,][]{grupe00}.  There are no high-ionization coronal
lines which sometimes appear in optical spectra of NLS1s
\citep[e.g.,][]{dck05}.  PHL 1811 is unusual in that it shows
prominent very low-ionization lines from \ion{Na}{1}D and \ion{Ca}{2}~H\&K,
but no \ion{He}{1}.

\subsection{The Near-UV Spectrum}

The near UV portion of the spectrum of AGN, between around
1600~\AA\ and 3500~\AA\/, generally exhibits strong and prominent
emission lines from \ion{C}{3}]~$\lambda 1909$ and
\ion{Mg}{2}~$\lambda 2800$.  A pseudocontinuum of \ion{Fe}{2} may
also be present over the entire 
region, and is most prominent between 2200 and 3050\AA\/. \ion{Fe}{3}
and \ion{Fe}{1} may also contribute.  Good signal-to-noise spectra
often reveal contributions from numerous weak semiforbidden lines of
oxygen, nitrogen, silicon, and carbon.  The near UV spectrum from
PHL~1811 is different from many AGN in that no semiforbidden or
forbidden lines are observed.  The spectrum is strongly dominated by
\ion{Fe}{2} and \ion{Fe}{3}.

\subsubsection{\ion{Mg}{2}, \ion{Fe}{2} and \ion{Fe}{3} in PHL 1811}

The properties of UV \ion{Fe}{2} and \ion{Mg}{2} in Active Galactic
Nuclei (AGN) are important diagnostics for several reasons.  Luminous
quasars are probes of the early Universe.  As discussed by
\citet{hf93} and others, the production of iron is thought to lag that
of the $\alpha$ elements including magnesium due to different
formation mechanisms: magnesium and about half of the iron
\citep{nnk92} are produced in supernovae of massive, rapidly-evolving
stars, while the remainder of the iron is produced largely in Type 1a
supernova which are triggered by accretion onto a white dwarf, and are
delayed with respect to core-collapse supernova.  Observation of an
evolution of the [Fe/Mg] ratio with redshift could constrain the time
of the first burst of star formation in the Universe.  The UV
\ion{Fe}{2} and \ion{Mg}{2} are found conveniently near one another in
the rest-UV bandpass, and they should both be produced in gas with
similar properties, so it was thought that the \ion{Fe}{2}/\ion{Mg}{2}
ratio could be an abundance diagnostic. No clear evidence for
\ion{Fe}{2}/\ion{Mg}{2} ratio evolution has been observed yet
\citep[e.g.,][]{dietrich03}.

As in the optical bandpass, the flux and equivalent width of
\ion{Fe}{2} are frequently measured using a template
\citep[e.g.,][]{cb96,forster01,dietrich02,lm06}. We developed a UV
\ion{Fe}{2} template from the {\it HST} spectrum of the prototypical
Narrow-line Quasar I~Zw~1, following \citet{vw01}: we subtracted a
power law identified at the relatively line-free regions near
2200\AA\/ and 3050\AA\/; absorption lines and prominent emission
lines not attributable to \ion{Fe}{2} were then subtracted.

Using this template, we recently measured the \ion{Fe}{2}/\ion{Mg}{2}
ratio in the spectra from 903 intermediate-redshift narrow-line
quasars  from the Sloan Digital Sky Survey \citep{lm06}.
In most of those spectra, the template modelled the iron 
quite well in the  2200--3090 \AA\/ bandpass.  However, when we try to
apply the template to PHL~1811, we cannot obtain a good fit.  While
the region of the spectrum around \ion{Mg}{2} matches the template
well, there is significant excess emission at shorter wavelengths.  To
illustrate this problem, we adopt a strategy frequently used by X-ray
astronomers: we fit the spectrum with a linear model, the iron
template, and two Gaussians to model the \ion{Mg}{2} doublet in the
$\sim 2650$--3090 \AA\/ bandpass.  Then, we extrapolate the model to
shorter wavelengths.  The result is shown in Fig.\ 5a for PHL~1811.
It is interesting that the template and continuum match the dip around
2250\AA\/ fairly well,  but there is significant excess emission
between 2300\AA\/ and 2500\AA\/ and shortward of 2250\AA\/. 

For comparison, we perform the same analysis on the spectra of several
other quasars and a composite spectrum.  In Fig.\ 5b we show I~Zw~1.
Naturally, the template fits the spectrum perfectly. Next, we show the
largest \ion{Fe}{2}/\ion{Mg}{2} composite from \citet[shown in their
  Fig.\ 6]{lm06}.  In this case and for the next case, we ignore the
wavelengths containing the prominent emission from \ion{Fe}{2} UV 62
and UV 63, low-lying transitions that sometimes produce strong
emission near 2760 \AA\/, and trim the fitted bandpass.  Finally,
we show the spectrum of SDSS J105023.68-010555.5.  The template
matches the data rather well for both of these spectra.  In Fig.\ 5a,
however, we compare PHL~1811 with the {\it HST} FOS spectrum of the
luminous NLS1 RX~J0134.2$-$4258 \citep[e.g.,][]{grupe00}, and the
spectrum of SDSS~J094257.80-004705.2.  In both of these objects, like
PHL~1811, extrapolation of the template leaves prominent excess
emission at short wavelengths.

What is the origin of the short wavelength excess in PHL~1811 and the
other objects?  In Fig.\ 6, we show the spectrum of PHL~1811 along
with the upper level energies of transitions of \ion{Fe}{2} and
\ion{Fe}{3} between 1000 and 3500 \AA\/; the atomic data was obtained
from Kurucz Atomic Line
Database\footnote{.http://cfa-www.harvard.edu/amdata/ampdata/kurucz23/sekur.html}.
Around 2000\AA\/, there are many \ion{Fe}{3} transitions with large
Einstein $A_{ij}$ coefficients.  These include the multiplets from
low-lying levels such as \ion{Fe}{3} UV 34 near 1900\AA\/ and
\ion{Fe}{3} UV 48 near 2060\AA\/, which are seen as excesses in the
spectrum.  But there are also many more transitions with upper level
energies between 15 and 19 eV. Thus, it may be that strong \ion{Fe}{3}
contributes to the excess shortward of 2250\AA\/.

Between 2300-2500\AA\/, there are no high $A_{ij}$ transitions of
\ion{Fe}{3}.  \ion{Fe}{2} is more prominent in this region.  The
principal low-lying transitions with high $A_{ij}$ values between 2300
and 2500 \AA\ are UV 2 near 2400 \AA\/, UV 3 near 2340 \AA\/, and UV
35 and UV 36 near 2380 \AA\/.  With upper level energy of only 5--6
eV, these are candidates for the short wavelength excess emission.
However, UV 1 near 2610 \AA\/, UV 62 and 63 near 2750 \AA\/, and UV 64 near
2570 \AA\/ also have upper level energies near 5--6 eV, and so one
might expect these to be enhanced also.  These transitions, and
especially UV 62 and UV 63 are sometimes quite strong in some
narrow-line quasars \citep[e.g., SDSS J105023.68-010555.5 in Fig.\ 5b;
  also][]{hall04}, but they do not appear to be excessively strong in
PHL~1811 or in the objects in Fig.\ 5a.  However, Fig.\ 6 shows that
there exists many higher-excitation \ion{Fe}{2} transitions with high
$A_{ij}$ values that produce emission in this region.  Especially
near 14 eV, there is a large concentration that is centered between
2300--2500 \AA\/, and we propose that it is this high-excitation
\ion{Fe}{2} emission that contributes to the \ion{Fe}{2}
pseudocontinuum so strongly in PHL~1811 and similar objects.  

Combined with the strong \ion{Fe}{3} emission, the strong \ion{Fe}{2}
emission from high energy levels suggests that the iron in PHL~1811
and similar objects is characterized by a relatively higher
ionization/excitation than in other objects. At first sight, this is a
puzzling result, because the soft spectrum implies that the energy per
ionizing photon is relatively low; using the PHL~1811 continuum
developed for {\it Cloudy} simulations, we obtain an average energy
per photon of only $19\rm \, eV$.  Also, the lack of hard X-rays
implies that the partially ionized zone, where \ion{Fe}{2} emission is
thought generally to be emitted, will be quite shallow.   But as we
show in \S 4.3, it turns out that  strong \ion{Fe}{2} and \ion{Fe}{3}
are consequence of the soft spectral energy distribution through
several mechanisms: the gas in the \ion{H}{2} region is dominated by
intermediate-ionization species, including Fe$^{+2}$, there is
strong continuum pumping of the Fe$^+$ ion, and there is strong
Ly$\alpha$ pumping associated with strong continuum emission in the
partially-ionized zone. 

The \ion{Mg}{2} is rather weak in PHL~1811, although in some sense it
appears strong because it is prominent in comparison with the other
lines.  We find that the equivalent width is just 12.9~\AA\/.  In
comparison, the equivalent widths from composite spectra are 50\AA\/
\citep{francis91}, 64\AA\/ \citep{zheng97}, and 34\AA\/
\citep{brotherton01}.  In \S 4.1.3 we show that the low equivalent width
for this collisionally excited line is plausibly a consequence of the
low temperature in the line-emitting gas.

Finally, we measure the UV \ion{Fe}{2}/\ion{Mg}{2} flux ratio, which,
as noted before, is used as an abundance diagnostic in quasars. As
shown above, the I~Zw~1 template does not fit the UV \ion{Fe}{2}
spectrum in PHL~1811 well.  Therefore, to measure the flux of this
component, we define the continuum, subtract a powerlaw model,
integrate over the remainder from 2000 to 3000\AA\/ \citep[the
\ion{Fe}{2} UV band defined by ][]{verner04}, and subtract the flux of
the \ion{Mg}{2} line.  Defining the continuum is difficult.  As noted
by \citet{verner04}, there is a gap in the \ion{Fe}{2} emission at
around 3050\AA\/, but then a pseudocontinuum of varying amplitude
continues to at least $\sim 1000$\AA\/.  That pseudocontinuum is
relatively low at 1800 or 2000 \AA\/, so to obtain a lower limit on
the \ion{Fe}{2} emission, we define two continua: from 1800--3050\AA\/
and from 2000--3050\AA\/.  The \ion{Fe}{2}~UV/\ion{Mg}{2} flux ratios
obtained this way are 14.9 and 13.7 for the 1800--3050 and
2000--3050\AA\/ continua, respectively.  These estimates are uncertain
because some \ion{Fe}{3} emission may be contained in the \ion{Fe}{2} flux
estimate, which would tend to decrease the ratio. In addition, there
is some uncertainty in the \ion{Mg}{2} emission-line flux, since 
\ion{Fe}{2} template that was used assumes no \ion{Fe}{2} emission
under the \ion{Mg}{2}; if there were \ion{Fe}{2} under \ion{Mg}{2},
the \ion{Fe}{2}/\ion{Mg}{2} flux ratio would be larger \citep[see][for
  a discussion of this point]{lm06}. 
For comparison, we found a maximum likelihood mean and $1\sigma$
dispersion  of the \ion{Fe}{2}/\ion{Mg}{2} ratio of $3.93 \pm 0.95$
in the 903 SDSS quasar spectra analyzed in \citet{lm06}. Note that
in that paper the \ion{Fe}{2} flux was integrated between 2200 and
3050\AA\/.  Thus, PHL~1811 shows an enhanced \ion{Fe}{2}/\ion{Mg}{2}
flux ratio, and enhanced UV \ion{Fe}{2} emission compared with other
quasars.   

\subsubsection{No Semi-forbidden Lines in PHL 1811}

Fig.\ 7 shows a portion of the near-UV spectrum from PHL 1811.  The
{\it HST} spectrum of the prototypical NLS1 I~Zw~1 is also shown.  The
positions of permitted, semiforbidden, and forbidden emission lines
commonly seen in AGN are marked.  Prominent \ion{Fe}{2} and
\ion{Fe}{3} lines are marked, although as discussed above, blends of
lines from the same ions produce a strong pseudocontinuum.  Absorption
lines that are present all originate in our Galaxy.

The most prominent difference between these two spectra is the
relative weakness of all emission features in the spectrum of
PHL~1811.  In most AGN, \ion{C}{3}]~$\lambda 1909$ is one of the
  strongest emission 
lines; the equivalent width of the 1900 \AA\/ feature in quasar
composite spectra is around 20\AA\/ (Francis et al.\ 1991; Brotherton et
al.\ 2001).  In I~Zw~1, \ion{C}{3}] is weak relative to \ion{Si}{3}],
perhaps because of a higher density in the broad-line region
\citep[e.g.,][]{laor97b}.  In the spectrum of PHL~1811, there is a bump
near the expected position of \ion{C}{3}].  However, as the inset
figure shows, the central wavelength of that feature is somewhat
longward ($\sim 1915$ \AA\/) of the expected wavelength for
\ion{C}{3}].  Its wavelength matches that of the \ion{Fe}{3}~$\lambda
1914.1$ emission line, part of the \ion{Fe}{3} UV34 triplet
(1894.8\AA\/, 1914.1\AA\/, 1926.3\AA\/) more closely.  As discussed in
\citet{lm04}, the optically thin ratios of the lines in this triplet
should be 1:1.4:1.8.  However, in some objects in the sample
considered by \citet{lm04}, the 1914\AA\/ component was strongly
enhanced.  The upper level energy of this transition corresponds to
1214.6\AA\/, which is just 1.1\AA\/ from Ly$\alpha$.  Therefore, it is
quite plausible that if microturbulence is present in the
line-emitting gas, this line may be strongly pumped by Ly$\alpha$ 
\citep[also][]{johansson00}.  We suspect that this phenomenon is
occurring in PHL~1811, so that the feature near 1914\AA\/ is
\ion{Fe}{3}, and \ion{C}{3}] is not present.

It is not straightforward to obtain an upper limit on \ion{C}{3}]
  because of the complicated iron emission in the region.  We assign
  the conservative upper limit recorded in Table 2 as follows: we
  assume that our identification of the 1915\AA\/ feature is mistaken,
  and it is really \ion{C}{3}].  We identify as pseudocontinuum on
  each side of that feature and integrate over it to obtain the flux.  

The lack of semiforbidden lines in PHL~1811 naturally suggests that
the emitting gas has a high density.  Of the six semiforbidden lines
(\ion{O}{3}]~$\lambda 1664$, \ion{N}{3}]~$\lambda 1750$, \ion{Si}{3}]
$\lambda 1892$, \ion{C}{3}]~$\lambda 1909$, \ion{N}{2}]~$\lambda 2142$,
and \ion{C}{2}]~$\lambda 2326$) in the region of the spectrum
shown in Fig.\ 7, \ion{Si}{3}~$\lambda 1892$ has the highest
critical density \citep[$3 \times 10^{11} \rm
cm^{-3}$;][]{hkfwb02}. However, as we show in \S 4, density indicators
break down for gas illuminated by a soft spectral energy
distribution (SED).  In fact, the lack of semiforbidden lines  can be
completely attributed to the soft SED.  Semiforbidden lines are
collisionally excited, and collisionally-excited lines are weak in gas
illuminated by a soft SED as a consequence of a low gas temperature.  

\subsubsection{Other Near-UV Line Emission}

We clearly see the \ion{Fe}{2} UV 191 feature in the spectrum.  This
triplet (1785.277, 1786.758, 1788.004\AA\/) originates in a
high-excitation state; the upper level energy is $9.8\rm \,eV$.  This
line is prominent in stellar spectra; the excitation mechanism is not
known. Dielectronic recombination has been suggested as a possible
mechanism \citep{jh88}.  That mechanism might be consistent with our
hypothesis, presented in \S 3.2.1, that the near-UV line emission is
dominated by Fe$^{+3}$ and high-excitation 
\ion{Fe}{2}.  We model this line with three Gaussian components fixed
at their expected optically thin relative fluxes of 2.0:1.5:1.0
\citep{johansson95}.  The best fitting $gf$ weighted wavelength of the
feature is 1783.2\AA\/.  Because this feature it so narrow, it can be
seen to be clearly blueshifted (by $570\rm \, km\, s^{-1}$) from the
expected $gf$ weighted wavelength of 1786.4\AA\/. A blueshift in a
narrow, low-excitation line is not expected.  Perhaps this is telling
us something about the emission mechanism?

We also identify \ion{Al}{3}~$\lambda\lambda 1854.7, 1862.8$.  We model
this feature with two Gaussians with equal widths, fluxes and a fixed
distance between.  We measure a relatively large width  (Table 2;
$3050\rm \, km\, s^{-1}$), and large redshift ($1340 \rm \, km\,
s^{-1}$). While the large widths and shifts may be real, it is also
quite possible that the line is blended with other emission lines.  An
attractive possibility are the ``unexpected UV'' lines at 1870\AA\/
and 1873\AA\/ resulting from Ly$\alpha$ pumping of \ion{Fe}{2}
\citep[e.g.,][]{rudy00}.    We note that there seems to be a small blue
shift from the marked rest-wavelength position in I~Zw~1 for both 
\ion{Fe}{2} UV 191 and \ion{Al}{3} \citep{laor97b}.

We also notice an excess in the spectrum at 1750\AA\/ that could be the
resonance transition of \ion{Ni}{2}.  In most AGN, that line would be
hidden under the \ion{N}{3}] multiplet; however, in this object, since
other semiforbidden lines are conspicuously weak, there is no
reason to suspect that \ion{N}{3}] is present.

As will be discussed in \S 3.3.2, we infer prominent \ion{Si}{2}
emission from multiplets near 1263\AA\/ and 1194\AA\/.  As discussed in
\citet{baldwin96}, if these emission lines are produced by collisional
excitation, we expect to see emission from lower-lying resonance
transitions near 1815\AA\/ and 1531~\AA\/, neither of which are
observed.  The 1531~\AA\/ line is difficult to identify,
because it is hidden under \ion{C}{4}.  We obtain an upper limit to
the 1815\AA\/ line by fitting with the \ion{C}{4} template, described
in \S 3.3.1, fixed at the expected wavelength.  We then vary the flux
of the template until we  obtain $\chi^2$  larger by 6.63, which
corresponds to 99\% confidence for one parameter of interest.  We note
that our upper limit is conservative; because the spectrum is complicated
in this region, we are forced to constrain the fit over only a narrow
bandpass and therefore cannot constrain the continuum very well.

\subsection{The Far-UV Spectrum}

The far UV continuum and emission lines are both unusual in PHL~1811.
Typical quasars show a break to a flatter spectrum at $\sim
1000$\AA\/, as is seen in a composite spectrum constructed from
{\it HST} spectra of quasars \citep{zheng97,telfer02}.  A break in
this vicinity is also observed in the spectral energy distributions of
individual low-redshift objects \citep{shang05}, which means that the
break cannot be a consequence of Ly$\alpha$ forest absorption.  Fig.\
8 shows the broadband continuum spectrum of PHL~1811 in comparison
with the Francis composite spectrum \citep{francis91}. We see no
evidence for a break down to 970\AA\/, the extent of the {\it HST}
spectrum.   The continuum is possibly a power law over the entire
optical-UV bandpass, with a large contribution from  \ion{Fe}{2} and
\ion{Fe}{3} between 2000 and 3200\AA\/.  Alternatively, a rollover
might be present at $\sim 2500$\AA\/, a longer wavelength than usual.

Ly$\alpha$ is generally a strong and prominent line in AGN spectra.
Since it is a recombination line of hydrogen, it is expected to be
strong in photoionized gas.  The equivalent width of Ly$\alpha$ plus
\ion{N}{5} in composite spectra is found to be $\sim 50$ to $\sim 90$
\AA\/ \citep{francis91,zheng97,brotherton01}.  In addition, \ion{C}{4}
is generally observed to be very strong as well. \ion{C}{4} is
expected to be strong because C$^{+3}$ is easily excited by collisions.
Composite spectra reveal a \ion{C}{4} equivalent width of 33--59\AA\/
\citep{francis91,zheng97,brotherton01}.  In PHL~1811, \ion{C}{4} and a
feature near Ly$\alpha$ are observed, but their equivalent widths are
small (Table 2; see below).

Although PHL~1811 has an unusual far-UV spectrum, it is not a unique
object.  Fig.\ 9 shows the spectra from two other
high-luminosity narrow-line quasars: RX~J0134.2$-$4258
\citep{goodrich00} and PHL~1092 (Leighly et al.\ in prep.).  All three
objects show low equivalent-width emission lines and blue far-UV
spectra.  For comparison, the {\it HST} composite spectrum
\citep{zheng97}, the spectrum from the NLS1 1H~0707$-$495
\citep{lm04,leighly04}, and a spectrum from the well-studied
broad-line Seyfert 1 galaxy, NGC~5548 \citep{korista95} are shown.

\subsubsection{The \ion{C}{4} Line, \ion{He}{2}~$\lambda 1640$, and
  1400 \AA\/ Feature} 

Fig.\ 10 shows the region of the spectrum that includes the \ion{C}{4}
line.  This line was somewhat difficult to analyze, because the
continuum was difficult to identify.  However, there is a relatively
flat region longward of the \ion{C}{4} line that stretches to
1800\AA\/ that we could plausibly identify as the continuum.  

As discussed in \citet{lhhbi01}, and in Paper I, PHL~1811 is X-ray
weak.  \citet{blw00} investigated soft X-ray weak objects among the
\citet{bg92} sample of PG quasars.  They found that soft X-ray weak
objects frequently show significant \ion{C}{4} absorption lines; in
fact, they showed that there is a trend for the most X-ray weak objects
to have the largest equivalent-width absorption lines.  They postulated
that soft X-ray weakness is caused by photoelectric absorption of the
X-rays by the same gas that is responsible for the \ion{C}{4}
absorption lines.  Indeed, in many cases, deep X-ray observations find 
that the signature of absorption is present in the X-ray spectra of
soft X-ray weak objects \citep{gall02}.  Since PHL~1811 is X-ray weak,
we might expect to observe \ion{C}{4} absorption lines, and based on
\citet{blw00}, we might expect an absorption line equivalent width of
up to 20\AA\/.  But as seen in Fig.\ 10, although the spectrum is
noisy, no evidence for a \ion{C}{4} absorption line is
seen.  There are negative fluctuations at the positions of resonance
lines from gas in our Galaxy, namely \ion{Si}{2}~$\lambda 1808$ (seen
at 1516.8\AA\/) and possibly from \ion{Al}{3}~$\lambda\lambda 1854.7$,
1862.8 (seen at 1556 and 1562.8 \AA\/).  There is another negative
fluctuation at 1529.3 \AA\/ that we identify as \ion{C}{4} absorption
from the $z(abs)=0.17651$ intervening system \citep{jenkins03}.
Thus PHL~1811 is not a typical soft X-ray weak object, and 
therefore we do not conclude that the soft X-ray weakness is a
consequence of absorption.

Fig.\ 10 shows the \ion{C}{4} emission line is blueshifted and
asymmetric.  Following the procedure discussed in \cite{lm04}, we
create a template from the \ion{C}{4} line. The template fit, and the
separate components, are shown overlaid on the \ion{C}{4} line.  The
shape of the PHL 1811 template is quite similar to the shape of the
templates obtained IRAS~13224$-$3809 and 1H~0707$-$495; given the
lower equivalent width of the \ion{C}{4} line, and noiser spectra,
they are essentially identical.  To illustrate this, we overlay the
IRAS~13224$-$3809 profile on Fig.\ 10.

The equivalent width of the \ion{C}{4} line is exceptionally low; we
measure it to be just 6.6\AA\/.  As discussed in \cite{lm04},
\ion{C}{4} equivalent widths are much larger in composite spectra;
\cite{francis91}, \citet{zheng97}, and \cite{brotherton01} measured
\ion{C}{4} equivalent widths of 37, 59, and 33\AA\/
respectively. Fig.\ 8 compares the PHL~1811 \ion{C}{4} line with that
from the Francis composite \citep{francis91}; it is a factor of $> 5$
smaller.  The two NLS1s, IRAS~13224$-$3809 and 1H~0707$-$495,
analyzed in \cite{lm04} had low equivalent width of 13--15\AA\/; the
\ion{C}{4} equivalent width in PHL~1811 is about half that.
\cite{leighly04} found that the weak \ion{C}{4} lines in those two 
NlS1s, along with other unusual emission-line ratios, could be
explained by a combination of high metallicity and an X-ray weak
continuum.  In \S 4 we show that the soft SED is responsible for the
weak \ion{C}{4} in PHL~1811. 

PG~1407$+$265 is a quasar that has low-equivalent width
high-ionization lines \citep{mcdowell95}.  \citet{blundell03} report
the observation of a weak relativistically-beamed radio jet, and infer
that the object is viewed face on. They then speculate that the
low-equivalent width emission lines are a consequence of the face-on
viewing angle, because they are diluted by the strong accretion disk
emission.  This reasoning has been recently applied to another
weak-line quasar, HE~0141$-$3932 \citep{reimers05}.

While PHL~1811 may well be observed face-on, that cannot be the reason
why the line equivalent widths are so low.  First of all, one may not
be able to achieve a very large boost in continuum emission by viewing
an object face on. \citet{ln89} show that the ionizing flux from an
accretion disk has nearly $\cos(\theta)$ dependence.  Thus, assuming
that  the typical viewing angle to the accretion disk is $\sim
30^\circ$, then  one gains only a factor of 1.15 by moving to a normal
viewing angle.  This is far from sufficient to explain the factor of
$\sim 5$ difference between the equivalent widths of \ion{C}{4} in
PHL~1811 and the  quasar composite spectrum.  Of course, this analysis
assumes that the continuum is not relativistically beamed; there is no
reason to expect that it is, because although it is a radio source, it
is a very weak one, and it is not radio-loud.  

Another argument against continuum boosting being responsible for the
low high-ionization line equivalent widths is illustrated in Fig.\
8. This figure shows that PHL 1811 has a continuum slope quite
similar to that of an average quasar; simultaneously the H$\beta$ line
has a normal equivalent width and the \ion{C}{4} has a very low
equivalent width. The reasoning goes as follows.  First, if the
continuum were uniformly boosted, one would expect both lines to have
low equivalent widths; yet we see that H$\beta$ has a normal
equivalent width, while the \ion{C}{4} equivalent width is low.
Second, if the continuum were boosted preferentially in the blue
(e.g., a super-Eddington disk may allow radiation to escape more
easily along optically thin pores or channels
\citep[e.g.,][]{begelman02} in the hotter regions of the disk), then
one would expect a bluer continuum than the average quasar.   Instead,
we see that PHL~1811 has the continuum of a normal quasar, at least
between \ion{C}{4} and H$\beta$.

We do not see any evidence for \ion{He}{2}~$\lambda 1640$.  To estimate
the upper limit of this line, we fit a \ion{C}{4} template shifted to
the appropriate wavelength for \ion{He}{2}. The best fit does not
detect this line, so we estimate the upper limit by increasing the
normalization until the $\chi^2$ has increased by 6.63 compared with
the best fit (99\% confidence for one parameter of interest).

The 1400\AA\/ feature is difficult to analyze.  First, the continuum
slope appears to change at $\sim 1470$\AA\/, although the continuum
blueward of the 1400\AA\/ feature appears to be flat and to
extrapolate to the continuum shortward of the Ly$\alpha$ feature.
Second, the signal-to-noise ratio in that part of the spectrum is not very
good.  Finally, absorption lines from Galactic \ion{Al}{2}~$\lambda
1670.79$ and from \ion{C}{4} in the z=0.08093 damped Ly$\alpha$ system
\citep{jenkins03} cut through the middle of the feature.  This feature
is difficult to model, regardless, because it is usually a
blend of the two multiplet lines of \ion{Si}{4} and the five multiplet
lines of \ion{O}{4}]. Here, based on the lack of semiforbidden lines in
the rest of the spectrum, it might be safe to assume that the entire
feature is \ion{Si}{4}.  However, in \citet{leighly04} it was inferred
that \ion{O}{4}] participated in the wind in IRAS~13224$-$3809 and
1H~0707$-$495, and therefore was emitted in a different component than
the intermediate-ionization lines such as \ion{C}{3}] that are so
clearly absent in PHL~1811.   To emphasize the uncertainty in the
identification of this feature, we refer to it as ``unidentified'' in
Table~2.    To estimate the flux and  equivalent width, we fit the
entire feature with a Gaussian.  The centroid is measured to be 1395
\AA\/, which possibly suggests an origin in \ion{Si}{4}, which has a
$gf$  weighted average wavelength of 1396.75 \AA\/.

\subsubsection{The Ly$\alpha$/\ion{N}{5} Region}

The far-UV spectrum of PHL 1811 is cluttered with absorption lines
that originate in intervening absorption systems \citep{jenkins03}.
In order to better see the shape of the far UV spectrum, we 
model out the absorption lines first.  Using \citet{jenkins03} as a guide,
we fit local continua around each absorption line, and then model each
absorption line as a Gaussian.  A comparison of the portion of the
spectrum around Ly$\alpha$ with and without the absorption lines
removed is seen in the top panel of Fig.\ 11.  

The Ly$\alpha$/\ion{N}{5} feature in PHL 1811 appears as a lump
stretching from $\sim 1140$ to $\sim 1270$\AA\/ superimposed on a
rising continuum that appears to be linear or a power law in the range
1090--1140 \AA\/ and 1320--1360 \AA\/.  The equivalent width of the
whole feature is only 24\AA\/, which is small compared with the
50--90\AA\/ measurements of Ly$\alpha$+\ion{N}{5} from composite
spectra mentioned above.

It is not obvious at first glance what the various bumps in the
Ly$\alpha$/\ion{N}{5} feature are.  There are peaks that could
correspond to \ion{Si}{2} multiplets near 1193, 1260 and 1304 \AA\/,
and peaks that may be \ion{N}{5} and Ly$\alpha$.  We apply the
approach previously used by \citet{lm04}, and fit these features using
the \ion{C}{4} template profile described in \S 3.3.1.  A difference
in the  analysis presented here is that we now use the IRAF
{\tt Specfit} program which can fit  an input profile, and properly minimizes 
$\chi^2$ to obtain the normalizations and spectral shifts.

 We first determine the continuum by fitting  the line-free bands
1086--1103, 1116--1137 and 1314--1350 \AA\/ with a linear model.   
Initially, we assumed that the feature is composed of the five lines
described above, and each one has the \ion{C}{4} template profile.
We construct profiles for each multiplet, assuming first that the
gas is optically thin and the multiplet ratios are proportional to the
$gf$ values, which were obtained from the NIST
database\footnote{http://physics.nist.gov/cgi-bin/AtData/main\_asd}.
Initially, we fix the wavelengths of the profiles at the rest 
wavelengths, this implicitly assuming that gas emitting all the lines
has uniform kinematics.  The resulting fit is shown in the middle panel of Fig.\
11.  Many of the bumps are modelled, but not well.  Specifically, the
\ion{N}{5} emission appears to be insufficiently blueshifted, and
there is extra emission blueward of \ion{Si}{2}~$\lambda 1193$.
Emission in this region has been previously identified as
\ion{C}{3}$*\lambda 1175$ \citep[e.g.,][]{laor97b}, so we assume that
as the tentative identification of this excess.

We also try profiles with all of the lines in the multiplets having
equal strength, as would be appropriate if the gas were very optically
thick. This model yielded a slightly larger $\chi^2$ but was not clearly
distinguishable from the optically-thin case.  We therefore proceeded
to use the optically-thin profiles.

We next apply the same model but add another template component around
1175 \AA\/ to model the suspected \ion{C}{3}$*$ emission, and allow the
central wavelengths of the profile components for each line to be
free in the fitting.  We obtain a much better fit, as seen in the
lower panel of Fig.\ 11.  The wavelengths, fluxes and equivalent
widths of this model are given in Table 2.  We find an excellent
correspondence between the laboratory wavelength and the observed
wavelength of the template fits of \ion{Si}{2}~$\lambda 1193$,
Ly$\alpha$ and \ion{C}{3}$*$; the remaining lines are somewhat
blueshifted.  The blueshift on \ion{Si}{2}~$\lambda 1308$ is
conceivably partially due to an incomplete or faulty subtraction of a
prominent absorption line.  The blueshift on \ion{N}{5} is plausibly
real, as it might be expected that higher ionization lines may be
emitted by gas with a higher velocity; this may be evidence that our
template model is too simple. 

The presence of \ion{C}{3}$*\lambda 1175$ is interesting.  This line
is emitted as a transition to a metastable state that is the upper
level for the \ion{C}{3}]~$\lambda 1909$ line.  \citet{laor97b} discuss
possible reasons for a strong \ion{C}{3}$*$ line in the prototype
Narrow-line Seyfert 1 galaxy I~Zw~1; we also tentatively identified
this line in two other NLS1s \citep{lm04}.  The presence of \ion{C}{3}$*$
suggests that there might be a similar transition from the metastable
upper level of \ion{Si}{3}]~$\lambda 1892$ line, since Si$^{+2}$ and
C$^{+2}$ are isoelectronic.  That transition should appear in the
region 1108.4--1113.2\AA\/ (the $gf$-weighted average 
wavelength is 1111.59\AA\/).  Indeed, there is a bump clearly present
at these wavelengths that we confidently identify as
\ion{Si}{3}$*\lambda 1112$.   Interestingly, it is not well fit by
the template profile; we model it using first a single Gaussian, and
then by the six components of the multiplet assuming optically thin
ratios.  For the second model, the width is $690 \pm 130 \rm \,
km\,s$ (Table 2), and it is marginally-significantly blueshifted by
$290 \rm \, km\, 
s^{-1}$.  

The successful deconvolution of the Ly$\alpha$/\ion{N}{5} lump into
these six components using the \ion{C}{4} profile as a template may be
fortuitous, and  we caution that it is certainly not unique
and may not be complete.  The presence of the metastable transitions
of C$^{+2}$ and Si$^{+2}$ led us to search for other similar
metastable lines.  On Fig.\ 1, we mark the positions of lines that
have lower levels that are metastable, so that the transitions to
ground state are semiforbidden, as described in \S 2.     In the
region of the Ly$\alpha$/\ion{N}{5} lump, there are a couple of
\ion{Si}{2}* transitions that could be strong.  We also mark
\ion{Si}{3}~$\lambda 1206.5$, a resonance line that may also
contribute; this line is analogous to \ion{C}{3}~$\lambda 977$.

Strong high-excitation  \ion{Si}{2} has been previously reported by
\citet{baldwin95}.  In \S 3.2.3, we obtained an upper limit on
\ion{Si}{2}~$\lambda 1814$, and using the results of the deconvolution
presented in this section, we measure a lower limit on the
\ion{Si}{2}~$\lambda 1263$/\ion{Si}{2}~$\lambda 1814$ of 8.5, even
larger than the value of 6 that was obtained by \citet{baldwin95} from
Q0207$-$398. \citet{baldwin95} showed that such high ratios are hard to
explain, if the gas is collisionally excited, as one expects that the
lower excitation line to be stronger.  They suggested that the
high-excitation transitions are selectively excited.  The key seems to
be that far UV lines are predominately excited by continuum pumping,
generally speaking. \citet{bottorff00} explored the effect of
microturbulence on line ratios in AGN, and showed that since pumped
lines are enhanced by turbulence, large \ion{Si}{2}~$\lambda
1263$/\ion{Si}{2}~$\lambda 1814$ ratios can be produced if   
the turbulent velocity is high.  Finally, \citet{clb05} investigated
\ion{Si}{2} emission as a function of the shape of the ionizing
continuum in their Appendix.  They note that high-excitation
\ion{Si}{2} emission is stronger for softer continua, and that the
{\it Cloudy} output information indicates that for soft 
continua, continuum pumping is important.  

In \S 3.3.1, we reported an upper limit on \ion{He}{2}~$\lambda 1640$,
and comparing with \ion{N}{5} measured from the deconvolution
presented here, we obtain a lower limit on the \ion{N}{5}/\ion{He}{2}
ratio of 11.6.  \citet{ferland96} reported \ion{N}{5}/\ion{He}{2} $>10$ for
``Component B'' (the outflowing component) in Q0207$-$398.  \ion{N}{5}
and \ion{He}{2} are both high-ionization lines from ions that require
similar ionization potentials (He$^{+2}$: 54.4 eV; N$^{+4}$: 77.5 eV);
therefore, they should be emitted by gas with similar properties.   A
thorough analysis carried out by \citet{ferland96} showed that the only way
that the high \ion{N}{5}/\ion{He}{2} ratio can be produced is if the
metallicity is enhanced by a factor of 5.  A large
\ion{N}{5}/\ion{He}{2} ratio in PHL~1811 may likewise imply a high
metallicity also, although we again note the uncertainty in the
deconvolution. 

\subsubsection{Other FUV lines}

There are a few relatively narrow, weak lines in the FUV that we try
to identify.  We were able to identify \ion{N}{2}~$\lambda 1085$,
which is a resonance line of \ion{N}{2}.  We model it with a single
Gaussian, and also with the six Gaussians that comprise the multiplet.
The latter fit yields a rather narrow line width, only $370 \rm \,
km\, s^{-1}$.   We also see another narrow line near 1034\AA\/,
although this line is uncertain because considerable reconstruction
had to be performed in the vicinity of this line.  We were unable to
confidently identify this line. It seems to be too centered and too
narrow to be associated with \ion{O}{6}~$\lambda 1032,1038$.  

\ion{C}{2}~$\lambda 1335$ is commonly seen in AGN, but we see no hint
of it although it occurs in a relatively line-free region of the
spectrum. To obtain an upper limit on the flux of this line, we add a
\ion{C}{4} template component for this line to the FUV model, and fit
over the entire region from 1086--1365\AA\/; we also include a short
strip of continuum between 1425 and 1462\AA\/. No significant emission
is detected, so we estimate the upper limit on the flux of \ion{C}{2}
by increasing the normalization until the $\chi^2$ is larger than the
best fitting value by 6.63 (99\% confidence for one parameter of interest). 

\section{Discussion}

The spectra and analysis presented in \S 3 reveal the
unusual emission-line properties of the narrow-line quasar
PHL~1811.  We briefly summarize the principal results here.   PHL~1811
has no forbidden or semiforbidden lines, and unusual very
low-ionization lines are present (e.g., \ion{Ca}{2} and \ion{Na}{1}D).
\ion{He}{1}~$\lambda 5876$ is very weak, and a very low
\ion{He}{1}/H$\alpha$ ratio is inferred.  The \ion{Fe}{2} and
\ion{Fe}{3} pseudocontinuum dominates the near-UV part of the
spectrum.  However, while the \ion{Fe}{2} in the majority of AGN is
fit well by the template developed from the high signal-to-noise {\it
 HST} observation of I~Zw~1, the pseudocontinuum in PHL~1811 has a
demonstratively different shape that we show is likely to be a
consequence of the presence of high-excitation \ion{Fe}{2} and strong
\ion{Fe}{3}.  Some of the other low-ionization lines are strong,
including  the high-excitation \ion{Si}{2} lines near 1194\AA\/ and
1263\AA\/, but other low-ionization permitted line emission (e.g.,
\ion{Mg}{2}) is weak or absent (e.g., \ion{C}{2}~$\lambda
1335$). While the hydrogen Balmer lines appear to have rather typical
equivalent widths, the equivalent widths of the high-ionization lines 
are small, and in addition, the high-ionization lines have a
blue-shifted profile.  Deconvolution of a feature near Ly$\alpha$ in
terms of a template developed from the blueshifted \ion{C}{4} line
reveals strong \ion{N}{5} and high-excitation \ion{Si}{2}, as well as
\ion{C}{3}*~$\lambda 1176$.   

The goal of this section is to interpret these results.  But first, we
summarize previous work that has addressed some of the results.  We
note that previous work has been mentioned throughout \S 3,
but we summarize the major points here also.  The production 
of \ion{Na}{1}D has been discussed by \citet{thompson91}, who inferred
that since the ionization potential of Na$^+$ is only 5.14~eV, this
line must be produced very deep in the partially-ionized region where
neutral sodium can survive. The production of  \ion{Ca}{2} has been
discussed by \citet{persson88, fp89} and \citet{joly89}.
Observationally, the infrared triplet is correlated with the optical
\ion{Fe}{2} \citep{fp89}, but the \ion{Ca}{2}~H\&K, the lines that we
report here, are rarely seen \citep{fp89}.  Both \citet{fp89} and
\citet{joly89} found that a high column density is necessary to produce
the infrared triplet.  Strong high-excitation \ion{Si}{2} was first discussed by
\citet{baldwin96}, who pointed out the difficulty of producing large
fluxes of the high-excitation lines such as \ion{Si}{2}~$\lambda
1263$, but at the same time producing very low fluxes of the resonance
line \ion{Si}{2}~$\lambda 1814$.  \citet{bottorff00} showed that this
problem may be solved if the emitting gas is turbulent, as the high
excitation \ion{Si}{2} lines are predominately pumped by the
continuum and will be selectively excited by turbulence.  Blueshifted
high-ionization lines have been observed in other NLS1s such as
IRAS~13224$-$3809 and 1H~0707$-$495 \citep[e.g.,][]{lm04}. The
\ion{C}{4} lines in these two objects also have relatively low
equivalent widths, especially compared with the \ion{N}{5} and the
1400\AA\/ feature. \citet{leighly04} showed that the gas emitting
these lines is optically thin to the continuum, and that the line
ratios can be explained by a somewhat X-ray weak spectral energy
distribution and elevated gas metallicity.  We do not detect
\ion{He}{2}~$\lambda 1640$ and thus infer a large
\ion{N}{5}/\ion{He}{2} ratio.  \citet{ferland96} discussed the large
\ion{N}{5}/\ion{He}{2} ratio in Q0207$-$398, and found that the only
way that this ratio can be explained is if the metallicity were enhanced
compared with solar by $Z\sim 5 Z_\odot$.  

What is the origin of the unusual emission-line properties in
PHL~1811?   PHL 1811 has one  property that is much different than an
ordinary quasar: it is intrinsically X-ray weak  (Paper I), and
therefore has an unusually soft spectral energy distribution. In the
next several subsections, we examine the physics of gas illuminated by
a very soft spectral energy distribution. We find that  gas
illuminated by a very soft SED has much different physical properties
than gas illuminated by a typical AGN SED, and that most of the unusual
observational features of PHL 1811 can be explained qualitatively by
the soft SED {\it alone}.  The low  \ion{He}{2}/H$\alpha$ ratio cannot
be explained by the soft SED, but we find that it can be explained by
a ``filtered'' continuum; we discuss this idea in \S 4.4.  Finally, in
\S 4.5 we discuss whether there may be other objects like PHL~1811,
and how we might identify them.   

\subsection{Physical Conditions in Gas Illuminated by a Soft SED}

In this section, we investigate the physical conditions and line
emission from gas illuminated by an  X-ray weak spectral energy
distribution.  Our goal is physical insight rather than a detailed
match to the line emission, so we begin by exploring the physics of
gas illuminated by a range of spectral energy distributions for a
representative value of the ionization parameter ($\log U=-1.5$) and  
density ($\log n=11$).  These correspond to a photoionizing flux of
$\log \Phi = 20$.  We use a hydrogen column density of
$\log(N_H)=24.5$, which corresponds to a thickness more than 100 times the
depth of the hydrogen ionization front.  We now define some terms.
We refer to the region between the illuminated surface 
of the gas and the hydrogen ionization front as the \ion{H}{2} region,
and the region at greater depths than the hydrogen ionization front as
the partially-ionized zone.   In addition, we discuss the ionization
potential to create emitting ions, and we refer to this as
``ionization potential'', whereas more correctly, it is the ionization
potential of the next lower charge state.

\subsubsection{The Continua}

We use a range of continua to explore the influence of the spectral
energy distribution.  We  start with the ones previously presented in 
in \citet{clb05} (hereafter referred to as the CLB
continua).  They are described in detail in the Appendix of that
paper; briefly, they constructed a number of spectral energy
distributions parameterized by the energy in eV of the UV bump
high-energy cutoff.  
They assumed that the temperature of the cutoff $T_{cut}$, which
corresponds roughly to the inner edge of the accretion disk, is
related to the luminosity by $L\propto T_{cut}^{-4}$.  Then, to
determine the X-ray power law normalization, they assumed that the UV
monochromatic luminosity at 2500\AA\/ is related to the X-ray
monochromatic luminosity at 2 keV by $\alpha_{ox}$, with $\alpha_{ox}$
dependence on UV luminosity given by the regression developed by
\citet{wilkes94}.  Finally, the proportionality  constants were
determined using the measured $T_{cut}$ from coordinated {\it IUE} and
{\it ROSAT} observations of the quasar 3C~273 \citep{walter94}.  For
more details, see the Appendix of \citet{clb05}.

We also use the PHL~1811 SED developed from
the {\it HST} and the brighter of the two {\it Chandra} observations,
supplemented by a {\it FUSE} point from a non-simultaneous 
observation, and using the {\it Cloudy} AGN continuum at higher and
lower energies where we have less information (Fig.\ 12).  We do not
know the shape of the extreme UV, so we simply join  the {\it FUSE}
point and the {\it Chandra} spectrum using a power law.     We also
use the {\it Cloudy} AGN continuum with parameters as used by Korista
et al. (1997); hereafter referred to as ``K97'', and also shown in
Fig.\ 12, this is taken to be a typical AGN continuum. 

Fig.\ 13 gives some information about the properties of these
continua.  The CLB continua are coded on the x-axis by their cutoff
temperature, and the other two files are named ``PHL1811'', and
``K97'' for the PHL 1811 and Korista et al.\ continua, respectively. 
The top panel shows the $\alpha_{ox}$ for the continua.   As noted in
Paper I, PHL~1811 is anomalously X-ray weak and has a very steep
$\alpha_{ox}=-2.3 \pm 0.1$.  In comparison, the $\alpha_{ox}$ for the
K97 spectrum is $-1.4$. 

In the next panel, we plot the continuum flux at 2500\AA\/ assuming
typical values of the input ionization parameter ($\log U=-1.5$) and
density ($\log n=11$).  These values correspond of course to the same
value of the  ionizing photon flux, but the continuum flux in the
optical or UV is much different. Specifically, an X-ray weak
continuum must be brighter in the optical and UV in order to produce
the same ionizing flux, since the ionizing flux is integrated from
13.6~eV toward shorter wavelengths.   This will influence the
equivalent widths of the lines produced; specifically, for objects
that have soft continua like PHL~1811, the lines will appear to have
lower equivalent width for the same global covering factor compared
with an ordinary AGN continuum. The PHL~1811 continuum is 3.5 times
brighter than the K97 continuum at  2500\AA\/.  This is certainly one
factor contributing to the low equivalent widths of the line emission
observed in PHL~1811.    

In the bottom panel of Fig.\ 13, we show the Compton temperature,
which is a measure of the mean photon energy of the continuum.  
The gas would reach the Compton temperature if it were in equilibrium 
with the radiation field.   The Compton temperature is
very low for the PHL~1811 continuum, around $5\times 10^4\rm K$; this
temperature is similar to the typical temperature of a photoionized
gas.   

\subsubsection{Temperature and Hydrogen Ionization}

We first examine the temperature structure of the ionized gas (Fig.\
14).  The drop in temperature at around 1--2$\times 10^{11}\rm \, 
cm$ occurs at the hydrogen ionization front; the kink at shallower
depths occurs at the helium (He$^{+2}$ to He$^+$) ionization front.
As might be expected, the very soft continuum from PHL~1811, and
accompanying low Compton temperature, produce a rather cool
temperature in the photoionized gas.  The temperature in the
partially-ionized zone for the PHL~1811 is exceptionally low; we will
return to this point later.  In contrast, the harder continua produce
significantly higher temperatures, especially in the \ion{H}{2} region.   

Next, we examine the fraction of ionized hydrogen as a function of
depth (Fig.\ 15).  The hydrogen ionization front is slightly deeper
for the harder SEDs, perhaps marking additional ionization associated
with the presence of soft X-rays, higher Compton temperature and
greater heating.  The fraction of ionized hydrogen in the partially
ionized zone follows the temperature for the harder SEDs; that is, at
a particular depth, the ionization fraction is higher for harder SEDs.
That breaks down at the softest SEDS, where the ionization fraction
rises again.  For  the PHL 1811 continuum, the ionization fraction is
higher than for many of the harder SEDs.  

To understand the origin of the high ionization fraction in the
partially-ionized zone of the gas illuminated by the PHL~1811
continua, we review the physical mechanisms that produce ionized
hydrogen in the partially-ionized zone.  There are two principal
mechanisms \citep[e.g.,][]{cd89, ferland99}.  The first mechanism
involves X-ray photons that photoionize heavy elements through
inner shell interactions creating secondary nonthermal electrons that
ionize  hydrogen.  These electrons also heat the gas.  The second  
mechanism involves photoionization of hydrogen in excited states
(e.g., $n=2$) by the direct continuum and the diffuse continuum.  The
hydrogen is in an excited state due to a circular
process. Recombining hydrogen produces Ly$\alpha$; the optical depth
to Ly$\alpha$ in the partially-ionized zone is very high, so the  
Ly$\alpha$ photons scatter many times before escaping the gas.  That
scattering may leave some hydrogen atoms in $n=2$ long enough that
they can be photoionized by the  Balmer continuum of the
incident quasar continuum.  
   
To investigate the importance of the latter process as a function of
spectral energy distribution, we examine the rate of photoionization
from $n=2$ per hydrogen atom in $n=2$ by solving the radiation
transport equation at any depth in the gas.  We compute the
photoionization rate from  the incident continuum, attenuated by
$e^{-\tau_\nu}$, where $\tau_\nu$ can be obtained using the extinction
coefficient  which can be output from {\it Cloudy}.  We also compute
the photoionization rate from the diffuse continuum, where the diffuse
continuum  $I_\nu$ is obtained by integrating the source function;
this can also be output from {\it   Cloudy}.  The resulting rates are
shown in Fig.\ 16.  For comparison, we show the collisional ionization rate.   

Fig.\ 16 shows first that collisional ionization is practically
negligible in the partially ionized zone.  This is not surprising,
since the temperature of the gas is a fraction of an eV.  Of the
photoionization rates, the contribution from the direct continuum is
more important than that from the diffuse continuum, but the
diffuse continuum contributes most to the harder continua, from, e.g.,
$\sim 15$\% for the K97 continuum to $\sim 40$--$60$\% in the very
hardest continua.  Considering the contribution from the direct continua alone, we
see that the photoionization rate from $n=2$ is highest for the
softest continua.   This is simply a consequence of the fact that for
a particular ionizing flux, the flux density in the optical and UV
will be higher for the soft continua (Fig.\ 13; \S 4.1.1).  By itself, this
predicts that the softest continua would produce the highest
ionization fractions in the partially ionized zone.  This is not what
we see, so the difference must lie in the number of hydrogen in $n=2$.  

The density of hydrogen in $n=2$ can be output from {\it Cloudy} also,
and we show it in left panel of Fig.\ 17.  We see indeed that in the
partially-ionized zone, this density is higher in objects with the
hardest SEDs. However, this difference is insufficient to explain the
dependence of the hydrogen ionization on the SED.  For example, at a
depth of $5\times 10^{11}\rm\, cm$, the hydrogen fraction in gas
illuminated by the PHL~1811 continuum is almost the same as in gas
illuminated by  the $kT=320\rm \,eV$ SED.  At that depth, the total
photoionization rate is a factor of five larger for PHL~1811, but the
density of hydrogen in $n=2$ is only a factor of 1.5 larger for the
$kT=320\rm \, eV$ SED.  Thus, the balance of hydrogen ionizations in gas
illuminated by the hardest continua come from the first process
mentioned above, the initial inner-shell X-ray photoionization and
resulting photoelectrons.   

Next, we investigate the origin of the hydrogen in $n=2$.  We plot
the ratio of the excitation temperature based on the 
populations in $n=1$ and $n=2$ to the electron temperature (Fig.\ 17).
The ratio exceeds 1 in the partially-ionized zone, except for the gas
illuminated by the hardest continua, which means that the temperature
is not sufficiently high to explain the population in $n=2$.
Therefore, the radiation field must be responsible, and indeed,
scattering of trapped Ly$\alpha$ mentioned above is likely to be the
origin of hydrogen in $n=2$ in gas illuminated by the PHL~1811
continuum.  

The surprising aspect of Fig.\ 17 is that the gas illuminated by the
PHL~1811 continuum exhibits a very large ratio, indicating an
exceptionally high ratio of hydrogen in $n=2$ compared with that
expected from a thermal gas.  This trend continues to even higher
$n$; in Fig.\ 18 we show the ratio of the density of hydrogen in $n=5$
from the {\it Cloudy} output to the ratio expected based on the
density in $n=1$ and the electron temperature; this parameter is
proportional to the departure coefficient for $n=5$.  This shows that
departure coefficient is higher in gas illuminated by the PHL~1811
continuum by nearly three orders of magnitude compared with gas
illuminated by the hardest continua.  

The origin of the excited-state hydrogen in PHL~1811 is the strong
diffuse continuum.  In Fig.\ 19 we show the integrated diffuse
continuum (i.e., energy in the diffuse continuum) multiplied by the
depth, as a function of depth.  The energy in the diffuse continuum is
much larger for gas 
illuminated by the PHL~1811 continuum.  Note that this does not
necessarily mean that we expect to see a strong continuum emission
from this object, as the diffuse continuum used in this calculation is
the emission at that depth, and the radiative transfer to the surface
has not been folded in.  This fact also explains the low electron
temperature in the partially-ionized zone; the energy is locked up in
the radiation field and excited states of hydrogen rather than
electron kinetic energy.  Thus, in PHL~1811, this strong continuum
scatters in the partially-ionized zone, causing the hydrogen to be in
high-excitation states.  We will see why the continuum is so strong in
the next section. 

To summarize this section, we find that the hydrogen in the
partially-ionized zone in gas illuminated by the PHL~1811 continuum is
rather highly ionized.  This is a consequence of the bright Balmer
continuum for a given photoionizing flux for soft SEDs, and the
relatively large density of hydrogen in $n=2$, which is a consequence
of strong diffuse continuum.  As will be shown in the next section,
the low temperature means that the gas cannot cool by emission of
collisionally-excited lines, so the energy is deposited in strong 
continuum and excited states of hydrogen, resulting in the relatively
low gas temperature.     

\subsubsection{Metal Ions: General Considerations}

The conditions in the line emitting gas can be understood further by
examining the behavior of the metal ions.  In Fig.\
20 we show the ratios of the column densities of the ions of interest
for the PHL~1811 continuum to the K97 continuum as a function of the
ionization potential to create each ion.   These ratios show that for ionization
potentials less than $\sim 10\rm \, eV$, the column densities are
approximately the same; the high point is Ca$^+$, in which the column
is a factor of 2.3 times greater in PHL~1811.  For ionization
potentials greater than 20 eV, the column densities are considerably 
lower in the gas illuminated by the PHL~1811 continuum. This is not
surprising because the PHL~1811 continuum lacks high energy photons
required to create these ions. For ionization potentials between 10
and 20 eV, the columns in PHL~1811 are higher than in K97.  This also
makes sense because these are the ions that occupy the \ion{H}{2}
region, instead of the high-ionization species that are usually
there. 

One might expect that if the column density of an ion is larger, the
line emission from that ion must also be larger.  Fig.\ 20 shows the
ratios of predicted emission-line strengths from gas illuminated by
the PHL~1811 continuum compared to those from gas illuminated by the
K97 continuum.  The ratio is smaller for the high-ionization lines, as
might be expected; those ions are few in the gas illuminated by the
PHL~1811 continuum, so the high-ionization line emission is
weak. However, for the low-ionization lines, the situation is much
different, and in fact lines from the same ion exhibit different
ratios. Some lines are stronger in PHL~1811; for example,
\ion{Si}{2}~$\lambda 1194$ is predicted to be 3.3 times larger for
PHL~1811.  Other lines from the same ion are weaker in PHL~1811; for
example, \ion{Si}{2}~$\lambda 1814$ is predicted to be 1.7 times
weaker.  This difference is a result of the different excitation
mechanisms for the lines.  \ion{Mg}{2}~$\lambda 2800$ and
\ion{Si}{2}~$\lambda 1814$ are transitions from the lowest permitted
levels; they are collisionally excited and their emission strengths
depend on the temperature of the gas.  Since the temperature is lower
in gas illuminated by the PHL~1811 continuum, the emission from these
lines is weaker.  In fact, it can be shown that the ratios of the
emission rates produced by the PHL~1811 continuum and the K97
continuum as a function of depth correspond fairly closely to the
ratios one would expect based on the difference in temperatures
through the Boltzmann excitation equation. This is demonstrated by the
close correspondence between the solid 
grey line and the dashed grey line for \ion{Si}{2}~$\lambda 1814$ in
Fig.\ 21 for large depths where Si$^+$ dominates.     In
contrast, high-excitation \ion{Si}{2}, such as \ion{Si}{2}~$\lambda
1194$, is predominately excited by pumping by the continuum.  As noted
in \S 4.1,1, the UV continuum will be relatively strong in PHL~1811
for the same ionization parameter and therefore, lines that are
excited by continuum pumping can be expected to be relatively
strong. This is demonstrated by the great discrepancy in the ratio
based on the {\it   Cloudy} emissivity output (solid black line) and
the ratio based on the Boltzmann equation (dashed black line)  in
Fig.\ 21.  

Fig.\ 20 shows that several bands of \ion{Fe}{2} behave differently
for the PHL~1811 continuum compared with the K97 continuum.  We
discuss the \ion{Fe}{2} further in \S 4.4.  In addition, we find that
the Fe$^{+2}$ ion column is elevated in PHL~1811, similar to the other
intermediate-ionization line columns, suggesting that \ion{Fe}{3}
should be strong.  We cannot check this suggestion, however, because
{\it Cloudy} does not model UV \ion{Fe}{3} lines, such as
\ion{Fe}{3}~UV~34, near 1900\AA\/, or \ion{Fe}{3}~UV~68, near
1950\AA\/.

It is important to note that common UV semiforbidden lines
(\ion{C}{3}]~$\lambda 1909$, \ion{Si}{3}]~$\lambda 1892$,
    \ion{N}{3}]~$\lambda 1750$, \ion{C}{2}]~$\lambda 2325$,
      \ion{N}{2}]$\lambda 2141$)  are suppressed
for the PHL~1811 continuum.  This is a consequence of two factors.  For
the twice-ionized ions, with the exception of Si$^{+2}$, the column of
the ion is lower in PHL~1811.  In addition, these lines are all
collisionally excited,  arising from ions that have two electrons in
their P-shell.  Thus, since the gas temperature is lower in PHL~1811,
these lines are weaker.  This result has a consequence for plasma
diagnostics. Semiforbidden lines are generally used as density
indicators, since they have critical densities in the range thought
typical of  broad-line region gas \citep[e.g., for the lines above,
  the critical densities range from $3.16 \times 10^9\rm \,cm^{-3}$
  for \ion{C}{2}~$\lambda 2325$ to $3.12 \times 10^{11}\rm \, cm^{-3}$
  for \ion{Si}{3}~$\lambda 1892$;][]{hkfwb02}.  This means that the
absence of these lines in objects with very soft continua like
PHL~1811 does not require the presence of high-density gas.

\subsubsection{Exploring the $\log U$, $\log n$ Parameter Space}

The discussion above considers only a single ionization parameter and
density;  next we examine the properties of the gas as a function of
these two parameters.    We perform simulations for the PHL~1811
continuum and for the K97 continuum for ionization parameter $-3.5
\leq \log U \leq 0.5$ and for density $8.5 \leq \log n \leq 12.5$.
In each case we  adjust the column density so that $\log N_H - \log U
= 26$; this ensures that the gas is very optically thick to the
continuum, and probes approximately the same thickness relative to the
hydrogen ionization front.

We first examine the column density of various ions as a function of
ionization parameter and density.  Since the total column density of
the gas varies with ionization parameter, we look at the column
density relative to the total column; this is equivalent to
$\log(N_{ion})-\log U$.    Generally speaking, the behavior of the
ionic column density for low and intermediate ionization metal ions
relative to the total column density was divided into two groups
according to ionization potential.  For the singly-ionized ions C$^+$
and N$^+$, the relative column was principally density independent,
and decreased as ionization parameter increased, as the ionization
state of the gas shifted to a higher ionization state.  O$^+$ looked
like H$^+$, most probably because the strong influence of the Bowen
effect.   In contrast, the ions  Mg$^+$, Al$^+$, Si$^+$ and Fe$^+$,
were almost independent of ionization parameter and density. 
In both cases, the ionic structure was nearly the same for the PHL~1811
continuum and the K97 continuum; this is not surprising because they
should both be roughly equally efficient in producing low-ionization
ions. 

For the twice-ionized ions, there are significant differences between the
gas illuminated by the PHL~1811 continuum and the K97 continuum.
First, there were differences in the total column densities.  For the
doubly-ionized ions with higher ionization potential (C$^{+2}$,
N$^{+2}$, and O$^{+2}$), the columns were higher for the K97
continuum.  In contrast, for the doubly-ionized ions with lower
ionization potential  (Al$^{+2}$, Si$^{+2}$, and Fe$^{+2}$), the
columns were higher for the PHL~1811 continuum.  This makes
sense, and it is something that we noted previously in CLB: for 
softer continua, the cooling in the region of intermediate ionization
shifts to ions that have lower ionization potential.  For these ions,
the dependence on ionization parameter and density were similar: there
is an ionization parameter at which the column is maximized, and at
that ionization parameter, the column decreases as a function of
density. This happens because at higher densities, recombination is
stronger, and the overall ionization state  of the gas is a little
lower.  At lower densities for the same ionization parameter, the
ionization shifted to the next higher state.  But for the
higher-ionization doubly-ionized ions, that structure was apparent
only for the K97 continuum; for the PHL~1811 continuum the column
increases with ionization parameter almost independent of density. 

More highly ionized ions (C$^{+3}$, N$^{+4}$, He$^{+2}$) were
significantly deficient over the entire parameter space for the
PHL~1811 continuum, as expected.  However, these ions were present at
highest ionization parameters.

We next look at equivalent widths of lines that are common in AGN.
Note that by discussing the equivalent width rather than the line
flux, we account for the fact that the optical and UV continuum
against which the lines are measured is brighter by a factor of $\sim
3.5$ in PHL~1811 for the same ionizing flux.  We first looked at lines
that had equivalent widths using the PHL~1811 continuum that were
predicted to be larger than or 
comparable to those obtained using the K97 continuum, both at our
fiducial point ($\log U=-1.5$ and $\log n=11$), and over the parameter
space.  The strongest lines were the UV \ion{Fe}{2} and the
high-ionization \ion{Si}{2}.  These are both among the strongest lines
in the observed spectrum, a fact that suggests that we are on the
right track.  But not all low-ionization lines were strong in
PHL~1811; notably weak ones were \ion{Mg}{2} and \ion{Si}{2}~$\lambda
1814$.  The difference again is the excitation mechanism: the
high-excitation silicon and iron lines are pumped by the continuum,
while the magnesium and low-excitation silicon are
collisionally-excited.  In addition, as expected, the high ionization
lines such as \ion{C}{4} were among the weakest in PHL~1811 compared
with K97 continuum.  We illustrate this in Fig.\ 22.  At the fiducial
point, the equivalent width of \ion{C}{4} is a factor of more than 100
times lower for the PHL~1811 continuum compared with the K97
continuum.

The soft continuum also affects the semiforbidden lines in interesting
ways.  Fig.\ 23 shows the equivalent width contours of
\ion{Si}{3}]~$\lambda   1892$ for the PHL~1811 continuum and the K97
  continuum.   This emission line has a critical density of $3 \times
  10^{11}\rm \,cm^{-3}$, so the decline toward  higher density at
  about this value for the K97 continuum is expected.  But the
  PHL~1811 continuum simulations show a different behavior; the line
  begins to decrease at much lower densities, around $10^{10}\rm \,
  cm^{-3}$.  Again, this means that the usual density indicators fail
  for soft continua, and the absence of semiforbidden lines does not
  require the presence of high-density gas. 

The weakness of the collisionally-excited metal lines in PHL~1811 has 
important consequences for the physical conditions in the gas.  If the
gas cannot cool by emission of collisionally-excited lines, and yet
there are plenty of photoionizing photons, the energy goes into the
production of hydrogen and other continua. This is what we see in
PHL~1811.  The situation is very similar to high-density gas
\citep[discussed by, e.g.,][]{rnf89}.  At high density, collisional
deexcitation of metal ions prevents the gas from cooling by
collisional-excitation of metal ions, and the high-density gas also
becomes dominated by continua.  In a sense, both types of gas are
``cooling challenged''.  They also have similar observational
properties, in particular, the weakness of the semiforbidden lines.  

To summarize this section, we find that properties of metal ions and
their emission in gas illuminated by the soft PHL~1811 continuum is
different from gas illuminated by the K97 continuum in the following 
ways.  The column densities and line emission from high-ionization
ions are lower in gas illuminated by the soft SED.  The column
density of the intermediate-ionization ions is larger, but the
emission from those ions is lower if the line is collisionally
dominated, or higher, if the line is pumped by the continuum.  Thus,
the strongest lines should be \ion{Fe}{2} and high-excitation
\ion{Si}{2}, as observed, while high-ionization lines and
semiforbidden lines are weak.  As a consequence, semiforbidden lines
are not good density diagnostics in gas illuminated by the soft SED.
Line emission from gas illuminated by a soft SED has many similarities
with high-density gas, and in principle may be mistaken for
high-density gas, as a consequence of inefficient cooling by line
emission in both cases.  

\subsection{A Locally Optimally-emitting Cloud Model}

In the previous section, we showed that the soft SED from PHL~1811
results in low emission line flux and low equivalent widths compared
to the more typical SED used by K97.  We first showed this
for the fiducial parameters ($\log U=-1.5$, $\log n=11$, $\log N_H =
24.5$) and then for a range of ionization parameters and densities,
with the column density adjusted so that $\log N_H+\log U=26$.  As a
final investigation into this point, we compute a Locally
Optimally-emitting Cloud (LOC) model, to ensure that our result is
robust against averaging.  

The LOC model was first presented by \citet{baldwin95}.  Motivated by
the fact that it is unrealistic to expect that the net emission from
the broad-line region could be from gas characterized by a single
ionization parameter, density, and column density, they postulated
that the observed emission should be an average of emission from gas
characterized by a range of properties.  In the standard LOC model,
the radial distribution (equivalent to photoionizing flux)  and
density distribution of the  line-emitting gas are characterized by
power laws, with the radial and 
density indices equal to $-1$, and the column density set equal to
$10^{23}\rm \, cm^{-3}$.  These parameters are rather arbitrary, but
they seem to give relatively good fits to most quasars, at least for
the high-ionization lines.  For example, \citet{clb05} computed an LOC
model for RE~1034$+$39, an object known for its hard SED, and found
that the ratio of the prominent lines with Ly$\alpha$ matched the
observed ones well; the exceptions were \ion{C}{4} and \ion{Mg}{2},
which were both predicted to be brighter with respect to Ly$\alpha$
than observed. 

To compute the LOC model, we run {\it Cloudy} models for the following
parameters: photon flux $17< \log \Phi < 24$, with $\Delta \Phi =
0.125$; density $7 < \log n < 14$ with $\Delta n=0.125$; column
density $N_H=10^{23}\rm \, cm^{-3}$ uniformly.  This is the range of
parameters computed by \citet{k97}, and extends the range originally used by
\citet{baldwin95} and \citet{clb05} to higher photon fluxes and lower
densities.  We compute the LOC assuming the power law indices of
$-1$ for the radial and density distributions; again, these are the
standard choices.  The outputs of the LOC program are the ratios of the
line fluxes to the continuum at 1215\AA\/.  We use the input
continua to convert these values to equivalent widths.  We choose a
covering fraction of 0.25 because that seems to give values for the
K97 continuum that correspond relatively well with equivalent widths
of generally bright lines from composite spectra compiled by
\citet{francis91} (LBQS quasars), \citet{zheng97} ({\it HST}),
\citet{brotherton01} ({\it FIRST} survey), and  \citet{vandenberk01}
(SDSS quasars).  The results are given in Table 3.

Examination of the table shows that the LOC model for the K97
continuum gives a reasonable match to composite spectra equivalent
widths overall, within a factor of two for most of the lines.  For
example, it predicts a \ion{C}{4} equivalent width of 47\AA\/, while
the four composite spectra yield measured values of \ion{C}{4} of 37,
59, 33, and 24\AA\/.  This good correspondence is expected, since the
K97 continuum is intended to be representative of the continuum of an
average AGN, and the LOC model was developed originally to explain
the emission lines of composite spectra.  

The equivalent widths of the lines observed in PHL~1811 are much
smaller than observed in the composite spectra.  For example, the
observed equivalent width of \ion{C}{4} in PHL~1811 is only 6.6\AA\/,
while the composite spectra yield 37, 59, 33, and 24\AA\/.
Correspondingly, the LOC model for the PHL 1811 continuum yields
consistently much smaller equivalent widths for all lines, in some
cases by large factors.   For example, the equivalent width of
\ion{C}{4} for the PHL~1811 LOC model is 0.9\AA\/, compared with
47\AA\/ for the K97 LOC model.  In fact, the PHL~1811 LOC equivalent
widths are generally smaller than the observed equivalent widths by
large factors; the exception is \ion{Mg}{2} in which the PHL~1811 LOC
model gives an equivalent width of 15\AA\/, while the observed
equivalent width is 13\AA\/.  The largest discrepancies are Ly$\alpha$
and \ion{N}{5}.  This may be evidence for a separate optically-thin
component that is not well described by the standard LOC parameters.
We note that difficulties in using the standard LOC distributions have
been reported for SDSS~J154651.75$+$525313.1, an object that also has
strong low-ionization lines \citep{dhanda07}.

Despite disagreement between the detailed predictions of the LOC
models and the data, we have shown that the equivalent widths of the
lines in gas illuminated by the PHL~1811 continuum remain low even
when we average over a large range of ionization parameters
and densities.  In the spirit of the locally optimally emitting cloud
concept, we note that the extremely soft continuum of PHL~1811 allows
almost no cloud to emit lines very optimally, since gas illuminated by
the soft SED has trouble producing line emission. However, even
though the continuum may be relatively strong in the gas illuminated
by the PHL~1811 continuum (as discussed in \S 4.1.2), we do not expect to see
strong continuum emission.  The ratio of the Balmer continuum flux to
the continuum at 3645\AA\/ (for the 0.25 covering fraction) is
310\AA\/ for the K97 continuum, but only 94\AA\/ for the PHL~1811.  

\subsection{\ion{Fe}{2} and \ion{Fe}{3} Emission}

In \S 3.2.1 we showed that the I~Zw~1 \ion{Fe}{2} template does not
describe the \ion{Fe}{2} and \ion{Fe}{3} pseudocontinuum in PHL~1811.
Thus, it is clear that understanding the \ion{Fe}{2} and \ion{Fe}{3}
spectrum of PHL~1811 is important for understanding the physics of the
line-emitting gas.  However, the tools that we can use are 
limited.  First of all, {\it Cloudy} does not include any near-UV
\ion{Fe}{3} lines.  Second, the Fe$^+$ ion in {\it Cloudy} has only
371 levels and the highest one has an energy of just $11.6\rm\, eV$;
in \S 3.2.1 we discussed why we believe that significant emission arises
from the levels between 13 and 14 eV. In addition, {\it Cloudy} uses
an approximation for the radiative transfer; this might be expected to
be a serious model weakness especially for the emission from the
partially-ionized zone where radiative transfer can influence the
thermal equilibrium.    Furthermore, it has been shown that this
\ion{Fe}{2} model does not explain the spectra of AGN well at all
\citep[e.g.,][]{baldwin04}.  These are clearly serious inadequacies
for understanding the \ion{Fe}{2} and \ion{Fe}{3} in this object;
nevertheless, we can learn some very useful things using the {\it
  Cloudy} \ion{Fe}{2} model.  

We start with our baseline parameters ($\log U=-1.5$, $\log n=11$, and
$\log N_H=24.5$), and run {\it Cloudy} models for the range of
spectral energy distributions discussed in \S 4.1.1.  We first examine
the \ion{Fe}{2} 
flux in each of several bands: 1000--2000\AA\/ (far UV),
2000--3000\AA\/ (UV), 4000-6000\AA\/ (optical), and
7800\AA\/--3 microns (infrared).   As can be seen in Fig.\ 24, in each
of these bands, the flux from the gas illuminated by the PHL~1811 is 
second-greatest, following and comparable to gas illuminated by the
$kT=10\rm \, eV$ continuum.  In order to 
conveniently plot all the results in one panel, we show the ratio of
the results with respect to the one from PHL~1811.  The \ion{Fe}{2}
flux varies with SED differently in the different 
bands.  In the far-UV band, the flux decreases smoothly
as the SED becomes harder.  For the other three bands, the flux first
decreases as the flux becomes harder, but then recovers for hardest
SEDs.  This effect is most pronounced for the optical band, in which
the flux for the $kT=320\rm\, eV$ SED is almost the same as that from
the PHL~1811 SED.  

As discussed in \S 3.2.1 and in previous papers
\citep[e.g.,][]{dietrich02}, the \ion{Fe}{2}/\ion{Mg}{2} ratio is used
as an abundance diagnostic in quasars, although as discussed in
\citet{lm06}, an observed range in this ratio over a limited range in
redshift observed in SDSS quasars makes this usage somewhat
questionable without a better understanding of \ion{Fe}{2} excitation.
In the second panel of Fig.\ 24 we show the ratio of the \ion{Fe}{2}
model emission between 2200 and 3050\AA\/ to \ion{Mg}{2}.  This ratio
is rather constant for harder SEDs, but is very high for PHL~1811, a
factor of 5.2 higher than that from K97.  This very high ratio is a
consequence of both the higher flux in \ion{Fe}{2} (a factor of 2.1
higher in PHL~1811 compared with K97) and a lower flux in \ion{Mg}{2}
(a factor of 2.4 lower in PHL~1811).  We have already discussed in \S
4.1.3 the fact that \ion{Mg}{2} is a collisionally-excited line and is
thus suppressed in PHL~1811 because of the low electron temperature.

As noted in \S 3.2.1, PHL~1811 does not have a particularly high ratio of
optical \ion{Fe}{2} to H$\beta$ compared with a previously-analyzed
sample of NLS1s; the optical \ion{Fe}{2} emission is not as
spectacularly prominent in PHL~1811 as it is in some AGN.   Indeed,
the model predicts only a factor of 1.7 higher flux in the optical
band in PHL~1811 compared with K97, in contrast to the factor of 2.2
higher flux in the UV band. Thus, the observed relatively high
UV-to-optical flux ratio is qualitatively predicted.  

\citet{baldwin04} discuss the properties of the 371-level \ion{Fe}{2}
atom currently implemented in {\it Cloudy}.  The model does not fit
the observed spectra from AGN at all well in the near UV.
Specifically, strong emission from resonance lines and low-lying
levels near 2400\AA\/ and 2600\AA\/ are predicted to produced strong
``spikes'' that are not seen in AGN spectra.  \citet{baldwin04}
parameterize this problem using a ``spike/gap'' flux ratio parameter.
The parameter is defined as the flux in the 2312--2328 and
2565-2665\AA\/ bands (the spikes) divided by the flux in the
2462--2530\AA\/ band (the region between the spikes).
\citet{baldwin04} note that the observed ratio is typically near 0.7
in AGN.  We plot the spike/gap parameter in the bottom panel of Fig.\
24.  We find that this parameter is minimized for the PHL~1811
continuum, so although it is still much larger than observed, the
trend is in the right direction.

We can understand these results using the supplemental output from {\it
  Cloudy} which includes the departure coefficients and populations of
  a few selected levels as a function of depth, the column density of the
  371 levels, and the emissivity of various spectral bins as a
  function of depth.  The departure coefficients show that the level
  populations are consistent with those predicted by the gas
  temperature for upper levels less $\sim 7\rm\, eV$.  Referring to the
  diagram shown in Fig.\ 6, these are the upper levels of the
  resonance lines and low-lying transitions.  For higher levels, the
  departure coefficient for PHL~1811 is consistently higher than those
  from the harder SEDs, and can be very large in the partially-ionized
  zone.  Fig.\ 25 shows one of the more extreme cases. High levels are
  apparently being selectively excited in PHL~1811, and since we see an
  inverse dependence with the hardness of the SED, it is most probable
  that the incident SED is responsible for  the pumping.

The departure coefficients are high in gas illuminated by the PHL~1811
continuum, but the temperature in the partially-ionized zone is low,
so we need to also examine the populations in the levels as well.
For levels with energies less than $\sim 7\rm\,eV$, the populations in
PHL~1811 were lowest, and the populations increased as the SED became
harder.  This is expected from the departure coefficient behavior, as
the temperature in the partially-ionized zone is highest for the
hardest SEDs.  But for higher levels, the populations reverse, finally
becoming much larger for PHL~1811 than for the other SEDS.  An extreme
example is seen in Fig.\ 25.  Thus the departure coefficient is high
for PHL~1811 both because the temperature is low, and because the
density of highly-excited ions is high.

Examination of the populations and departure coefficient shown in
Fig.\ 25 reveals another important feature of gas illuminated with the
PHL~1811 continuum.  As will be discussed in \S 4.4, the transition from
He$^+$ to neutral He coincides with the hydrogen ionization front in gas
illuminated by a typical AGN continuum.  At the same time, Fe$^+$ is
almost exclusively found in the partially-ionized zone.  However, for
soft continua, the He$^+$ to neutral He transition occurs at a 
shallower depth than the hydrogen ionization front.  Thus, there
exists a region in which hydrogen is fully ionized and He is neutral.
Fe$^+$ is present in this region where hydrogen is fully ionized.
This is the origin of the shelf between $8\times 10^{10}$ 
and $1\times 10^{11}\rm\, cm$ seen in Fig.\ 25.  These Fe$^+$ can
contribute significantly to the emission.  While this is not a very
extensive region compared with the whole partially-ionized zone, it is
worth remembering that the UV \ion{Fe}{2} is highly trapped, and the
bulk of the emission comes from the vicinity of the
hydrogen-ionization front.  The density of Fe$^+$ in the
H$^+$--neutral He zone is small, and thus they may be less
trapped. Examination of the emissivity times depth confirms that a
substantial fraction of the emission comes from this region (Fig.\
26).

We are now in the position to understand the results shown in Fig.\
24.  High-energy lines in the far UV are purely pumped; thus they are
strongest for the PHL~1811 continuum and other soft SEDs, and are
weakest for the hardest SEDs.  The UV lines have contributions from
pumping in the higher-energy transitions, and contributions from
collisional excitation for the resonance and low-lying transitions.
Thus, they are strong for the soft SEDs, where pumping dominates, and
strong for the hardest SEDs that  produce the highest temperatures,
allowing collisional excitation to become equally important.  Optical
emission comes from transitions from the UV levels near 5 eV down to
the metastable levels near 3eV; these low-lying levels are
predominately collisionally excited. Recall that the optical 
emission is not trapped like the UV emission, so the hotter
partially-ionized zone produced by gas illuminated by the hard SED
produces plenty of optical \ion{Fe}{2}. Finally, the infrared emission
is produce by Ly$\alpha$ pumping; it is thus approximately
proportional to the diffuse continuum, which is shown in \S 4.1.2 to be
strongest for the soft SEDs, becoming weaker for intermediate SEDs,
and finally increasing again for the hardest SEDs.

Overall, the strong UV \ion{Fe}{2} emission seems to come from
continuum and Ly$\alpha$ pumping and a significant column of Fe$^+$ in
the H$^+$--neutral He zone.  All levels of the 371-level atom are more
highly populated in the gas illuminated by the PHL~1811 SED.  At the
same time, because of the lower temperature, collisional excitation is
less important, producing less optical emission relative to UV in the
partially-ionized zone.  \ion{Mg}{2} is also weak, as discussed in \S
4.1.3, and therefore the observed relatively high UV
\ion{Fe}{2}/\ion{Mg}{2} and UV/optical ratios are explained.  Finally, 
the ``Spike/Gap'' result is explained.  The spikes are predominately
collisionally excited, and are thus very strong for the hardest SEDs,
while the gap is predominately emission from high levels, and are thus
stronger for the softer SEDs.  Thus the softer SEDs produce a lower
``Spike/Gap'' ratio, although for the fiducial choice of parameters is
it still much larger than typically observed in AGN. 

So far we have only examined the \ion{Fe}{2} and \ion{Fe}{3} emission
for a single value of the ionization parameter and density.  Next, we
investigate the effect of varying the ionization parameter and
density.  As was done in \S 4.1.3, we compute the {\it Cloudy} models for
the PHL~1811 continuum and the K97 continuum for $-3.5 <  \log U <
0.5$, and $8.5 < \log n < 12.5$.  The column density was adjusted so
that $\log N_H + \log U = 26$.  First, as we noted above, {\it Cloudy}
does not output any \ion{Fe}{3} lines; however, we find that the
column density of Fe$^{+2}$ ions is consistently larger in the gas
illuminated by the PHL~1811 continuum by a factor of $\sim 2.5$, with
the largest column densities produced at the highest densities and
$\log U=-0.5$.  This is evidence that the \ion{Fe}{3} flux should be
greater from gas illuminated by the PHL~1811 continuum.    

We next examine the dependence of the \ion{Fe}{2} properties
considered in Fig.\ 24 as a function of ionization parameter and
density for gas illuminated by the PHL~1811 and the K97 continua.
For the \ion{Fe}{2} flux, we see the same trends as seen in Fig.\ 24:
the flux is uniformly larger for the gas illuminated by the PHL~1811
continuum.  The largest PHL~1811/K97 flux ratio is seen in the FUV
band,  ranging from $\sim 1$ to $4.2$ with a mean of $2.9$, with the
largest ratios observed at the highest ionization parameter and
density.  The same pattern is seen in the UV band, ranging from $\sim
1$ to $2.6$, again with the largest values seen at highest ionization
parameters and densities.  The optical flux ratio ranges from $\sim 1$
to 1.65, with a mean of 1.3, and the IR flux ratio ranges from $\sim
1$ to 2.7 with a mean of 1.8.  The maximum ratios for the optical and
IR ratios occurs however at $\log U=-1.5$ and $\log n=11$.   However,
perhaps a more relevant parameter is the equivalent width, since for
any photon flux, the PHL~1811 continuum will be brighter than the K97
(Fig.\ 13).  The equivalent widths in the far UV and IR bands are
almost the same, and they are a factor of $\sim 2$ lower in the UV and
optical.   

We next look at the behavior of the ratio of the UV \ion{Fe}{2} flux in the
2200--3050\AA\/ band to the \ion{Mg}{2} flux as a function of
ionization parameter and density (Fig.\ 27).  Generally speaking, this
ratio is larger for gas illuminated by the PHL~1811 continuum compared
with the K97 continuum by a factor of $\sim 3$ or more.  As can be
seen in Fig.\ 27, the range in the ratio is large with a strong
dependence on ionization parameter; however, the mean
obtained from the log of the ratio is $\sim 2.8$ for the K97
continuum,  which is not far from the value measured from composite
spectra from the SDSS \citep[3.9 or 5.3, depending on \ion{Fe}{2}
  template,][]{lm04}.  The mean ratio for the PHL~1811 continuum is
much higher, around 7.9.   The ratio is larger for the PHL~1811
continuum, both because the UV \ion{Fe}{2} flux is higher by a factor
of $\sim 1.8$, and because the \ion{Mg}{2} flux is lower by a factor
of $\sim 1.5$.  

Finally, we investigate the dependence of the ``spike/gap'' ratio as a
function of ionization parameter and density (Fig.\ 28).  This ratio is 
consistently smaller for gas illuminated by the the PHL~1811
continuum; the mean for the PHL 1811 continuum is 3.5, and the mean
for the K97 continuum is 4.6.  Thus the excitation differences
discussed are maintained as the ionization parameter and density is
varied.  We note, however, that even though the spike/gap ratio is
lower for PHL~1811, the minimum is 2, which is still much larger than
the observed value of $\sim 0.7$ \citep{baldwin04}. But as discussed
above, the relatively larger populations at high levels in the gas
illuminated by the PHL~1811 is an indication that a more complete atom
would produce emission from higher levels, plausibly filling in the
gap further.  

To summarize this section, we note that the prominence of the
\ion{Fe}{2} and \ion{Fe}{3} in PHL~1811, and the unusual shape of the
pseudocontinuum mean that understanding the emission of these lines is
important for understanding PHL~1811 in general.  Although the tools
at our disposal are limited ({\it Cloudy} does not predict UV
\ion{Fe}{3} emission lines, does not do radiative transfer properly,
and its Fe$^+$ model atom only goes up to 11.3~eV), we show that the
soft SED can qualitatively explain features of the \ion{Fe}{2} and \ion{Fe}{3}
emission.  A higher predicted column density of Fe$^{+2}$ predicts
stronger \ion{Fe}{3}, as observed.  \ion{Fe}{2} emission is predicted
to be generally stronger in gas illuminated by the soft SED, as a
consequence of the continuum pumping and interaction with the strong
diffuse continuum.  This trend is important in the near UV, resulting
in a high \ion{Fe}{2}/\ion{Mg}{2} ratio, as observed.  The optical is
less affected, resulting in a high UV-to-optical \ion{Fe}{2} ratio,
as observed.  High excitation and departure coefficients, as well as a
low \citet{baldwin04} ``spike'' to ``gap'' ratio, show that emission
from high levels is stronger in gas illuminated by the soft SED,
suggesting that the observed strong emission from very high levels may
be predicted if a larger model atom were available. 

\subsection{``Filtering'' and \ion{He}{1}, H$\alpha$, and narrow Ly$\alpha$}

As we show in \S 3.1, we do not detect \ion{He}{1}~$\lambda 5875$ in
PHL~1181.  Assuming that \ion{He}{1}~$\lambda 5875$ has the same
profile of H$\alpha$, we obtain a conservative upper limit of
0.0062 on the \ion{He}{1}/H$\alpha$ ratio in PHL~1811.  As noted
before, \citet{crenshaw86} tabulated the fluxes of \ion{He}{1} and
H$\alpha$  for 9 Seyfert galaxies; the mean \ion{He}{1}/H$\alpha$
ratio is 0.061. \citet{thompson91} tabulated \ion{He}{1}/H$\alpha$
ratio for 6 quasars, and the mean ratio is 0.025.  Thus, PHL~1811 has
abnormally weak \ion{He}{1} emission.   

Before going any farther, it is worthwhile reflecting that \ion{He}{1}
and H$\alpha$ are recombination lines and are thus considerably
different from the lines that we have discussed already.  Also, in 
gas illuminated by a typical AGN SED, the transition from He$^+$ to
neutral He occurs at approximately the same depth as the transition
from H$^+$ to neutral H -- the hydrogen ionization front. In a gas
illuminated by a soft SED, the continuum runs out of photons that can
ionize helium before it runs out of photons that can ionize hydrogen.
The result is that there is a region in which hydrogen is ionized and
helium is neutral.  \ion{He}{1} is produced in this region, which
means a low \ion{He}{1}/H$\alpha$ ratio is more difficult to produce
for soft SEDs than for hard ones.  

We first use the {\it  Cloudy} simulations using the CLB continua to
explore whether we can produce a low \ion{He}{1}/H$\alpha$ ratio,
especially for soft continua. In this set of simulations we examine
the ratio of the simulated flux to the observed flux evaluated at
Earth.  Representative results as a function of $kT_{cut}$, ionization
parameter, density and depth 
$D$, where depth is measured in terms of the depth of the hydrogen
ionization front are shown in Fig.\ 29.  We find that H$\alpha$ is
almost independent of $kT_{cut}$ and increases with column density.
H$\alpha$ decreases at high density at a given ionization parameter
because of thermalization.  We find that we can 
produce the H$\alpha$ emission that we see for $kT_{cut}=10$ and
$D=10$ when $\log(U)=-2$ and $\log(n)=11.5$, although we note that we
have assumed 100\% covering fraction, and including a more realistic
covering fraction would reduce the model flux.  In this region of
parameter space, and every other where we can match the H$\alpha$
emission, the predicted \ion{He}{1} emission is far too strong.
\ion{He}{1} decreases with density; however, H$\alpha$ decreases more
quickly, and at very high densities and ionization parameters,
H$\alpha$ is predicted to be too weak.  As discussed above, we see
that at a given depth the \ion{He}{1} flux increases as the SED
becomes softer.  

It is possible to obtain a low \ion{He}{1}/H$\alpha$ ratio using the
concept of ``filtering'' that was first introduced in
\citet{leighly04}.  In that paper, it was shown that the intermediate-
and low-ionization lines of two NLS1s could be better explained if the
continuum were transmitted through the wind creating the blueshifted
high-ionization lines observed from these objects {\it before} it
illuminates the intermediate- and low-ionization line-emitting gas
\citep[see][for details]{leighly04}.  

PHL~1811 has blueshifted high-ionization lines, and perhaps filtering
can help with the \ion{He}{1}/H$\alpha$ problem.  Specifically, the
continuum may be transmitted through a wind so that the emerging
spectrum lacks photons in the \ion{He}{1} continuum and yet the column
density is such that the hydrogen continuum remains relatively strong.
In fact, it is easier to produce such a filtered continuum in gas
illuminated by a soft SED, because of the presence of the region of gas
in which hydrogen is ionized and helium is neutral.  The presence of
this region means that the continuum will run out of photons in the
\ion{He}{1} continuum before it runs out of photons in the hydrogen
continuum, yielding the transmitted continuum that we need.

We test this scenario using {\it Cloudy}.  We first show that a
continuum with the necessary features can be produced.   We use 
the  PHL~1811 continuum shown in Fig.\ 12.  The high
\ion{N}{5}/\ion{He}{2} ratio implies that the metallicity in PHL~1811
may be enhanced; as in \citet{leighly04}, we assume metals augmented
by a factor of five, and nitrogen enhanced by a factor of ten.  We
first locate the column density at which the ratio of the He$^+$
column to the H$^+$ column is the largest for a range of densities
$7.875 \leq \log(n) \leq 12.75$ and ionization parameters $-3.0 \leq
\log(U) \leq -1.125$.  We then run {\it Cloudy} models using these
column densities, and output the transmitted continua.  We identify
SEDs with weak He$^+$ continua by taking the ratio of the transmitted
plus diffuse emission at 23.9~eV, just below the \ion{He}{1} edge at
24.6~eV, and at 28.8~eV, just below the \ion{N}{2} edge at 29.6~eV,
which seems to be the most prominent feature toward higher energies
from the \ion{He}{1} edge.  We find a very large range in
this ratio, from 2.6 to larger than $10^6$.  In fact, there is a
plateau in the ionization parameter/density plane such that the ratio
high for densities less than about $10^{11.5}$, and for ionization
parameters higher than $-2.0$ (Fig.\ 30).   We choose a representative
continuum with $\log(U)=-1.5$ and $\log(n)=11.5$, where the column
density at which He$^+$/H$^+$ is maximized is $10^{21.3}\rm \,
cm^{-2}$.  We note that any continuum from the afore-mentioned plateau
should produced the same result.  The continuum is shown in Fig.\ 30. 

The gas that the continuum is transmitted through will produce line
emission.  This gas may be the wind that emits the blueshifted lines,
as was assumed and required in \citet{leighly04}.  The {\it Cloudy}
simulations show that the strongest predicted line relative to the
observed emission lines is Ly$\alpha$.  A covering fraction of 10\%
ensures that it is predicted to be no stronger than observed.   Also,
if the gas filtering the continuum is a wind, the line emission may be
broadened and blueshifted, and would be more difficult to distinguish
from the continuum than a narrow line. As shown in Fig.\ 30, much of
the energy of the absorbed continuum is emitted as bremsstrahlung in
the infrared. This emission will join the transmitted continuum in
illuminating the low-ionization line-emitting gas, where it may
enhance  the low-ionization line emission through free-free absorption
\citep{fp89}.   

Finally, we run {\it Cloudy} simulations to determine whether the
predicted \ion{He}{1} is reduced as expected; the results for
\ion{He}{1} are shown in Fig.\ 31, and the results for H$\alpha$ are
shown in Fig.\ 32.  Each figure shows the predicted emission compared
with observed as 
a function of ionization parameter and density for the unfiltered and
filtered continuum shown in Fig.\ 30.   We assumed a covering fraction
of 10\%, noting that the covering fraction must be equal to or
smaller than the one required by the line emission in the filtering
gas, above.  These figures show that H$\alpha$ is predicted to be
nearly the same for the direct PHL~1811 continuum and for the
transmitted continuum shown in Fig.\ 30.  This is not surprising
because the Lyman continuum, responsible for photoionizing the
hydrogen, is not much different for the direct and transmitted
continua.  Including the covering fraction means that we produce
sufficient H$\alpha$ for moderate-to-low ionization parameters,  high
column densities and moderate densities, although we are a factor of
two lower than the observed.  In contrast, the \ion{He}{1} is much
weaker for  the transmitted continuum, and is now in the vicinity of
the measured upper limit for the same parameter choices.  

To summarize this section, we used {\it Cloudy} models to try to
understand the very low \ion{He}{1}/H$\alpha$ ratio observed from
PHL~1811.  We found that a soft spectral energy distribution
exacerbates the problem; \ion{He}{1} should be stronger in objects
with soft SEDs because of the large H$^+$/neutral He zone.   We
postulated that transmitting the continuum through a wind may yield a
``filtered'' continuum with a large \ion{He}{1} edge, noting that  a
similar scenario was successful in explaining the intermediate- and
low-ionization lines in IRAS~13224$-$380 and 1H~0707$-$495
\citep{leighly04}.  To test this idea, we showed that for 
column densities chosen to maximize the He$^+$ column to
the H$^+$ column ratio, a large area of $\log(U)$--$\log(n)$ parameter
space would produce transmitted spectral energy distributions with
the \ion{He}{1} continuum suppressed by a factor of up to $10^6$.
That we can create such a continuum is a direct consequence of the
presence of a region in gas illuminated by a soft spectral energy
distribution where fully ionized hydrogen and He$^+$ coexist; it means
that the column density of the filtering gas can be chosen to lie in
this region, producing a large \ion{He}{1} edge in the transmitted
continuum.   We show also that transmission through such gas would not
be predicted to produce strong line emission that is not observed if
we assume a 10\% covering fraction. Rather, for a representative case,
we see that most of the energy goes to  bremsstrahlung emission in the
infrared.  We illuminate  gas with the filtered continuum, and find
that while H$\alpha$ remains very similar to the unfiltered case,
\ion{He}{1} is significantly suppressed.  Thus, we have demonstrated
again that some line emission in AGN can be explained by illumination
by continua that have been transmitted through a wind.   

\subsection{Is PHL 1811 Unique?}

PHL 1811, being intrinsically X-ray weak, and lacking strong emission
lines, is a very unusual AGN.  Are there any others like it?  We do
not know of any exactly like it, but there are a few that are similar.

As seen in Fig.\ 9, PHL~1092 and RX~J0134.2$-$4258 have similar far UV
spectra; they both also lack strong and prominent Ly$\alpha$ and \ion{C}{4}
lines.  PHL 1811 is also similar to these objects in that they all
have extreme Eigenvector 1 properties; specifically, they have very
high \ion{Fe}{2}/[\ion{O}{3}] ratios, because they all lack or have
relatively weak [\ion{O}{3}] emission (PHL~1811: Fig.\ 2; PHL~1092:
e.g., \citet{leighly99b}; RX~J0134.2$-$4258: \citet{grupe00}).
They also have unusual X-ray 
properties.  RX~J0134.2$-$4258 had one of the softest X-ray spectra of
any AGN observed in the {\it ROSAT} All Sky Survey, which implies a
steep $\alpha_{ox}$; later {\it ROSAT} PSPC and {\it ASCA}
observations found it to have a more typical X-ray spectrum and
$\alpha_{ox}$.  It showed rather high amplitude X-ray variability
during the {\it ROSAT} pointed observation and the {\it ASCA}
observation, although typical for a luminous NLS1.  The {\it ROSAT}
spectrum of PHL~1092 revealed a very steep spectrum \citep{fh96},
similar to RX~J0134.2$-$4258.  It also showed rather high amplitude
short and long time scale X-ray variability for such a luminous quasar
\citep{fh96, leighly99a, bbfr99, gallo04}.  There
may be other objects that are similar for which we have less
information, such as RX~J2219.9$-$5941, a luminous NLS1 which shows
high amplitude X-ray variability \citep{grupe01, grupe04}.

The common properties among these three objects are their unusual UV
spectrum, their high luminosity for NLS1s (e.g., PHL~1811 is more
luminous than all but 2 of the 87  quasars in the \citet{bg92} subsample
of the PG quasars), their extreme X-ray properties, and their extreme
Eigenvector 1 properties.  As noted by \citet{laor00}, high luminosity
NLS1s should have high Eddington ratios.  We previously speculated
that X-ray weakness in NLS1s is a characteristic of a high Eddington
ratio \citep{leighly01}.  {\it XMM-Newton} observations of four
luminous NLS1s found that one (RX~J1225.7$+$2055) was X-ray weak with
$\alpha_{ox}\sim -2.0$ \citep{mlk04}.  Furthermore, a compilation of 
$\alpha_{ox}$  values from NLS1s from the literature indicated that X-ray
weakness may be more common in luminous NLS1s. In Paper I, we
speculated that a high Eddington ratio leads to a weak corona; the
quenching mechanism proposed by \citet{proga05} may be applicable.

In the Sloan Digital Sky Survey, some unusual high-redshift QSOs have
been discovered that have very blue rest-frame UV spectra, but no
discernible emission lines.  They are identified as quasars, and their
redshifts are obtained from the onset of Ly$\alpha$ forest absorption.  The
prototype object is SDSSp~J153259.96$-$003944.1 \citep[][hereafter
  referred to as SDSS~J1532$-$0039]{fan99}.  Featureless continua are 
typically found in BL Lac objects.  However, SDSS~J1532$-$0039 does
not appear to have other characteristic properties of BL Lac
objects; it was not detected in a deep radio observation and was not
strongly optically polarized.  It has recently been found to be
optically variable \citep{ss05}; however, this may not be conclusive
evidence that it is a BL Lac object, since one luminous NLS1
\citep[PHL~1092;][]{gallo04} has been observed to be tentatively
variable in the UV on short time scales.  Furthermore, follow-up
observations with {\it Chandra} failed to detect the quasar,
indicating an $\alpha_{ox} < -1.74$ \citep{vignali01}.  We thought
that the lack of broad emission lines and the X-ray weakness suggest a
similarity to PHL~1811 \citep{lhj04}.

A few other objects appear equally puzzling.  The
z$\sim 1$ quasar PG~1407$+$265 is possibly the first known weak-line
object \citep{mcdowell95}.  Two more such high redshift objects were found
in SDSS \citep[SDSS~J130216.13$+$003032.1 and
  SDSS~J144231.73$+$011055.3;][]{anderson01}. The z=0.494 quasar
2QZJ215454.3$-$305654 
\citep{londish04} was found in the 2dF/6dF QSO redshift survey.  It
has a blue continuum, no detected radio or X-ray emission, and is not
polarized.  In the Hamburg/ESO survey, the z=1.8 quasar HE~0141$-$3932
was found to have unusually weak Ly$\alpha$ emission; It also has
blueshifted high-ionization lines, similar to PHL~1811
\citep{reimers05}.

Because of their steep spectra and high amplitude X-ray variability,
it has been proposed that NLS1s are characterized by a high accretion
rate.  Luminous NLS1s should have exceptionally high accretion rates
compared with their Eddington value.  Luminous high redshift quasars
should have large black holes, and in order to grow so large in the
short amount of time implied by the standard cosmology, they should be
accreting at a rapid rate.  If the weak X-ray emission and UV line
emission is characteristic of a high accretion rate in a large black
hole, it may mean that PHL~1811 is the local prototype of the high-$z$
lineless QSOs \citep[see also][]{lhj04}.  

The observations, however, do not unequivocally support this argument.
First of all, not all of the lineless quasars found in the SDSS have
weak X-ray emission; some have normal $\alpha_{ox}$.  For example,
PG~1407$+$265 has $\alpha_{ox}$  $-1.44$ \citep{wilkes99}, which is
actually flat for a quasar of its 2500\AA\/ luminosity
\citep[$\log(l_{o})=32.01$;][]{wilkes94}.  Also,
SDSS~J144231.73$+$011055.3 has a relatively flat $\alpha_{ox}$ of
$-1.37$ \citep{schneider05}. Furthermore, it is quite possible that
some line-less objects are radio-quiet BL Lac objects \citep{col05},
i.e., a hypothetical class of objects in which the beamed continuum
overwhelms the lines, but unlike ordinary BL~Lacs, there is for some
reason weak radio and X-ray emission.  Thus, it could be that lineless
quasars are a heterogeneous   population. 

Although there are a number of other objects that are quite similar to
PHL~1811, the group comprises just a handful, insignificant
compared with, for example, the thousands of quasars identified in the
SDSS.  But can we be confident that we would be able to identify all
of these objects?    X-ray emission is considered to be an identifying
feature 
of AGN, and it has been shown to be useful for identifying AGN in deep
fields \citep[e.g.,][]{bh05}, and is even thought to be a key for
constraining the AGN contribution to the power in ultraluminous {\it
IRAS} galaxies \citep[e.g.,][]{ptak03}.  PHL~1811 is a powerful quasar
with a large black hole, but if it were embedded in a ULIRG it would
not be identified as such because of its weak X-ray emission.  Quasars
are also identified by their radio emission, and PHL~1811 was
rediscovered as a member of the FIRST Bright quasar survey.  However,
it is a weak radio source, and had it been further away, it would not
have been found.  AGN are commonly identified by their blue color in
photometric surveys, and PHL~1811 would have been easily identified as
a PG quasar, if it has been in the region of the sky covered by that
survey.  However, if it were embedded in a dusty starburst host, the
continuum would be reddened, and it would be much more difficult to
find.  One way that we might be able to identify these objects is
through their rest-frame infrared properties.  It has been 
shown that AGN can be distinguished from starbursts on the basis of
the $F_{25\mu m}/F_{60\mu m}$ flux ratio, which measures the
temperature of the dust \citep[e.g.,][]{degrijp87}.  Otherwise, if we
speculate that X-ray weakness, and weak line emission is a
characteristic of a high accretion rate, one might imagine legions of
PHL~1811s lying, unidentified, in high redshift galaxies.

\section{Summary}

In this paper, we report the results and analysis of {\it HST} STIS
optical and UV spectroscopic observations that were
coordinated with the {\it Chandra} observations
discussed in Paper I, and contemporaneous ground-based optical
observations of the unusually luminous, nearby narrow-line quasar
PHL~1811.  Here, we summarize the primary results of the paper.

\begin{itemize}

\item  The optical spectrum indicates that PHL~1811 is a luminous
  example of a   Narrow-line Seyfert 1 galaxy, but the emission lines
  are not typical of any type of AGN.  Overall, the line emission 
  is weak.  There is no evidence for forbidden or semi-forbidden line
  emission.  Very low ionization  lines are present, including
  \ion{Na}{1}D and  \ion{Ca}{2} H\&K, that are not typically observed
  in AGN.  These lines appear to be narrower than H$\beta$.  The
  near-UV region of the spectrum is dominated by \ion{Fe}{2} and 
  \ion{Fe}{3}. The UV \ion{Fe}{2} emission is strong with respect to
  both the optical \ion{Fe}{2} and \ion{Mg}{2}.  We find evidence that
  the UV \ion{Fe}{2} and \ion{Fe}{3} pseudocontinuum has a different
  shape in  PHL~1811 compared  with other AGN including the prototype
  NLS1 I~Zw~1; this appears to be a consequence of a higher
  ionization/excitation state of the iron in PHL~1811.   The
  \ion{C}{4} line is remarkably weak, with an equivalent width of only
  $6.6$\AA\/, about 5 times weaker than \ion{C}{4} in quasar
  composite spectra; it is also broad and blueshifted, similar to
  the NLS1s IRAS~13224$-$3809 and 1H~0707$-$495, indicating an outflow
  \citep{lm04}.   The region around Ly$\alpha$ is strongly   blended,
  but can be successfully deconvolved using a template created   from
  the \ion{C}{4} profile into Ly$\alpha$, \ion{N}{5}, \ion{Si}{2}
  and \ion{C}{3}*.   We also find that \ion{Si}{3}* is clearly present
  near 1112\AA\/.   

\item In Paper I, we showed that PHL~1811 is intrinsically X-ray weak,
  and therefore we infer that it has an unusually soft spectral energy
  distribution.  Can a soft SED by itself produce the observed unusual line
  emission?   We explore this question using  {\it Cloudy}, and find
  that it can.  A soft spectral energy distribution is stronger in
  the UV for a particular photoionizing flux than a hard SED; this is
  one factor that reduces the observed equivalent widths, but it also
  enhances lines that are pumped by the continuum, such as
  high-excitation \ion{Si}{2}.  The gas is cooler than gas illuminated
  by a hard SED, but has a higher hydrogen ionization fraction in the
  partially-ionized zone than might be naively expected.  The low 
  temperature means that collisionally-excited line emission,
  including semiforbidden lines and \ion{Mg}{2}, is weak.  Without the
  collisionally-excited line emission to cool the gas, the
  photoionization energy goes to excited states of hydrogen and
  continuum emission, rather than to the  electrons, resulting in the cool
  temperature. We conclude that the soft SED alone can qualitatively
  explain the line emission: high-ionization lines are weak because
  the continuum lacks photons to create the highly-ionized ions,
  semiforbidden and low-ionization collisionally-excited lines are
  absent because the gas is too cool to excite them, but lines excited
  by continuum pumping are strong because a soft SED continuum is
  stronger for a given photoionizing flux than a typical SED.  We also
  compute an LOC model and show that the above results are robust to
  averaging.

\item \ion{Fe}{2} and \ion{Fe}{3} are very important in the near-UV
  spectrum.  The {\it Cloudy} iron ion models are too limited to
  sufficiently   model this line emission in detail.  Specifically, we
  see strong   \ion{Fe}{3} in the spectrum, but no UV \ion{Fe}{3}
  lines are output   by the model.  In addition, we infer \ion{Fe}{2}
  emission from   highly excited levels near 14~eV, but the highest
  level treated by   the model is only 11.3~eV.  Finally, {\it Cloudy}
  uses a approximation for radiative transfer, and that may be a
  significant shortcoming in the partially-ionized zone.
  Nevertheless, {\it Cloudy} simulations reveal evidence that the
  \ion{Fe}{2} and  \ion{Fe}{3} emission is qualitatively explained by
  soft SED.  The 
  Fe$^{+2}$ columns are larger in gas illuminated by the soft SED, for
  a range of ionization parameters and densities, due to the fact that 
  the \ion{H}{2} region is dominated by intermediate-ionization ions,
  since the usual high-ionization ions are not produced by the soft
  SED.  The predicted \ion{Fe}{2} emission is uniformly higher in gas
  illuminated by the soft SED.  This is a consequence of several
  factors including strong continuum  and Ly$\alpha$ pumping, as
  mentioned above, as well as the presence of a region of gas in which
  hydrogen is ionized but helium is neutral and where Fe$^+$ is
  also present.  The excitation of Fe$^+$ is higher in the gas
  illuminated by the soft SED, and  relatively high departure
  coefficients are seen in the highest levels treated by the model,
  suggesting that emission from even higher levels might be strong.
  Overall, as far as we can tell given the limitations of the model,
  the soft SED is responsible for the unusual \ion{Fe}{2} and
  \ion{Fe}{3} observed.   

\item PHL~1811 has a very low ratio of \ion{He}{1}~$\lambda
  5876$/H$\alpha$.  This is not easy to explain using the soft SED
  because a soft SED predicts a larger value of this ratio as a
  consequence of the presence of a region in which hydrogen is ionized
  but helium is neutral.  
  We observe blueshifted high-ionization lines in this object and
  postulate that if the continuum were transmitted through
  optically-thin gas, the resulting ``filtered'' continuum may be
  strong in the hydrogen continuum, necessary to create H$\alpha$, but
  absorbed in the \ion{He}{1} continuum, with a large \ion{He}{1}
  edge. \citet{leighly04} showed that a ``filtered'' continuum was
  able to better explain the intermediate- and low-ionization line
  emission in the NLS1s IRAS~13224$-$3809 and 1H~0707$-$495. We locate
  the regions of parameter space for the filtering gas where the
  required    large   \ion{He}{1} edge will be produced, and
  demonstrate that   illuminating   gas   with the resulting continuum
  will produce a low   \ion{He}{1}/H$\alpha$ ratio as observed.   
\end{itemize}

In this paper, we present a detailed analysis of a single unusual
object, PHL~1811.  However, a number of things that we have learned
may carry over to our understanding of AGN in general.  For example,
the PHL~1811 spectrum lacks forbidden and semiforbidden lines. 
Generally, one would interpret weak semiforbidden lines to be a
consequence of high density.  We showed, however, that weak
semiforbidden lines are expected from gas illuminated by a soft SED,
which means that semiforbidden lines do not perform well as density
indicators when the illuminating SED is soft.  How many times have
weak semiforbidden lines been taken to be evidence for high density,
when in reality, a soft SED is the culprit?  The similarity between
emission from a high-density gas and gas illuminated by a soft SED is
not accidental; they are both a consequence of the inefficiency of
collisionally-excited line emission.  In the case of
PHL~1811, the soft SED is intrinsic, but it is also possible that the
soft SED may be the result of filtering.  Whether or not a filtered
continuum can mimic a high density gas will be the topic of a future
paper.  

PHL~1811 is a high-luminosity NLS1, and according to the
currently-accepted paradigm, these objects should be accreting at the
largest rates. Therefore, we may expect that they are present at high
redshift. However, they may be difficult to identify as AGN, and there
plausibly be a large number embedded in dusty galaxies.  The line
emission is weak, and in spectra with low signal-to-noise ratio, the
optical and UV spectra may appear not to have any emission lines at
all.  The continuum colors are blue, but that would not be seen if
embedded in a dusty galaxy.   The X-ray emission is a hallmark of AGN,
but it is weak in PHL~1811.  The best way to identify similar objects
at high redshift may be through their infrared properties.  



\acknowledgments
We thank Bob Becker for observing SDSS J094257.80$-$004705.2 and SDSS
J105023.68$-$010555.5.  KML thanks Eddie Baron and Joe Shields for
useful discussions.  Part of the work presented here  was done while
KML was on sabbatical at the  Department of Astronomy at The Ohio
State University, and she thanks the members of the department for
their hospitality.  Support for proposal \# 09181 was provided by
NASA through a grant from the Space Telescope Science Institute, which
is operated by the Association of Universities for Research in
Astronomy, Inc., under NASA contract NAS5-26555. This research has
made use of the NASA/IPAC Extragalactic Database which is operated by
the Jet Propulsion Laboratory, California Institute of Technology,
under contract with the National Aeronautics and Space
Administration.  

{\it Facilities:} \facility{HST (STIS)}, \facility{KPNO:2.1m ()}

\appendix

\section{Appendix}

In \S 4 we showed that gas illuminated by the soft spectral energy
distribution has rather unique properties that have not been discussed
in the literature in the context of AGN emission lines, and these
properties produce the unusual line emission in PHL 1811.  These
unique properties include weak high-ionization lines,  low
temperatures that result in weak collisionally-excited lines, a strong
UV continuum for a given value of photoionizing flux that results in
strong pumped lines, and high level of hydrogen ionization in the
partially-ionized zone due to the inability of the gas to cool via
collisionally-excited line emission.  In this appendix we further
investigate aspects of the properties of the gas illuminated by the
soft spectral energy distribution that are not directly related to
interpretation of the spectra, but are nevertheless interesting and
are included for completeness.  In particular, we are interested in
precisely what characteristics of the continuum result in the observed
properties; is it the weak X-ray or the weak extreme UV, or both?
This issue is examined in \S 6.1.  We are also interested in whether
we can destroy these characteristic properties by including turbulence
and changing the metallicity; these issues are discussed in \S 6.2.

\subsection{Effect of the  Spectral Energy Distribution}

The PHL~1811 spectral energy distribution shown in Fig.\ 12 differs
from a typical quasar in two respects: it is weaker in the extreme UV,
and it is weaker in the X-ray.  In fact, we do not know the shape of
the spectrum in the extreme UV, so in principle it could be brighter
in that band.  
In this section, we examine the influence of these two factors by
varying the extreme UV cutoff and the $\alpha_{ox}$ independently.  

The spectral energy distributions were constructed from the PHL~1811
SED shown in Fig.\ 12, with a few changes.  First, the PHL~1811
UV continuum extends to $\log \nu=15.53$ before dropping to the hard
X-rays.  We round this value down to  $\log \nu = 15.5$.  We then
create another break in the spectrum, in the extreme UV, and vary the
frequency of that break between $\log \nu=15.5$ and $16.4$. The upper
limit corresponds to about 100~eV.  These new points are extrapolated
from $\log \nu=15.5$ along a power law with a slope (in $log \nu
F_{\nu}$) of 0.5.  Then, they are joined to the X-ray, but while
the PHL~1811 continuum shown in Fig.\ 12 joins the X-ray at $\log
\nu=16.76$, we instead join at $17.1$, as that better approximates the 
position of the break in the K97 spectrum.  Finally, the X-rays are
renormalized to produce an $\alpha_{ox}$ between $-2.25$ 
(approximately the observed value for weaker X-ray observation of
PHL~1811) and $-1.4$ (approximately the observed value for K97).   The
spectral energy distributions are shown in Fig.\ 33. {\it
Cloudy} models for these 324 SEDs are run for the fiducial parameters
($\log U=-1.5$, $\log n=11.0$, $\log N_H=24.5$), yielding a grid of
results as a function of $\alpha_{ox}$ and the extreme UV cutoff.
PHL~1811 is characteristic of steep $\alpha_{ox}$ and low EUV cutoff
(lower left corner of the plots that follow) and K97 is characteristic
of flat $\alpha_{ox}$ and high EUV cutoff (upper right corner of the
plots that follow).

We first look at the physical characteristics of these spectral energy
distributions as a function of $\alpha_{ox}$ and EUV cutoff.  The flux
at 2500\AA\/ is nearly purely dependent on the EUV cutoff, with the
largest values being found for the lowest cutoff energies; this makes
sense because the majority of the photoionizing photons have energies
just larger than 13.6~eV due to the steep dependence of photoionizing
flux on energy.  Thus, the larger the EUV cutoff, the lower the
overall normalization for the same photoionizing flux.  In contrast,
the Compton temperature is nearly purely dependent on $\alpha_{ox}$.  

We next look at gas properties as a function of depth and normalized
to the depth of the hydrogen ionization front.  In the \ion{H}{2}
region, the temperature depends on both $\alpha_{ox}$ and the EUV
cutoff, with the lowest temperatures found for the steepest
$\alpha_{ox}$ and lowest-energy EUV cutoff, and the highest
temperatures for the flattest $\alpha_{ox}$ and highest-energy EUV
cutoff.  Then, for lines produced in the \ion{H}{2} region, the
behavior of the column density of the ion producing the line and the
lines themselves depends on the ionization potential, primarily, so
that there is dependence on $\alpha_{ox}$ if the ionization potential
is high, and limited to dependence on EUV cutoff if the ionization
potential is low, roughly speaking.  Contours of flux for the
high-ionization line \ion{C}{4} and the intermediate-ionization lines 
\ion{C}{3}]~$\lambda 1909$ and \ion{Si}{3}]~$\lambda 1892$ are shown
in Fig.\ 34.   

Turning to the properties of the partially-ionized zone, we examine
the temperature, the hydrogen ionization fraction, and the ratio of
the excitation temperature to the electron temperature at $n=2$ at the
representative depth of ten times the depth of the hydrogen ionization 
front (Fig.\ 35).  The temperature depends primarily on $\alpha_{ox}$
for flatter $\alpha_{ox}$ with some dependence on EUV cutoff for
steeper $\alpha_{ox}$.  The lowest temperatures occur, not
surprisingly, for objects with the steepest $\alpha_{ox}$ and lowest
energy EUV cutoff.  The hydrogen ionization fraction depends primarily 
on the EUV cutoff, which shows that photoionization from $n=2$ by the
input continuum is the dominant process.  However, at the flattest
values of $\alpha_{ox}$, the hydrogen ionization fraction increases
somewhat, due to the additional ionization by nonthermal electrons
from photoionization by soft X-rays.  That this depends on
$\alpha_{ox}$ is not surprising, because the break to the X-ray power
law, at 100~eV, is in the soft X-ray band so that the X-ray power law
includes a region of high opacity to photoelectric absorption by
metals.  Finally, the ratio of the excitation temperature to the
electron temperature of $n=2$ hydrogen is a convolution
between the temperature and the hydrogen fraction, and the highest
values occur for the steepest $\alpha_{ox}$ and lowest EUV cutoff,
similar to the PHL~1811 continuum.  

The low-ionization lines, excluding \ion{Fe}{2}, divide naturally into
those that are pumped and those that are collisionally excited.
Collisionally excited lines are weakest for the steepest $\alpha_{ox}$
and the lowest values of the EUV cutoff.  Examples of such lines are
\ion{Mg}{2}, \ion{C}{2}~$\lambda 1335$, and \ion{Si}{2}~$\lambda
1814$.   The pumped lines follow the UV flux and are strongest for the
lowest values of the EUV cutoff.   Examples of such lines are the
high-ionization \ion{Si}{2} lines including those at 1260 and
1194\AA\/. Contours for \ion{Mg}{2}, \ion{Si}{2}~$\lambda 1814$, and
\ion{Si}{2}~$\lambda 1194$ are shown in Fig.\ 34.  The \ion{Fe}{2}
lines are all stronger for low values of the EUV cutoff, but in some
cases, they increase for flat $\alpha_{ox}$ at a particular value of
the EUV cutoff.  In this way they resemble the hydrogen ionization
fraction shown in Fig.\ 35. These may be responding to Ly$\alpha$
pumping.  Examples of \ion{Fe}{2} that appears to be pumped
are UV~191, the infrared lines between 7300\AA\/ and 3 microns, which
are thought arise predominately in cascades after Ly$\alpha$ pumping,
and the ``gap''.  Example of lines that may be responding to
Ly$\alpha$ pumping are the ``spike'' lines and other low-lying levels,
and the near-UV  band between 2000 and 3000\AA\/.  Fig.\ 34 shows the
results for UV~191, the ``spike'' and the ``gap''.  

We also computed the equivalent widths, but since the continuum is
much stronger for continua with low EUV cutoffs, they are almost all
weaker for low EUV cutoffs and steep $\alpha_{ox}$.  Lines that are
purely pumped by the continuum have equivalent widths independent of
SED shape.  

In this section, we have investigated simulations for only single 
values of the ionization parameter and the density; computing an
average or LOC would provide more robust results but is clearly beyond
the scope of this paper and will be pursued in the future.
Nevertheless, it seems that we can differentiate between SEDs by the
behavior of the high-, intermediate- and low-ionization lines.
Intermediate-flux high-ionization line emission will be produced for
either high EUV cutoff or flat $\alpha_{ox}$, or both.  Weak
high-ionization line emission requires both a low EUV cutoff and a
steep $\alpha_{ox}$.  The same is true for intermediate- and
collisionly-excited low-ionization line emission.  Pumped lines are
stronger for low EUV flux.  

\subsection{Effect of Turbulence and Metallicity}

In the simulations presented in \S 4 and in \S 6.1, we assumed solar
metallicity and no microturbulence.  Both of these factors influence
the cooling of the gas, and can therefore potentially disrupt the
effects of the soft spectral energy distribution discussed previously.   

The most extensive investigation of the influence of microturbulence
on the line emission in quasars was performed by \citet{bottorff00}.
They  noted that microturbulence has two main effects on the spectrum.
First, the line optical depths decrease as the line profile is spread
over a larger range in wavelength.  This causes the lines to be
brighter overall as they can escape more freely.  The second effect is
that continuum pumping becomes more important, again because the line
profile has a larger width.  As far as the gas properties are
concerned, \citet{bottorff00} indicated that the temperature will be
lower for very large $V_{turb}$ as the photons escape more readily and
cool the gas better, but higher for intermediate values due to
continuum pumping of lines.  They found that most lines become brighter
as $V_{turb}$ is increased, but by the same amount, which means that
they would not be good diagnostics of the influence of turbulence.
Some lines, however, become a lot brighter, including
\ion{Si}{2}~$\lambda 1260$, \ion{Si}{2}~$\lambda 1305$,
\ion{C}{2}~$\lambda 1335$, and \ion{Fe}{2}~$\lambda 1787$.  Thus,
\cite{bottorff00} suggested that ratios of these lines with respect to
intercombination lines and \ion{Si}{2}~$\lambda 1814$ would be good
diagnostics of turbulence.   

There have been a number of publications that discuss the influence of
enhanced metallicity.  \citet{ferland96} discussed the effects of
increasing the metallicity on production of the high-ionization
lines, especially \ion{N}{5}, with the goal of explaining the high
velocity component of Q0207$-$398.  \citet{sg99} discussed the fact that
enhanced cooling by higher metallicity gas will result a somewhat
higher characteristic ionization parameter, allowing a smaller
characteristic radius and higher characteristic density, thus
alleviating discrepancies between photoionization models and
reverberation mapping.  \citet{hkfwb02} investigated the behavior of
various lines as a function of metallicity, with the aim of
determining the line ratios that are most robustly sensitive to
metallicity variations to use as diagnostics.  Finally,
\citet{leighly04} showed that enhanced metallicity is required to
explain unusual high-ionization line ratios in two NLS1s that have very
blueshifted high-ionization lines.  

In this section, we discuss the effects of the metallicity and
turbulence on the properties of the gas illuminated by a soft spectral
energy distribution.  Our goal is qualitative insight, so we limit
ourselves to the fiducial gas parameters ($\log U=-1.5$, $\log
n=11.0$, and $\log N_H=24.5$).  It would be interesting to explore the
full parameter space, but that is beyond the scope of this paper.  

We first investigate the temperature of the gas as a function of
depth normalized to the hydrogen ionization front, shown in Fig.\ 36.
Addition of turbulence results in a higher temperature in both the
\ion{H}{2} region, and the partially-ionized zone.  In the \ion{H}{2}
region, most of the temperature increase occurs in the neutral
He/ionized hydrogen region, and is a consequence of the disappearance of
that region; that is, the transition from He$^+$ to neutral
helium moves closer to the hydrogen ionization front.  There is an
overall increase in the level of ionization in the \ion{H}{2} region,
including a decrease in the Fe$^{+2}$ column.  This is probably a
consequence of the fact that ions are better able to absorb continuum
photons, and so the energy of the continuum is released into the gas.
In the partially-ionized zone, the temperature increase is a
consequence of the loss of the synergy described in \S 4.  Once the
temperature becomes high enough that metal ions can be collisionally
excited, the amount of energy in the continuum decreases, the
excitation temperature decreases, and the energy previously locked up
in the excited hydrogen and continua is released to heat the gas.  

Turning to the metallicity, we find that the temperature is lower for
higher metallicities only for the highest ionization regions of the
gas.  In this region, the optical depth is relatively low and the
photons escape, cooling the gas.  In the intermediate-ionization
regions, and the partially-ionized zone, the temperature is higher for
higher metallicity.  In this case, the region where hydrogen is
ionized and helium is neutral becomes larger as the metallicity
increases.  Practically all metal-ion columns increase along with
metallicity, although with different slopes.  Photoionization of a
metal ion releases more energy than a hydrogen ion, and in regions
where the gas is starting to have appreciable optical depth, those
photons may not be able to readily escape to cool the gas.  In the
partially-ionized region, the temperature increases because the
availability of more metal ions to excite means that not so much
energy goes into the continuum and excited-state hydrogen, so it is
released into the gas.

Next we turn to the lines.  The behavior of the lines as a function of
turbulence has already been investigated by \citet{bottorff00}; we
look at only a few lines that behave differently for the PHL~1811 and
K97 continuum for this particular choice of ionization parameter.  As
discussed in \S 4.1.3, \ion{Mg}{2} is weak in PHL~1811 because the
temperature is low.  As seen in the top left panel of Fig.\ 37, \ion{Mg}{2}
increases more quickly with increasing turbulence in gas illuminated
by the PHL~1811 continuum compared with the K97 continuum as a result
of the increasing temperature in the line emitting gas.  This is also
true for the collisionally excited line \ion{Si}{2}~$\lambda 1814$.
In contrast, the \ion{Si}{2}~$\lambda 1263$, which is predominately
excited by continuum pumping, increases at the same rate in both
cases.  Next, we turn to the semiforbidden lines \ion{Si}{3}]~$\lambda
1892$ and \ion{C}{3}]~$\lambda 1909$.  The transitions producing
these lines have low $gf$ values, so they do not become optically
thick, and should not be influenced by lower opacity of the turbulent
gas.  Indeed, for the gas illuminated by the K97 continuum, there is
very little dependence on turbulence.  However, there is fairly strong
increase in the flux of these collisionally-excited lines in gas
illuminated by the PHL~1811 continuum that is a consequence of the
increase in temperature.

The results when the metallicity is varied are somewhat different
(Fig.\ 37, right).  Generally speaking, the generally prominent
high-ionization metal lines (e.g., \ion{C}{4}) do not vary much with
metallicity; their emission is primarily a function of gas
temperature.  The intermediate- and low-ionization metal lines
increase with metallicity, generally speaking, although with a range
of slopes.  \ion{Mg}{2} increases more quickly with metallicity for
the gas illuminated by the PHL~1811 continuum compared with the gas
illuminated by the K97 continuum; this is probably a consequence of
the increase in gas temperature in the partially-ionized zone as the
metallicity is increases. Both \ion{Si}{3}] and \ion{C}{3}] increase
with metallicity, although with different slopes, again pointing to a
lack of robustness as a density indicator.  Also, the increase is more
rapid for the K97 continuum for many of the intermediate ionization
lines.  This seems to be a consequence of changes in the ionization
structure of the gas as well as the change in temperature.  Finally,
we examine the ``spike'' and ``gap'' \ion{Fe}{2} bands
\citep{baldwin04}.  We see that both increase more rapidly for the
PHL~1811  continuum, but with the ``spike'' (resonance lines and
low-lying levels) increasing more rapidly than the ``gap''
(high-excitation \ion{Fe}{2}).  Recall that the model spike/gap ratio
is much larger than observed; thus a high metallicity only exacerbates
the problem.

To summarize this section, we have investigated the effects of
turbulence and metallicity changes on the gas properties and
emission.  In order to obtain a qualitative understanding, we have
limited ourselves to a single representative ionization parameter and
density.  We find that as the metallicity and turbulence increase, the
temperature increases and the energy in the continuum decreases, and
this generally causes an increase in the emission line fluxes.
Naturally, lines that are collisionally excited are more sensitive to
the temperature increase.  Weak collisionally-excited line emission
is a characteristic of the PHL~1811 spectrum, so these results may
mean that a high degree of turbulence and high metallicity are not
present in this object.  




\clearpage

\begin{deluxetable}{lccccc}
\tablewidth{0pt}
\tablecaption{Observing log}
\tablehead{
\colhead{Spectrometer} & \colhead{Date} &  
 \colhead{Exposure} & \colhead{Bandpass} &
 \colhead{Slit Width} & 
 \colhead{Resolution} \\
&& \colhead{(seconds)}}
\startdata

HST STIS FUV MAMA (G140L) & 2001-12-03 & 2640 & 1119--1715 \AA\/  &
$0.2^{\prime\prime}$  & 1.2\AA\/ \\
HST STIS NUV MAMA (G230L) & 2001-12-03 &  1667 & 1576--3159 \AA\/ &
$0.2^{\prime\prime}$ & 5.2\AA\/\\ 
HST STIS Optical CCD (G430L) & 2001-12-03 & 240 & 2894--5702 \AA\/ &
$0.2^{\prime\prime}$ & 5.5\AA\/ \\
KPNO 2.1m Goldcam & 2001-07-23 & 2700 & 6080--9160\AA\/ &
$1.9^{\prime\prime}$ & 4\AA\/ \\
KPNO 2.1m Goldcam & 2001-10-24 & 900 & 3938--7597 \AA\/ &
$1.9^{\prime\prime}$ & 5\AA\/ \\
KPNO 2.1m Goldcam & 2001-10-25 & 900 & 5466--8532 \AA\/ &
$1.9^{\prime\prime}$  & 4\AA\/ \\
\enddata

\end{deluxetable}

\clearpage

\begin{deluxetable}{llllccl}
\tabletypesize{\scriptsize}
\rotate
\tablewidth{0pt}
\tablecaption{Emission-line properties}
\tablehead{
\colhead{Emission Line} & \colhead{Line Model\tablenotemark{a}} &  
\colhead{Measured Wavelength\tablenotemark{b,c}} &  
 \colhead{Lab. Wavelength\tablenotemark{b,c}} & \colhead{Flux} &
 \colhead{Equivalent Width} & 
 \colhead{Velocity Width\tablenotemark{d}} \\
& & \colhead{(\AA\/)} & \colhead{(\AA\/)} & \colhead{($\rm 
   10^{-14} erg\, cm^{-2}\, s^{-1}$)} & \colhead{(\AA\/)} &
 \colhead{($ \rm km \,s^{-1}$)}} 
\startdata
\ion{He}{1} & L (UL) & 7067.2 (f\tablenotemark{e}) & 7067.2 & $<0.048$ & $<0.17$ &
$1752$(f) \\
H$\alpha$\tablenotemark{f} & L & $6564.12 \pm 0.04$ & 6564.7 &
$100.6 \pm 0.2$ & 302 & $1752 \pm 6$ \\
H$\beta$\tablenotemark{g} & L & $4864.2 \pm  0.09$  &  4862.7 & $29.4
\pm 0.2$ & 50 & $1943 \pm 19$ \\
H$\gamma$\tablenotemark{g} & L & $4347.1 \pm 0.3$ & 4341.7 & $9.5 \pm  0.3$ &
12 & $1716 \pm 60$ \\ 
\ion{Na}{1}D & L(2,thick) &  $5893.5 \pm 0.6$ & 5893.58  & $1.95 \pm 0.06$ &
4.9 & $1590 \pm 125$ \\ 
\ion{Na}{1}D & L(2,thick,+\ion{He}{1}) &  $5894.5 \pm 1.1$ & 5893.58
& $1.61 \pm 0.12$ & 4.1 & $1365  \pm 180$ \\ 
\ion{He}{1} & L(+\ion{Na}{1}D; UL)&  $5879.2 \pm 0.7$ & $5877.45$  &
$0.63$ & 1.6 & $1752$(f) \\  
\ion{Ca}{2}H & G &  $3936.6 \pm 0.4$ & 3934.8 & $1.30 \pm   0.12$ &
1.3  & $993 \pm 82$ \\  
\ion{Ca}{2}K & G &  3971.4 & 3969.6 &  $1.02 \pm 0.10$ & 1.1  & $993$ \\ 
\ion{Mg}{2} & G(2,thick) &  $2800.3 \pm  0.4$ &  2800.0  & $28.2 \pm
0.5$ & 12.9 & $2550  \pm 110$ \\
\ion{C}{3}] & integrated flux UL & 1915.8 & 1908.7 & $<7.3$ & $<1.7$ \\
\ion{Al}{3} & G(2,thick) & $1850.5 \pm 0.8$ & 1858.8 &  $8.3 \pm 0.4$
& 1.9 & $3210 \pm 340$ \\
\ion{Fe}{2}191 & G(3) & $1783.18 \pm 0.25$ & 1786.3 & $3.27 \pm 0.13$ &
0.74 & $1140 \pm 120$ \\
\ion{Si}{2} & T (UL) & 1815.02 (f) & 1815.02 & $< 6.1$ & $<1.4$ & \\
\ion{He}{2} & T (UL) & 1640.41 (f) & 1640.41 &  $<5.2$ & $<1.11$ & \\
\ion{C}{4} & T & 1549.5 & 1549.5 & $32.0 \pm 0.7$ & 6.6 \\
unidentified\tablenotemark{h} & G & $1395.2 \pm 1.7$ & & $28.0 \pm 3.0$ &
5.0 & $8470 \pm 710$ \\ 
\ion{C}{2} & T (UL) & 1335.31 (f) & 1335.31 & $<7.9$ & $<1.3$ & \\
\ion{Si}{2} & T & $1302.0 \pm 0.95$ & 1307.62 & $14.9 \pm 0.8$ & 2.4 \\
\ion{Si}{2} & T & $1259.9 \pm 1.8$ & 1263.32 & $52.0 \pm 1.1$ & 7.9 \\
\ion{N}{5} & T & $1236.3 \pm 0.4$ & 1240.15 & $60.5 \pm 1.3$ &  9.0 \\
Ly$\alpha$ & T & $1216.0 \pm 0.5$ & 1215.67 & $41.5 \pm 1.2$ & 6.0 \\
\ion{Si}{2} & T & $1193.6 \pm 0.6$ & 1194.12 & $25.6 \pm 1.0$ & 3.6  \\
\ion{C}{3}$*$ & T & $1175.4 \pm 0.5$ & 1175.67 & $13.8 \pm 0.9$ & 1.9 \\
\ion{Si}{3}$*$ & G(1) & $1110.9 \pm 0.5$ & 1111.59 & $1.5 \pm 0.3$ &
0.20 & $1160 \pm 230$ \\ 
\ion{Si}{3}$*$ & G(6) & $1110.5 \pm 0.3$ & 1111.59 & $1.7 \pm 0.03$ &
0.22 & $690 \pm 130$ \\ 
\ion{N}{2} & G(1) & $1084.2 \pm 0.2$ & 1085.12 & $0.90 \pm 0.20$
& 0.11 & $460 \pm 110$ \\ 
\ion{N}{2} & G(6) & $1084.2 \pm 0.2$ & 1085.12 & $0.95 \pm 0.02$ &
0.12 & $370 \pm 150$ \\ 
\enddata

\tablenotetext{a}{Codes for line models are as follows: L=Lorentzian
  profile; UL=upper limit ($\Delta\chi^2=6.63$); ``thick'' means that we
  set multiplet ratios equal to one; G=Gaussian line profile;
  T=\ion{C}{4} template was used for the profile; a number gives the
  number of components in the fit, usually the number of multiplets,
  or ``1'' if fitted with a single line.}
\tablenotetext{b}{For multiplets, we list the $gf$ weighted values,
  unless marked ``thick'' in column 2, in which case we give the
  average wavelength.} 
\tablenotetext{c}{Vacuum wavelengths.}
\tablenotetext{d}{Uncorrected for detector resolution.}
\tablenotetext{d}{(f) means that the parameter was fixed at the value
  given.} 
\tablenotetext{f}{Using the V\'eron-Cetty, Joly, \& V\'eron
  \ion{Fe}{2} template  subtraction.}
\tablenotetext{g}{Using the Boroson \& Green \ion{Fe}{2} template
  subtraction.}  
\tablenotetext{h}{This 1400\AA\/ feature is likely to be a blend of
  \ion{Si}{4} and \ion{O}{4}]; since the identification is ambiguous, 
  it is labeled as unidentified.}  

\end{deluxetable}

\clearpage

\begin{deluxetable}{lccccccc}
\tablewidth{0pt}
\tabletypesize{\scriptsize}
\tablecaption{LOC Model Results}
\tablehead{\colhead{Emission Line} & \multicolumn{4}{c}{Composite
    Spectra Equivalent Widths\tablenotemark{a}} & \colhead{K97 LOC\tablenotemark{b}} & \colhead{PHL 1811
    LOC\tablenotemark{b}} & \colhead{PHL 1811 Observed} \\
\cline{2-5} \\
& \colhead{Francis} & \colhead{Zheng} &  
 \colhead{Brotherton} & \colhead{Vanden Berk} }
\startdata
Ly$\alpha$+\ion{N}{5} feature & 52 & & 87 & & 94.6  \\
Ly$\alpha$ & & 85 & &  92.9 & 90.2 & 26.5 & 6.0 \\
\ion{N}{5} & & 10 & & 1.11 & 4.4 &  0.06 & 9.0 \\
1400 \AA\ feature & 10 & 7.3 & 8 & 8.1 & 5.9 & 0.3 & 5.0 \\
\ion{C}{4} & 37 & 59 & 33 & 23.8 & 47.2 & 0.9 & 6.6 \\
\ion{He}{2} & 12\tablenotemark{c} & 3.9 & 7.0\tablenotemark{c}  & 0.51
& 12.7 & 0.24 & $<1.1$ \\ 
1900 \AA\ feature & 22 & & 17 & & 14.2 & 3.0 & \\
\ion{Al}{3} & & 3.5 & & 0.40 & 1.3 & 0.14 & 1.9 \\
\ion{Si}{3}] & & 3.5 & & 0.16\tablenotemark{d} & 6.4 & 1.1 &  \\
\ion{C}{3}] & & 17.0  & & 21.2\tablenotemark{e} & 6.4 & 1.7 & $<1.7$\\
\ion{Mg}{2} & 50 & 64 & 34 & 32.3 & 107 & 15.0 & 12.9  \\
\enddata
\tablecomments{All values are equivalent widths in \AA\/.}
\tablenotetext{a}{References for composite spectra are
  \citet{francis91}, \citet{zheng97}, \citet{brotherton01}, and
  \citet{vandenberk01}.}
\tablenotetext{b}{A covering fraction of 0.25 was assumed.}
\tablenotetext{c}{\ion{He}{2}~$\lambda 1640$+\ion{O}{3}]~$\lambda 1663$.}
\tablenotetext{d}{Includes also part of \ion{Fe}{3} UV 34.}
\tablenotetext{e}{Includes also part of \ion{Fe}{3} UV 34, UV 68, and
  UV 61.}
\end{deluxetable}

\clearpage


\begin{figure}
\epsscale{0.8}\plotone{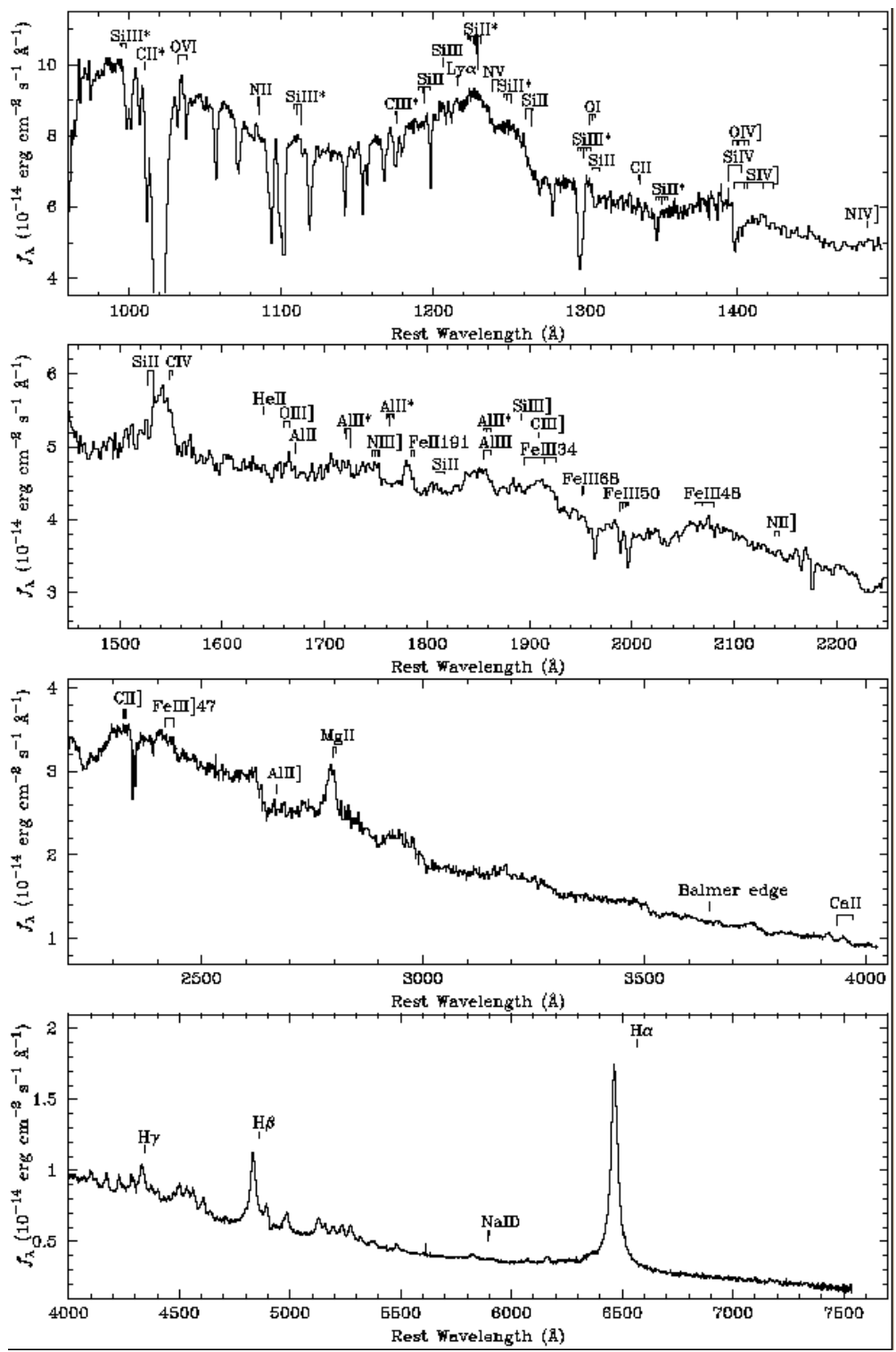}
\caption{Merged UV and optical spectra.  Prominent lines expected in
  active galaxies are labeled, along with a number of metastable
  lines. Length of tick marks is proportional to  $gf$ for individual
  multiplets.  Absorption lines, almost all of which are identified
  as originating in our Galaxy or in intervening absorption systems,
  are identified in   \citet{jenkins03}. 
\label{fig1}} 
\end{figure}

\clearpage 

\begin{figure}
\epsscale{1.0}\plotone{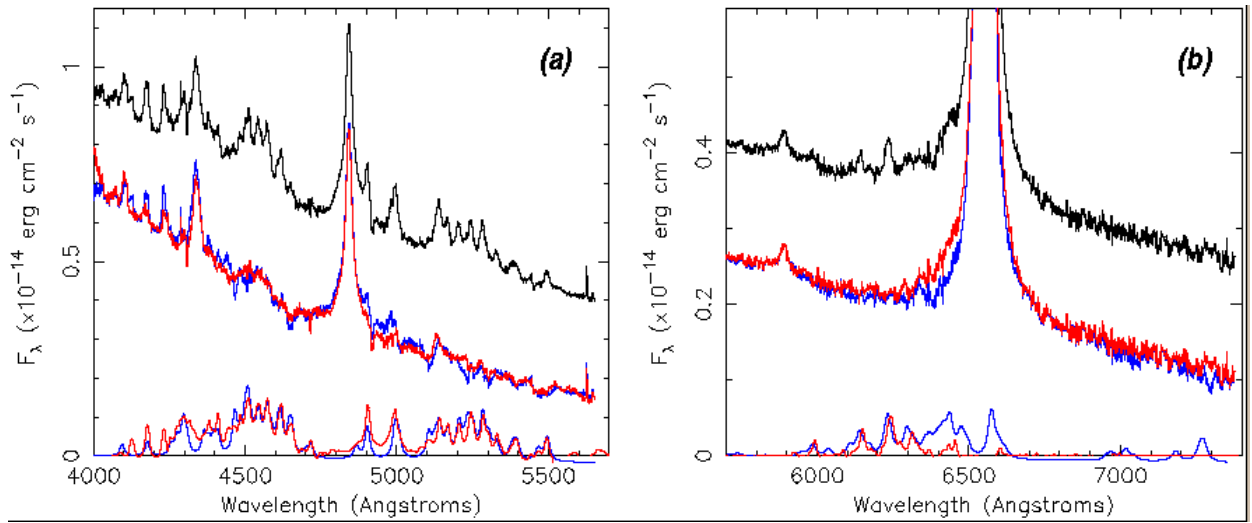}
\caption{Optical spectra near the H$\beta$ (a) and H$\alpha$ regions
  (b).  The   top lines show the original spectra.  The offset middle
  lines show the   \ion{Fe}{2} subtracted spectra, where the blue and
  red lines show the   results using the \citet{vjv04} and
  \citet{bg92} iron emission   templates, respectively (bottom lines).
  Note that   the  \citet{vjv04} template subtraction yields a more
  symmetric   H$\alpha$ line. 
\label{fig2}} 
\end{figure}

\clearpage 

\begin{figure}
\epsscale{1.0}\plotone{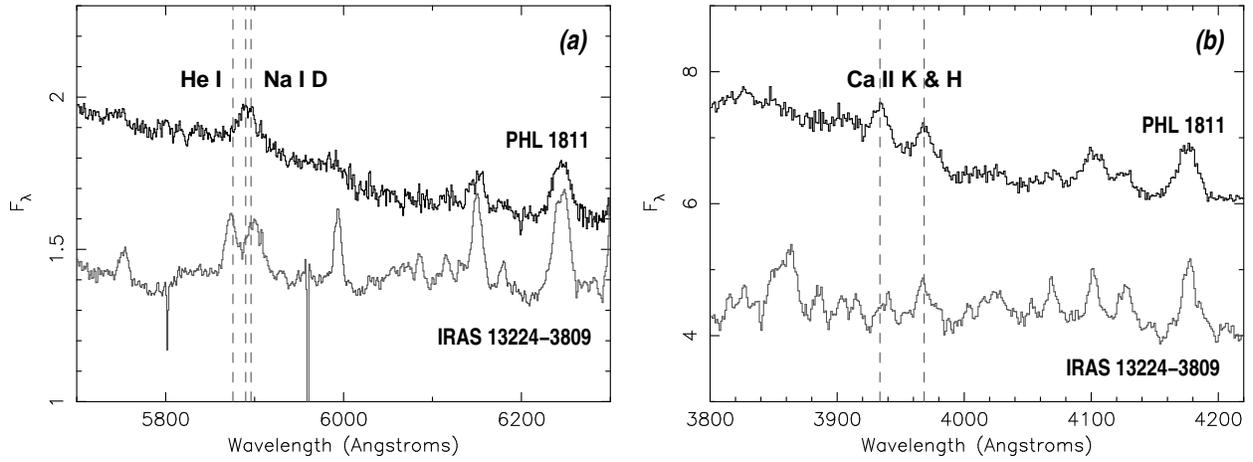}
\caption{We observe unusual very low-ionization line emission from
  \ion{Na}{1}D (a) and \ion{Ca}{2} K\&H (b) in  PHL~1811, but we observe no
  convincing evidence for \ion{He}{1}.    For comparison, the
  offset spectrum from the narrow-line Seyfert 1 galaxy
  IRAS~13224$-$3809 is shown.  
\label{fig3}} 
\end{figure}

\clearpage 

\begin{figure}
\epsscale{1.0}\plotone{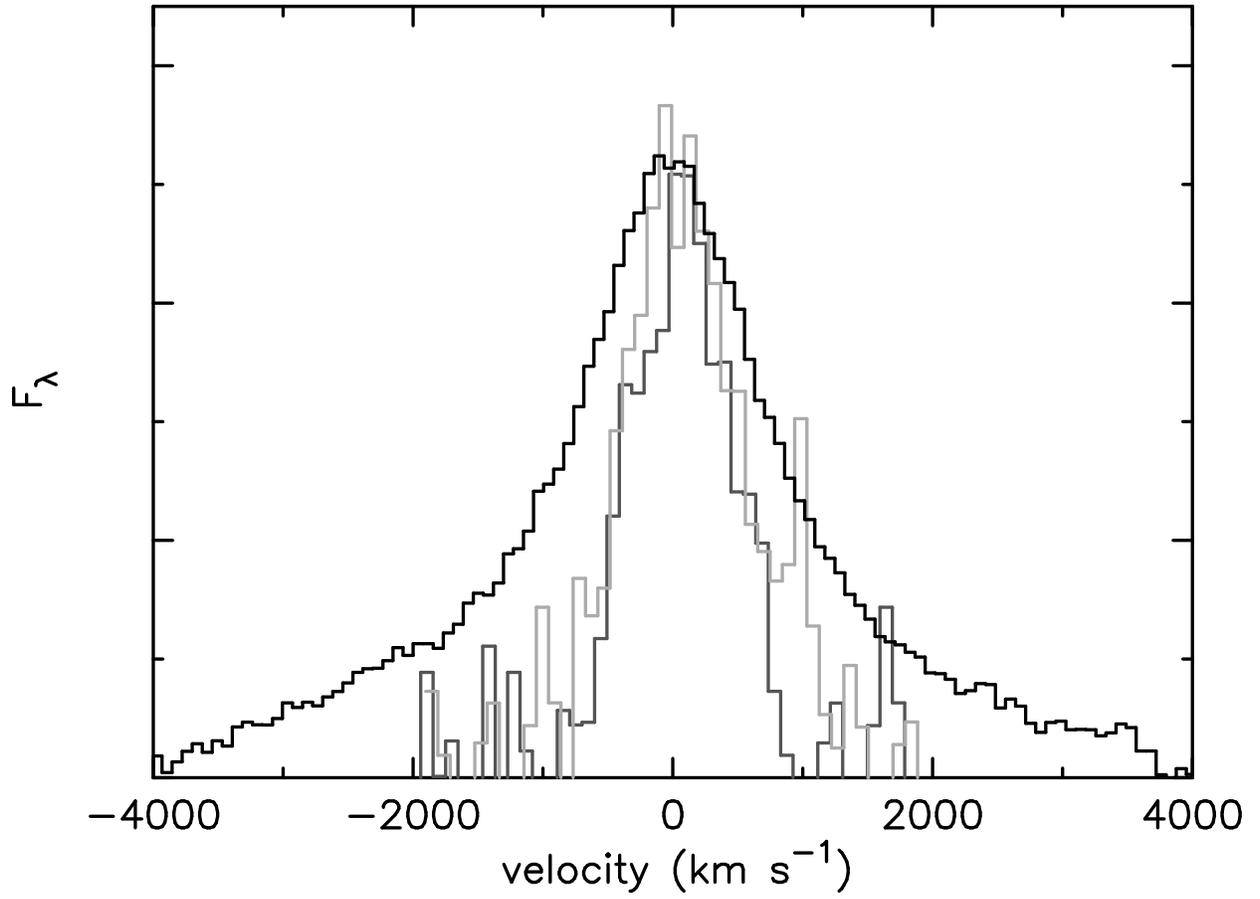}
\caption{A comparison of the \ion{Ca}{2} H\&K (grey lines) and H$\beta$
  (black line) profiles in velocity space shows that the \ion{Ca}{2}
  lines are distinctly narrower.
\label{fig4}} 
\end{figure}

\clearpage

\begin{figure}
\epsscale{1.0}\plotone{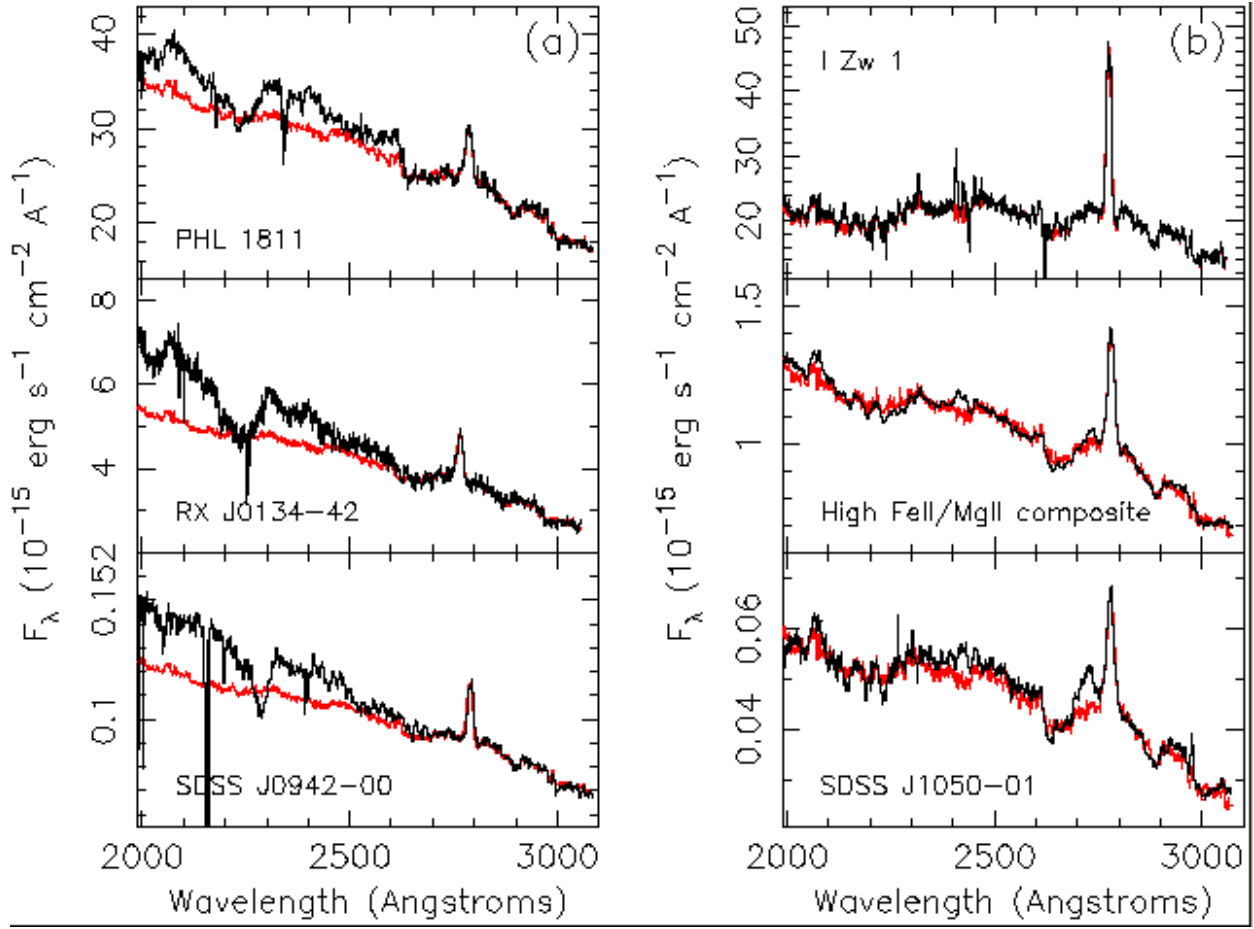}
\caption{UV \ion{Fe}{2} in PHL 1811 and other quasars.  The spectra
  shown were fitted between 2650 and 3090 \AA\/ with a model that
  included two Gaussians to model the \ion{Mg}{2} emission at
  2800\AA\/ and the \ion{Fe}{2} template plus a line to model the
  continuum.  Then the model was   extrapolated toward shorter
  wavelengths.  ({\it a}) The extrapolated   model leaves excesses between
  2300 and 2500 \AA\/, originating in   high-excitation \ion{Fe}{2},
  and shortward of 2200 \AA\/, originating   in \ion{Fe}{3}.  PHL~1811
  falls into this category.  ({\it b}) the   extrapolated model fits these
  spectra well.   
\label{fig5}} 
\end{figure}

\clearpage 

\begin{figure}
\epsscale{1.0}\plotone{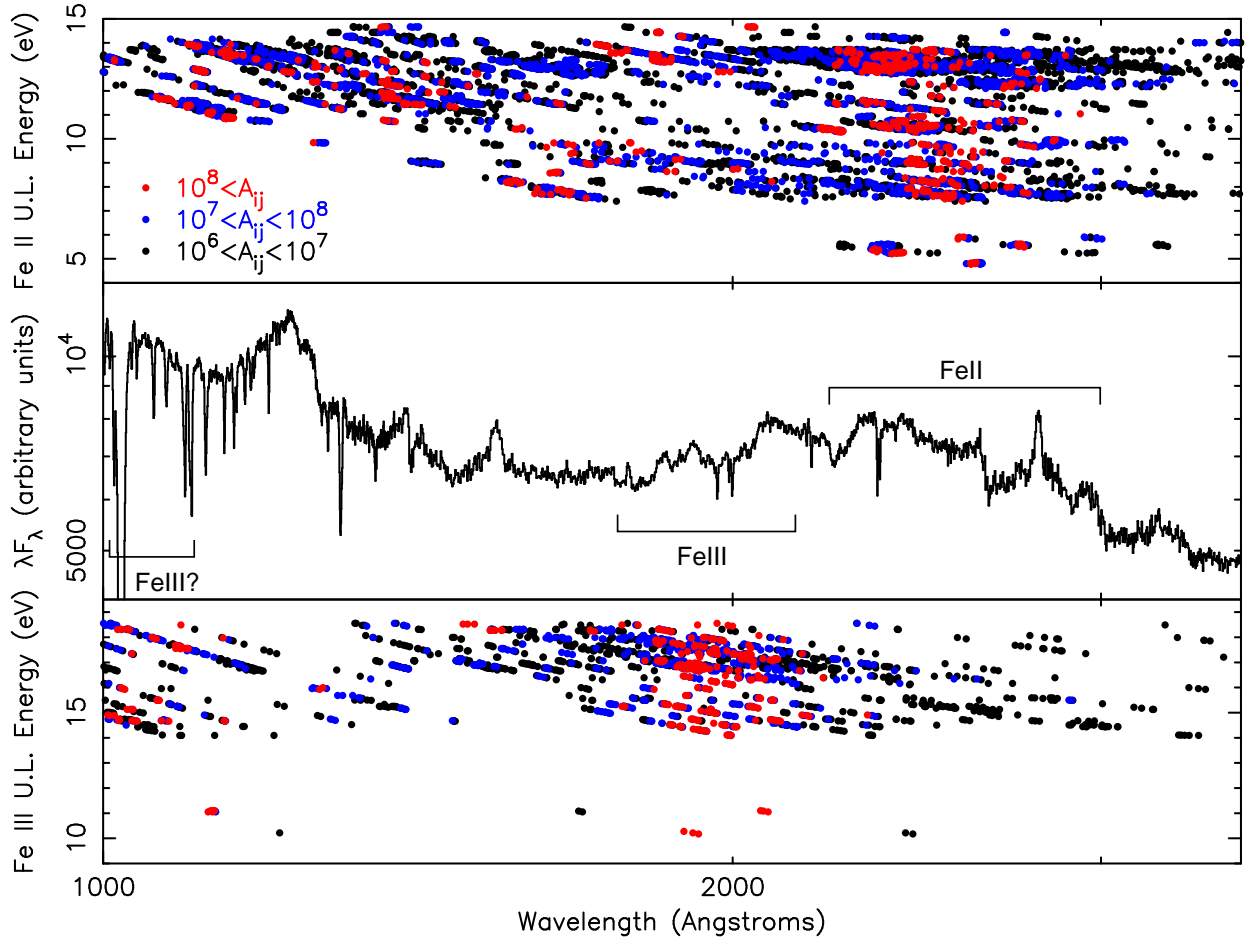}
\caption{{\it Top}: The upper level energies for \ion{Fe}{2} emission
  lines, color coded according to their $A{ij}$ value.  {\it Middle}:
  the $\lambda F_\lambda$ spectrum of PHL~1811.  {\it Bottom}: Same as
  the top panel, for \ion{Fe}{3}.  
\label{fig6}} 
\end{figure}

\clearpage 

\begin{figure}
\epsscale{1.0}\plotone{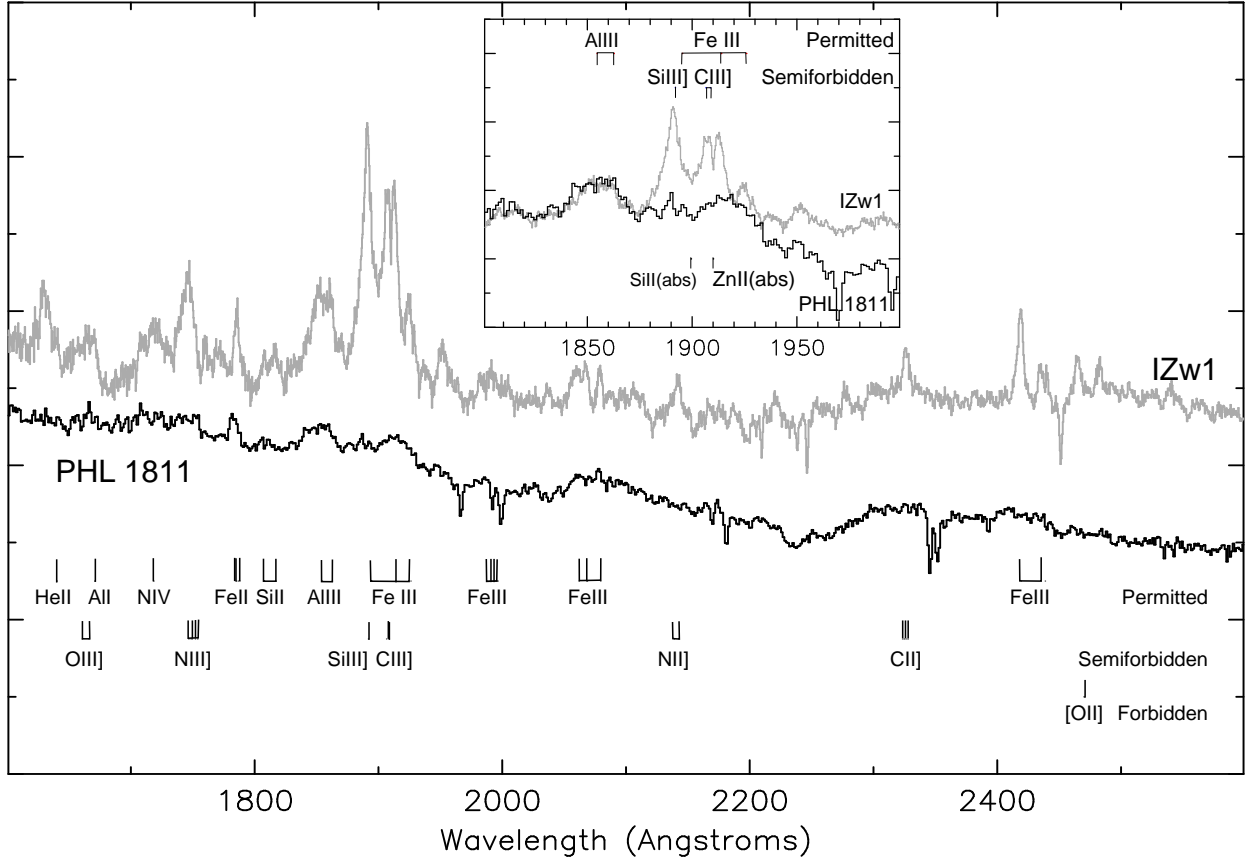}
\caption{The near-UV spectrum of PHL 1811.  Positions of permitted,
  semiforbidden and forbidden lines 
  commonly seen in AGN spectra are marked.  All of these lines are
  present in the  I~Zw~1 spectrum, but forbidden and semiforbidden
  lines are absent   in PHL~1811.  A feature near 1915\AA\/ is near
  the expected   position of  \ion{C}{3}]~$\lambda 1909$.  The inset
  shows that the   centroid of this   feature is redward of the
  expected position of   \ion{C}{3}], and it   is more likely to be
  the 1914.1\AA\/ component   of \ion{Fe}{3} UV34   (see text for
  details).  
\label{fig7}} 
\end{figure}

\clearpage 

\begin{figure}
\epsscale{1.0} \plotone{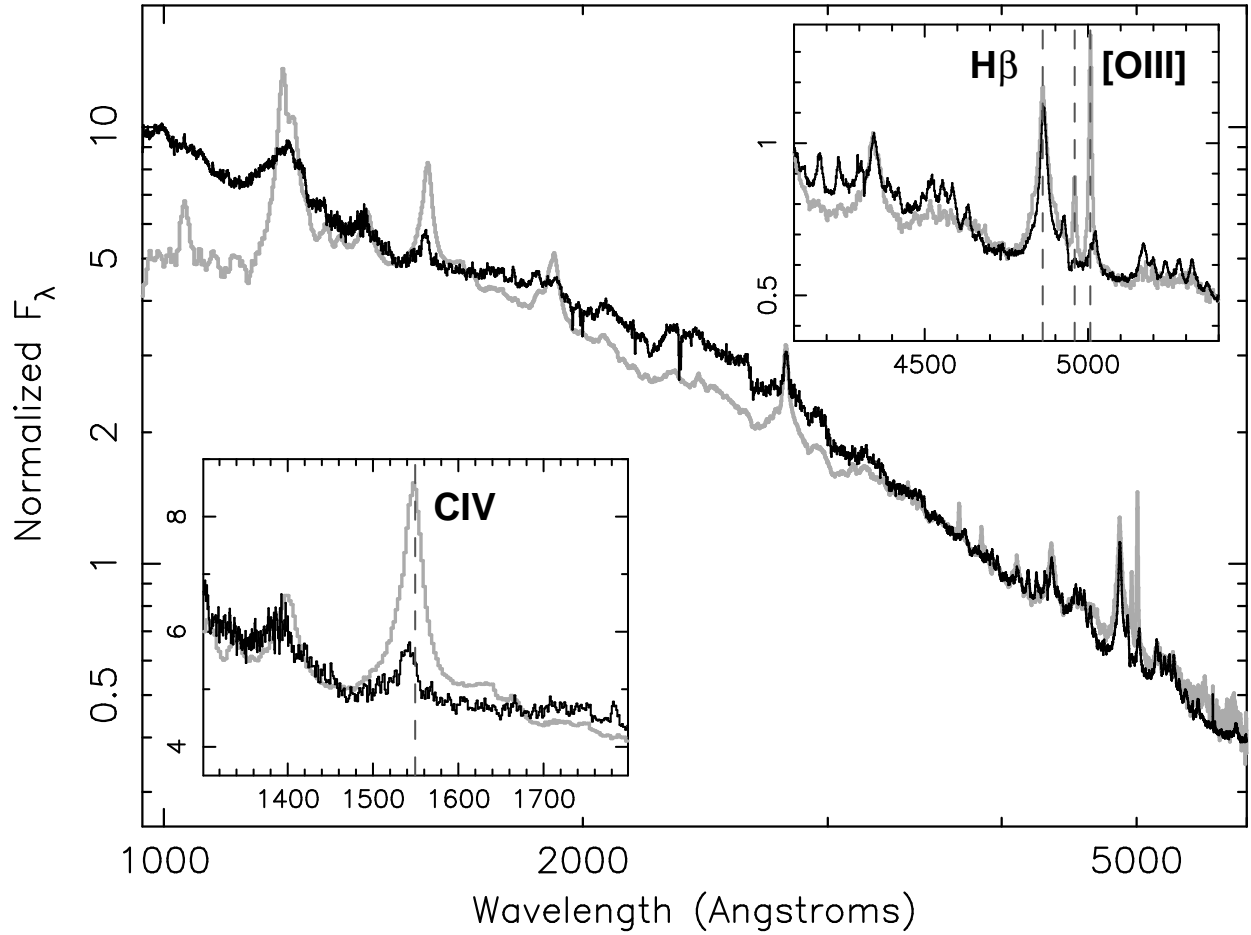}
\caption{A comparison of the PHL~1811 continuum (black line) with the
  Large Bright   Quasar Survey composite spectrum \citep[light grey
  line;][]{francis91}.   Notable   features are the lack of a break in the 
  PHL~1811 continuum toward short wavelengths in the far UV, and the
  large excess iron emission in the near UV.  Insets show a comparison
  of the \ion{C}{4} lines and the H$\beta$ region.
\label{fig8}} 
\end{figure}

\clearpage 

\begin{figure}
\plotone{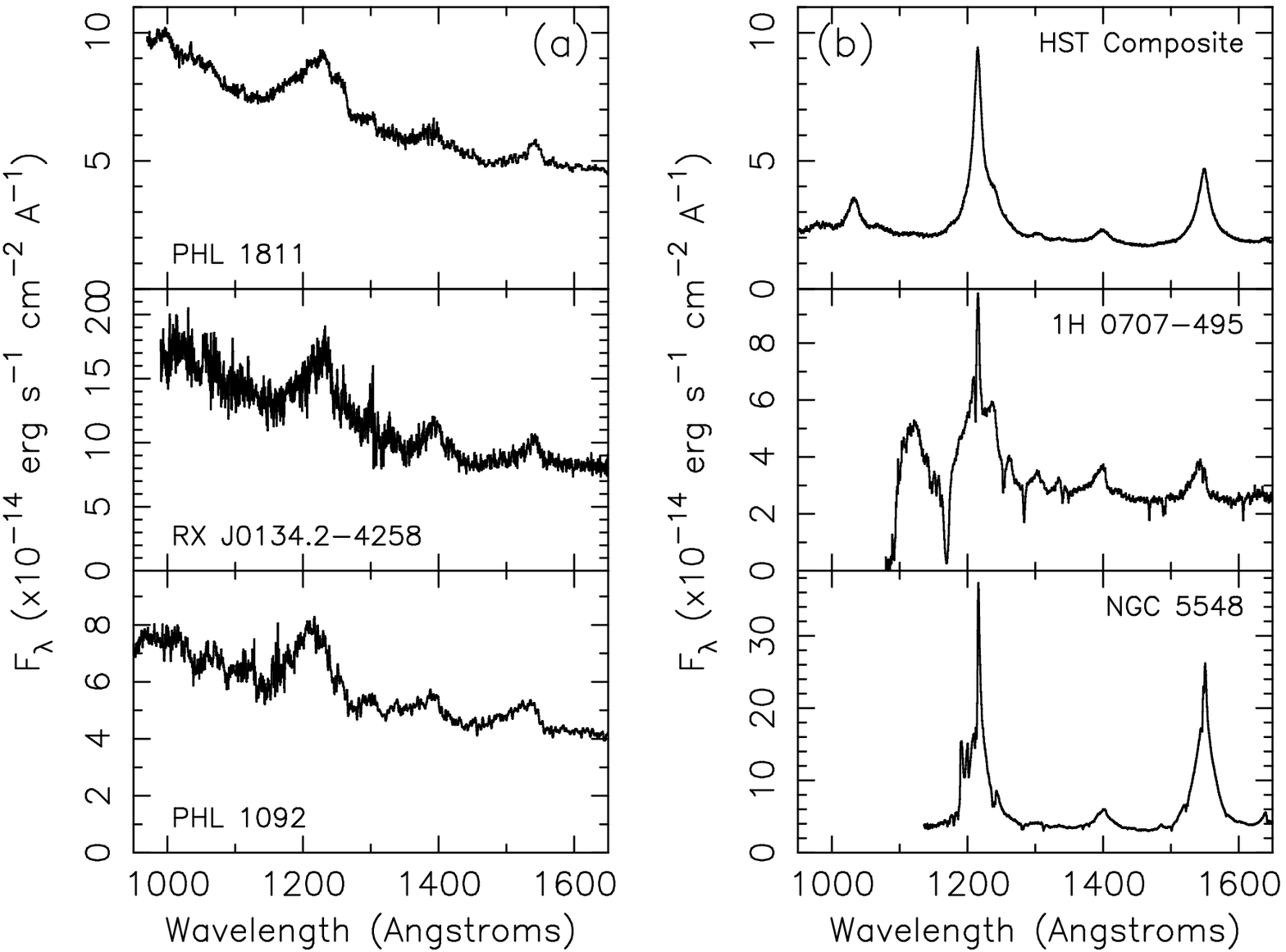}
\caption{A comparison of the far UV spectra of AGN.  {\it Left:} 
  PHL~1811 and similar high-luminosity narrow-line quasars
  RX~J0134.2$-$4258, and PHL~1092. {\it Right:} Comparison spectra
  include an {\it HST} composite spectrum \citep{zheng97}, the
  NLS1 1H~0707$-$495 \citep{lm04,leighly04}, and a broad-line Seyfert
  1 galaxy, NGC~5548 \citep{korista95}.  PHL~1811 and similar objects
  have unusually blue spectra shortward of $\sim 1450$\AA\/, very weak
  \ion{C}{4} emission, and a severely blended low equivalent width
  Ly$\alpha$/\ion{N}{5} feature.
\label{fig9}} 
\end{figure}

\clearpage 

\begin{figure}
\epsscale{0.8} \plotone{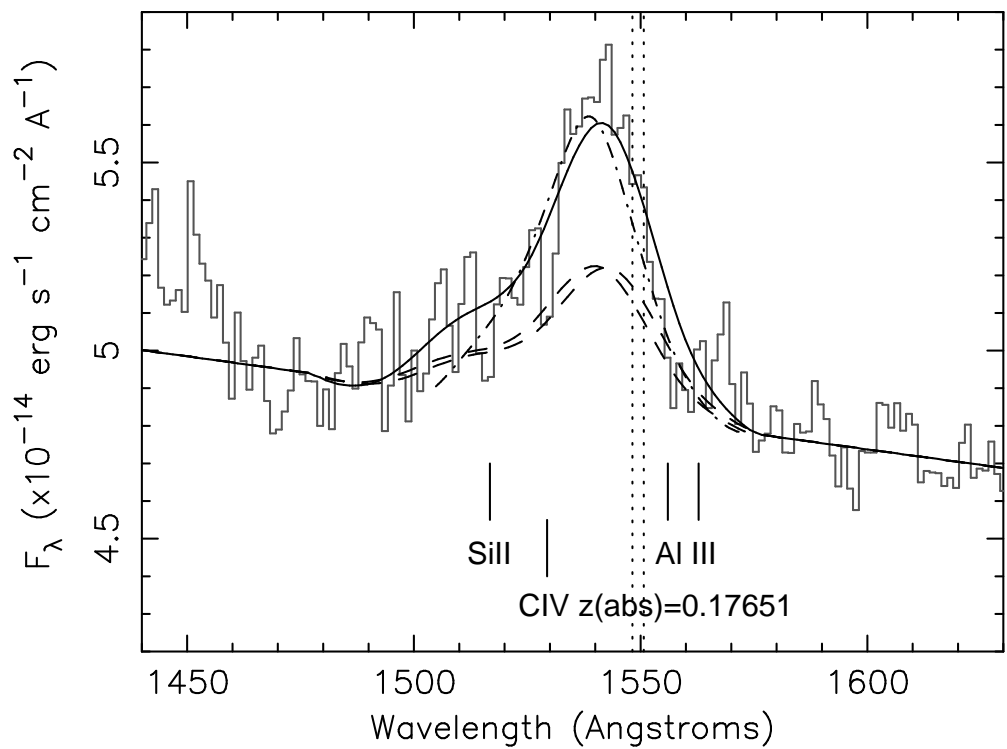}
\caption{The region of the spectrum near \ion{C}{4}.  The rest
  positions of the \ion{C}{4} doublet lines are marked by dotted
  lines.  Also marked   are the positions of plausible weak Galactic
  absorption lines, and   an absorption line from \ion{C}{4} in the
  $z(abs)=0.17651$   intervening system reported in
  \citet{jenkins03}. The template   model for the \ion{C}{4} profile
  is shown; the dashed lines show the   doublet components, derived
  from the sum profile (solid line)  using   the   method
  described in \citet{lm04}.  The dot-dashed line shows   the template
  obtained for the NLS1 IRAS~13224$-$3809   \citep{lm04}.  
\label{fig10}} 
\end{figure}

\clearpage 

\begin{figure}
\epsscale{0.7} \plotone{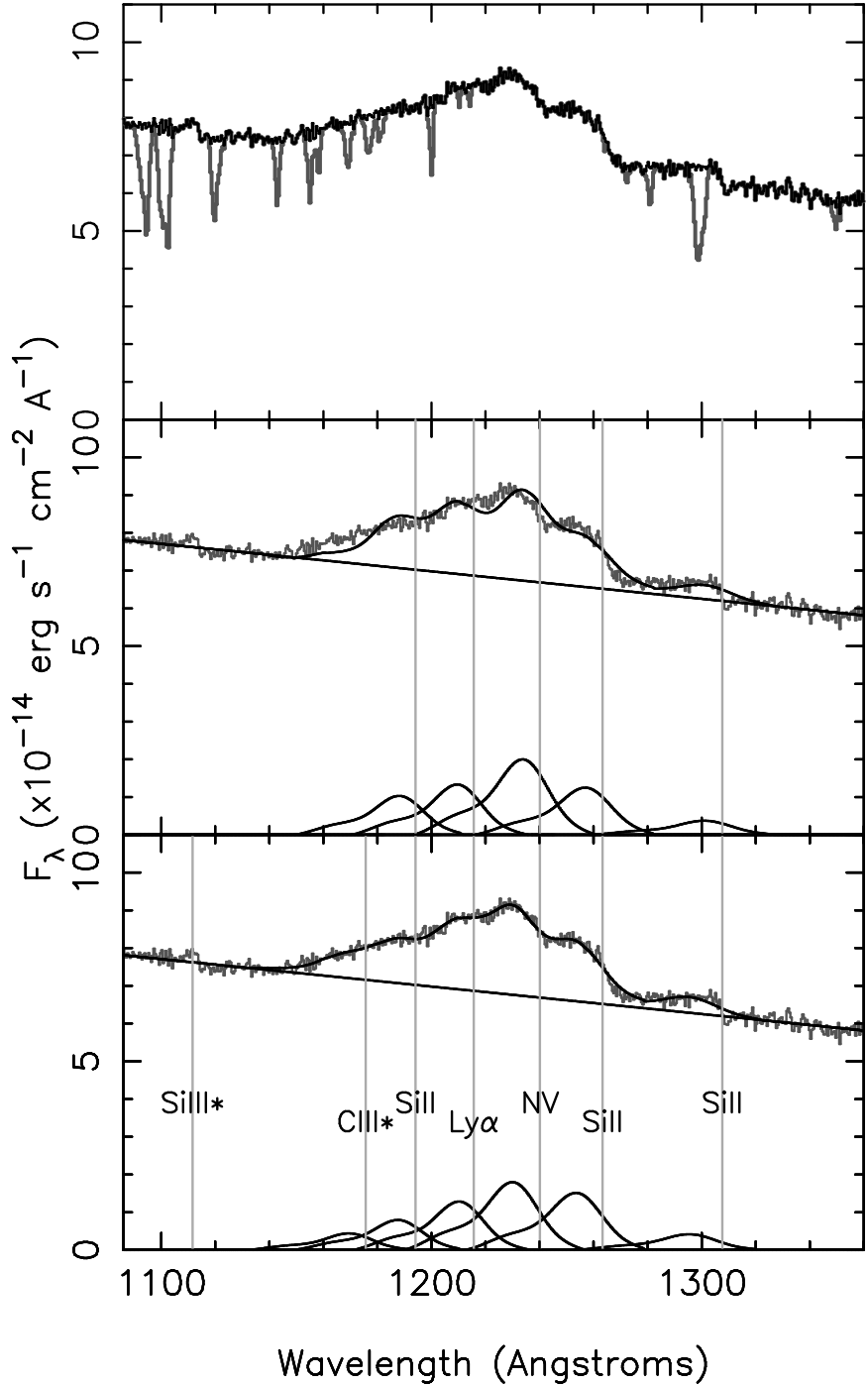}
\caption{Modeling the Ly$\alpha$/\ion{N}{5} feature.  {\it Top:} The far UV
  spectrum with the absorption lines (grey) modeled out. {\it Middle:}
  The fit using five templates corresponding to Ly$\alpha$, \ion{N}{5},
  \ion{Si}{2}~$\lambda 1308$, \ion{Si}{2}~$\lambda 1263$, and
  \ion{Si}{2}~$\lambda 1194$, fixed at their rest wavelengths shown by
  the vertical grey lines.  {\it
  Bottom:}  The fit using the five templates above, and one for
  \ion{C}{3}*~$\lambda 1176$ as well, in which the wavelengths of the
  templates were free parameters.  The detected metastable
  \ion{Si}{3}*~$\lambda 1112$ is marked as well.  See \S 3.3.2 of the text
  for   details.  
\label{fig11}} 
\end{figure}

\clearpage 

\begin{figure}
\epsscale{1.0} \plotone{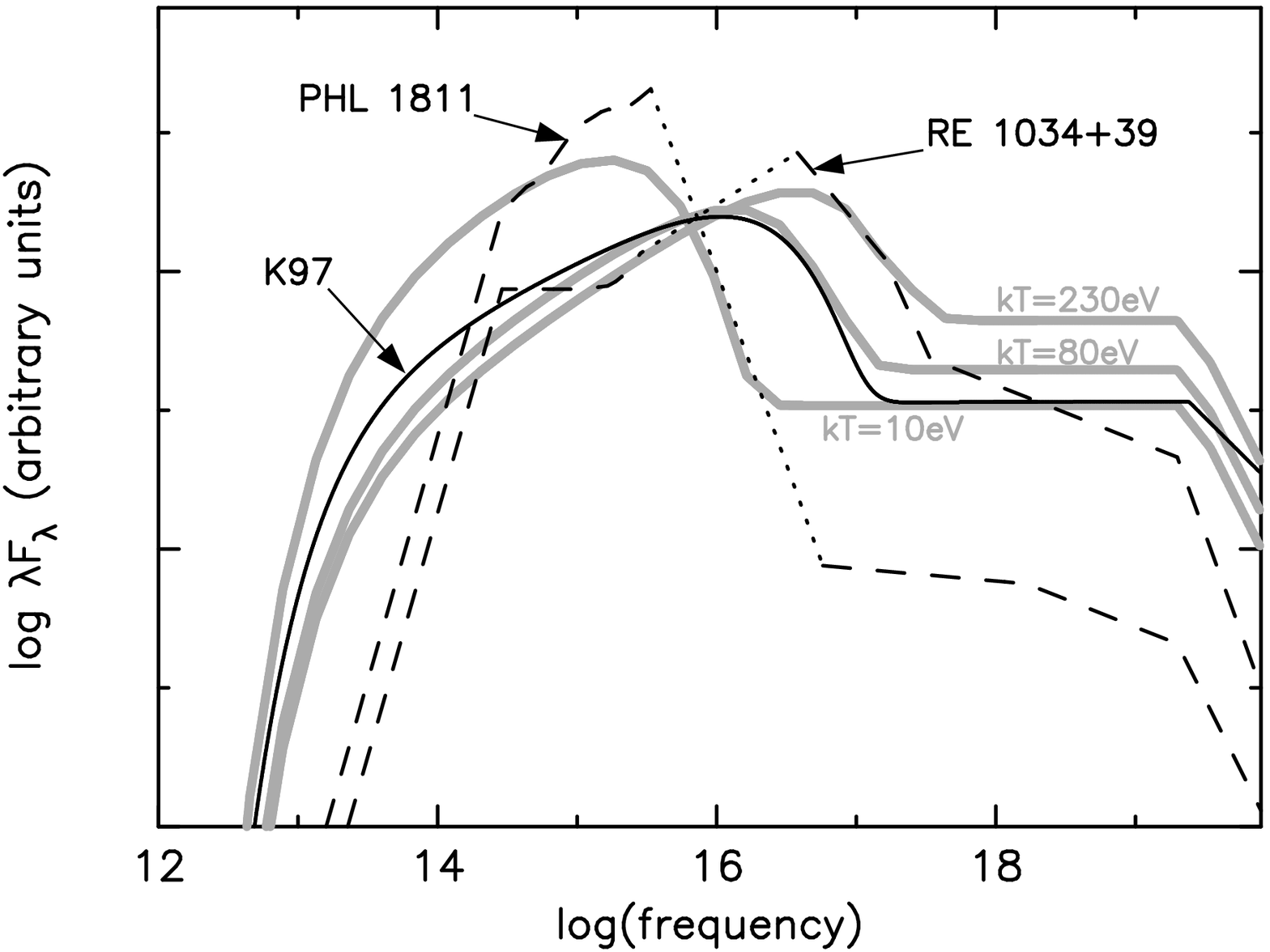}
\caption{Comparison of spectral energy distributions.  The dashed
  lines (dotted lines in the unobservable EUV) show SEDs derived from
  coordinated observations of PHL~1811 and the narrow-line Seyfert 1
  galaxy  RE~1034$+$39 \citep{clb05}.  The solid black line shows the
  K97 continuum intended to represent the SED of a typical AGN
  \citep{k97}.  The thick grey lines show the semiempirical SEDs used
  for {\it Cloudy}  simulations \citep{clb05}, labeled by temperature
  of the  big-blue-bump rollover, that correspond most closely to the
  observed  SEDs ($kT_{cut}=10\rm\, eV$ for PHL~1811 and $kT_{cut}=230
  \rm \,  eV$ for RE~1034$+$39, respectively), and to the K97 SED 
  ($kT_{cut}=80\rm \, eV$).  The SEDs used for the {\it
  Cloudy} simulations correspond reasonably well with the observed
  SEDs and K97 SED, at least in terms of the position of the peak of
  emission. 
\label{fig12}} 
\end{figure}

\clearpage 

\begin{figure}
\epsscale{0.6} \plotone{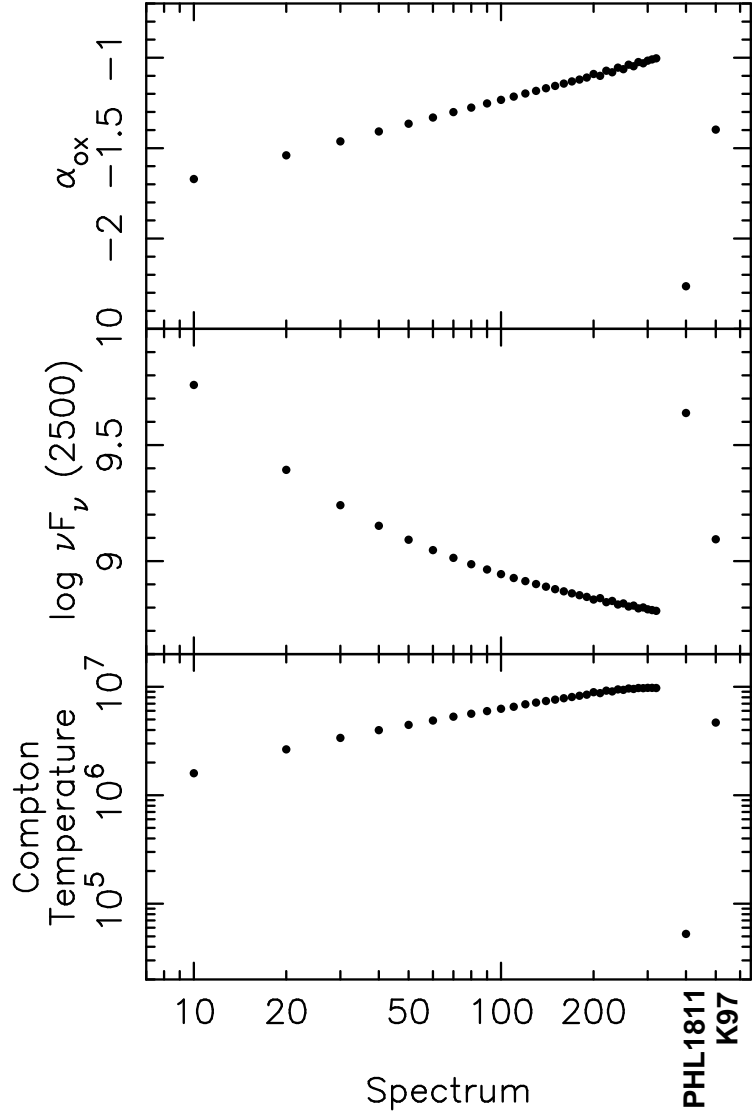}
\caption{Gross properties of the SEDs used for simulations.
  Information for each SED is plotted as a function of the SED, where
  the CLB continua are coded by the big blue bump cutoff temperature
  \citep[leftmost 32 points;   see][for details]{clb05}, and the  two
  points on   the right-hand side   are for the observed   PHL~1811
  continuum, and   the AGN continuum   used by \citet{k97}. The    top
  panel shows the $\alpha_{ox}$, the 
  middle panel shows the   continuum flux at 2500\AA\/ for a
  photoionizing flux of $10^{20}\rm   \, photons\, s^{-1}\, cm^{-2}$,
  and the lower panel shows the   Compton temperature.  The PHL~1811
  continuum is characterized by a   very steep $\alpha_{ox}$, and
  requires a relatively bright optical and UV in order to produce the
  same photoionizing flux as the other continua.  This will reduce the
  line equivalent widths.   The soft spectral energy distribution
  produces a very low Compton   temperature of around  $5\times
  10^4\rm \,K$.
\label{fig13}}   
\end{figure}

\clearpage 

\begin{figure}
\epsscale{1.0} \plotone{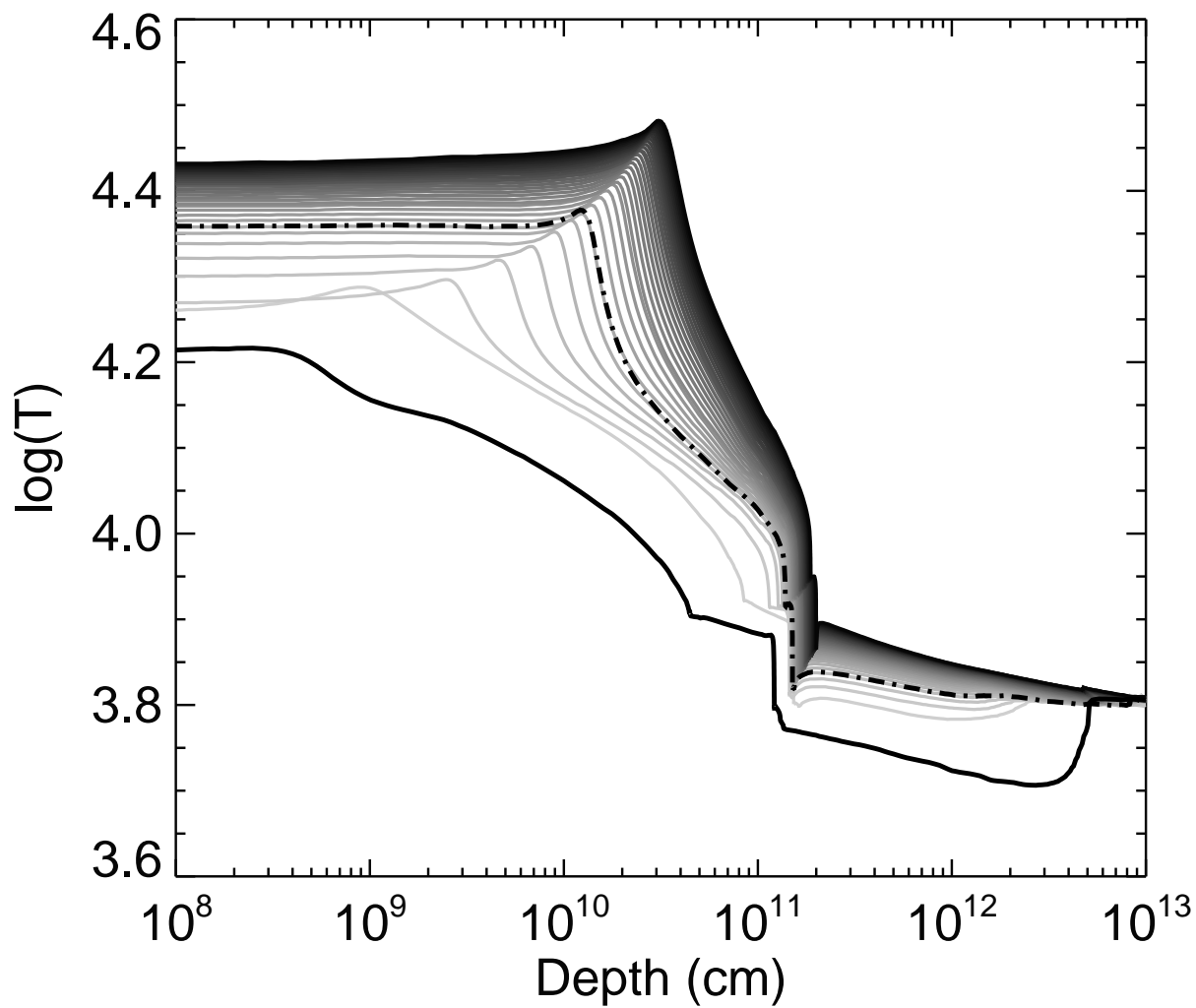}
\caption{The electron temperature as a function of depth for our 
  range of spectral energy distributions and the fiducial ionization
  parameter $\log U=-1.5$ and density $\log n=11$.  The thin lines
  show the CLB continua with light to dark showing the progression of
  soft SEDS ($kT_{cut}=10\rm \, eV$)  to hard ones ($kT_{cut}=320\rm
  \, eV$).   The
  solid thick line and the dash-dot line shows the result for the
  PHL~1811 continuum  and the K97 continuum, respectively.  The
  electron   temperature is lower for the PHL~1811 continuum than any
  other   continuum.
\label{fig14}}  
\end{figure}

\clearpage 

\begin{figure}
\epsscale{1.0} \plotone{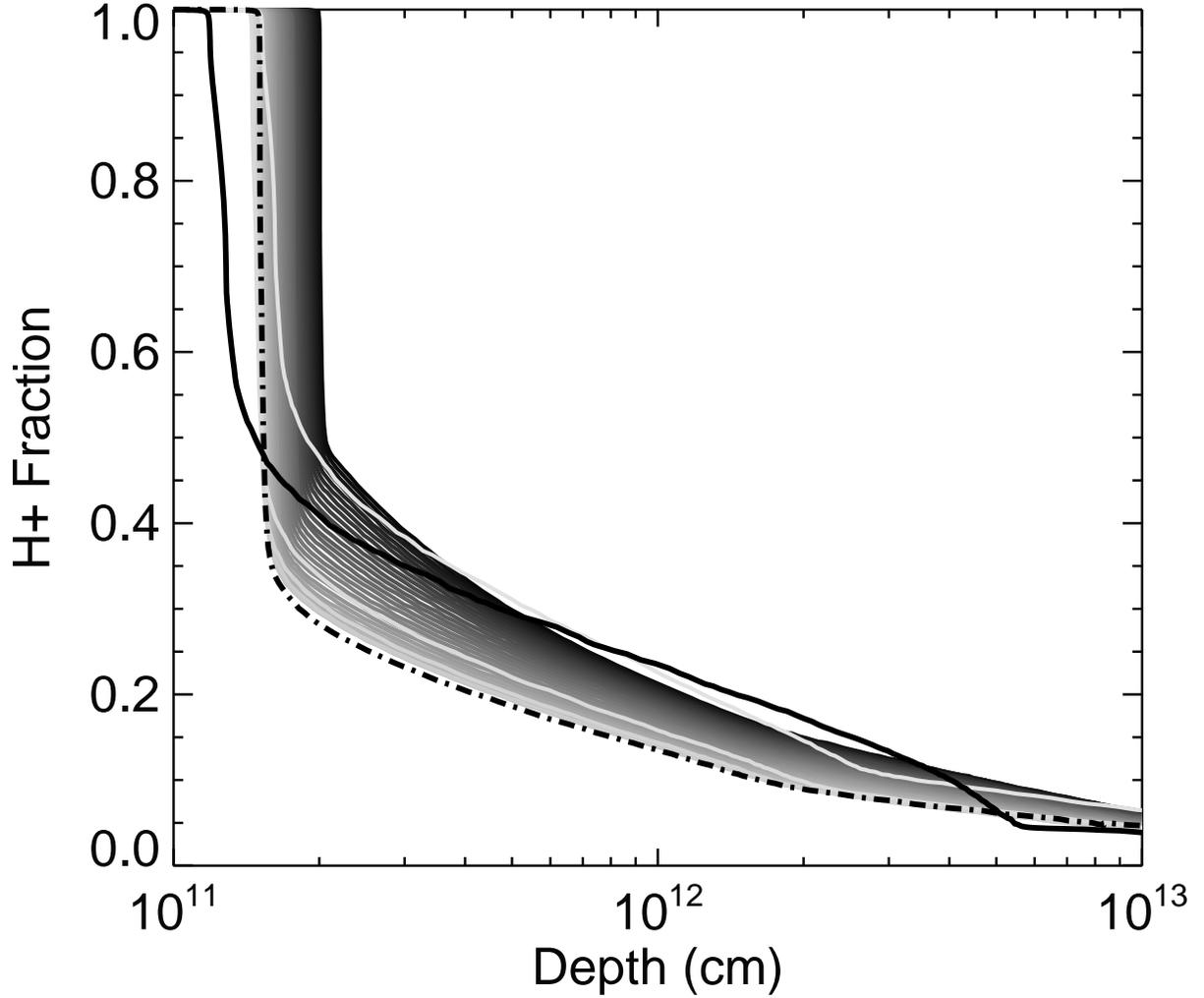}
\caption{The hydrogen ionization fraction in the vicinity of the
  hydrogen ionization front and into the partially-ionized zone for
  the range of spectral energy distributions and the fiducial
  ionization   parameter $\log U=-1.5$ and density $\log n=11.0$. The
  lines have the same meaning as in Fig.\ 14.  For the CLB continua,
  the fraction of ionized hydrogen 
  at a particular depth decreases as the continuum becomes softer
  ($kT_{cut}$ decreases), but then increases again for the softest
  continua.  Likewise, the hydrogen ionization fraction is higher for
  the PHL~1811 continua.  The resulting total ionized hydrogen column
  is 25\% greater for the PHL~1811 continuum than for the K97
  continuum.  
  \label{fig15}}  
\end{figure}

\clearpage 

\begin{figure}
\epsscale{1.0} \plotone{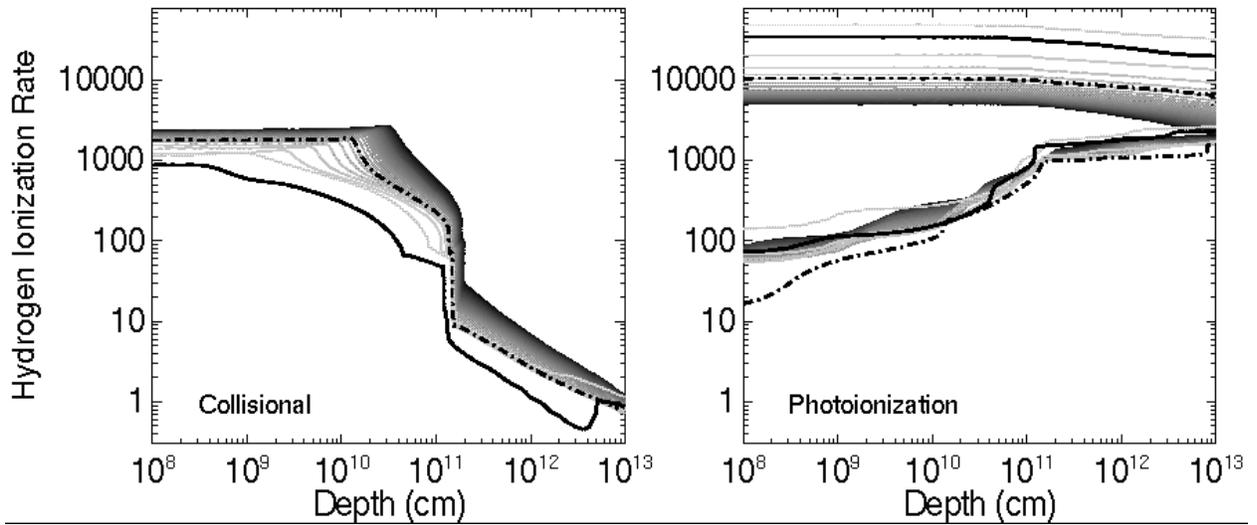}
\caption{The rate of ionization of hydrogen from $n=2$ from collisional
  ionization (left panel) and photoionization (right panel).  The
  upper group of lines in the photoionization plot shows the rate from
  the direct continuum, and the lower group of lines shows the rate
  from the diffuse emission.  The lines and shading have the same
  meaning as in Fig.\ 14.   The rates are dominated by
  photoionization from the direct continuum, and are very large for
  PHL 1811 and other soft SEDs.  
   \label{fig16}}  
\end{figure}

\clearpage 

\begin{figure}
\epsscale{1.0} \plotone{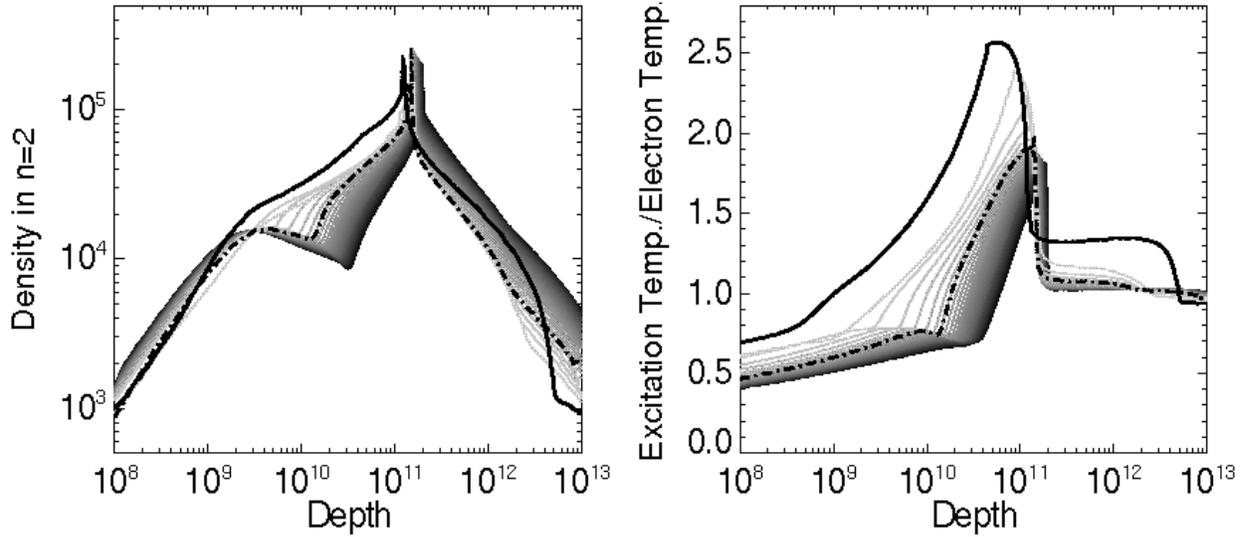}
\caption{{\it Left:} The density of hydrogen in $n=2$ as a function of
depth for the fiducial parameters ($\log U=-1.5$ and $\log n=11$).
The lines have the same meaning as in Fig.\ 14.
In the partially ionized zone, the density of excited hydrogen is
larger in gas illuminated by hard SEDs than in gas illuminated by soft
SEDs, but recovers for PHL~1811.  {\it Right:} The ratio of the
excitation temperature to the electron temperature for the fiducial
parameters.  The ratio is near one for the hard SEDs, implying that
the number of hydrogen in $n=2$ are consistent with that expected
from the electron temperature of the gas, and therefore the excitation
mechanism is collisions.  In contrast, the excitation of PHL~1811 is
higher than predicted by the rather low temperature.   
\label{fig17}} 
\end{figure}

\clearpage 

\begin{figure}
\epsscale{1.0} \plotone{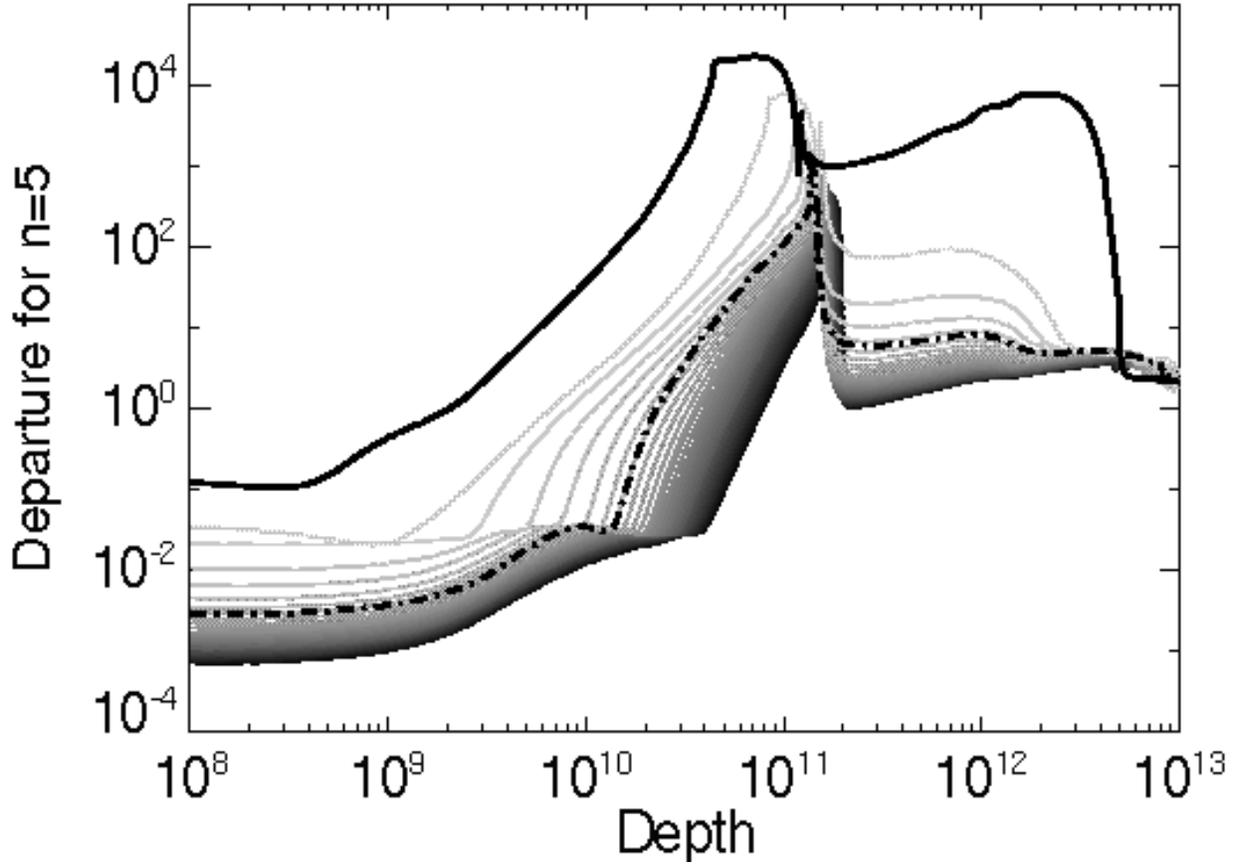}
\caption{The ratio of the density of hydrogen in $n=5$ from the {\it
    Cloudy} output to the density expected based on the density in
    $n=1$ and the electron temperature for the fiducial gas parameters
    ($\log U=-1.5$ and $\log n=11$); this parameter is proportional
    to the departure coefficient for $n=5$.   The lines  have the same
  meaning as in Fig.\ 14.  The departure coefficient is very large for
  PHL~1811. 
\label{fig18}} 
\end{figure}

\clearpage 

\begin{figure}
\epsscale{1.0} \plotone{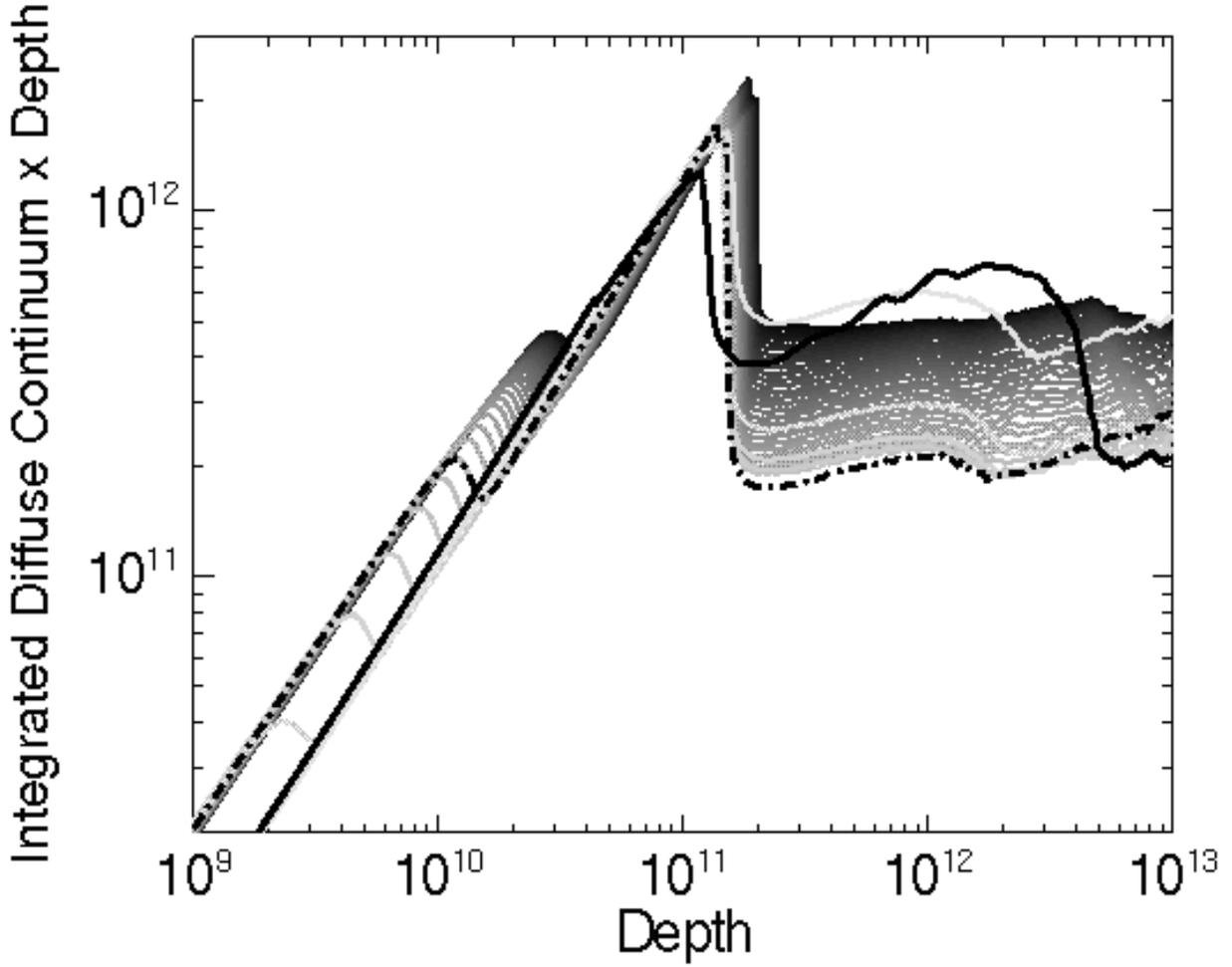}
\caption{Integrated diffuse continuum times the depth (in $\rm
  erg\,s^{-1}\,cm^{-1}$) as a function of depth for  the fiducial parameters
  ($\log U=-1.5$ and $\log n=11$).  The lines  have the same
  meaning as in Fig.\ 14. The energy in the diffuse continuum is
  greater in gas illuminated by the hard continuum compared with the
  gas illuminated by the softer continua, but is even higher in
  PHL~1811.  The energy in the continuum is large in objects with hard
  SEDs because of the greater amount of energy injected into that
  region by the soft X-ray photoionization.  The energy in the
  continuum is large in PHL~1811 because the gas is so cool that usual
  channels of cooling (e.g., excitation of metal ions) are less
  important, and the gas has no options besides emission of
  continuum.  The gas illuminated by the hard continuum
  resembles gas with a high density in that cooling options are
  limited.     
\label{fig19}}  
\end{figure}

\clearpage 

\begin{figure}
\epsscale{0.5} \plotone{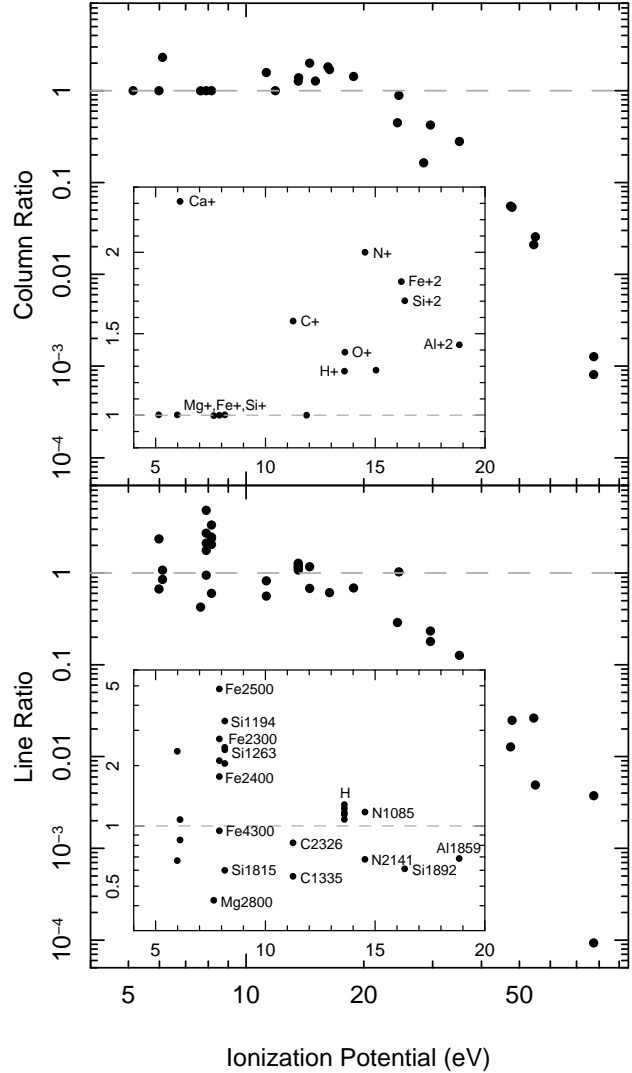}
\caption{A comparison of the predicted ion and emission-line
  properties for the   PHL~1811 continuum and the K97 continuum for
  the fiducial gas   parameters ($\log U=-1.5$ and $\log n=11.0$).   In  both cases, the
  column density is $10^{24.5}$, which is sufficiently large to be
  very   optically thick to the   hydrogen continuum.    The embedded
  small panels show the same information as the larger panels on an
  expanded scale.  The top plot shows the ratio of the predicted ion columns
  for PHL~1811 to K97 as a function of the ionization potential.  For
  the lowest ionization potentials, the columns are nearly the same.
  For the highest ionization potentials, the column is much lower for
  PHL~1811 continuum, as it is too soft to create or excite
  highly-ionized ions.  Intermediate-ionization ions fill the region
  above the hydrogen ionization front, and thus their levels are
  elevated in PHL~1811.  The bottom panel shows the ratio of resulting
  common emission lines.  The disperson for a lines emitted by a
  particular ion arises from the fact that lines that are
  collisionally excited are generally weaker for PHL~1811 because the
  temperature is lower, whereas lines that are excited by continuum
  fluorescence are brighter because the UV continuum is relatively
  brighter for the same ionization parameter and density in PHL~1811
  compared with K97.   
\label{fig20}} 
\end{figure}

\clearpage 

\begin{figure}
\epsscale{0.9} \plotone{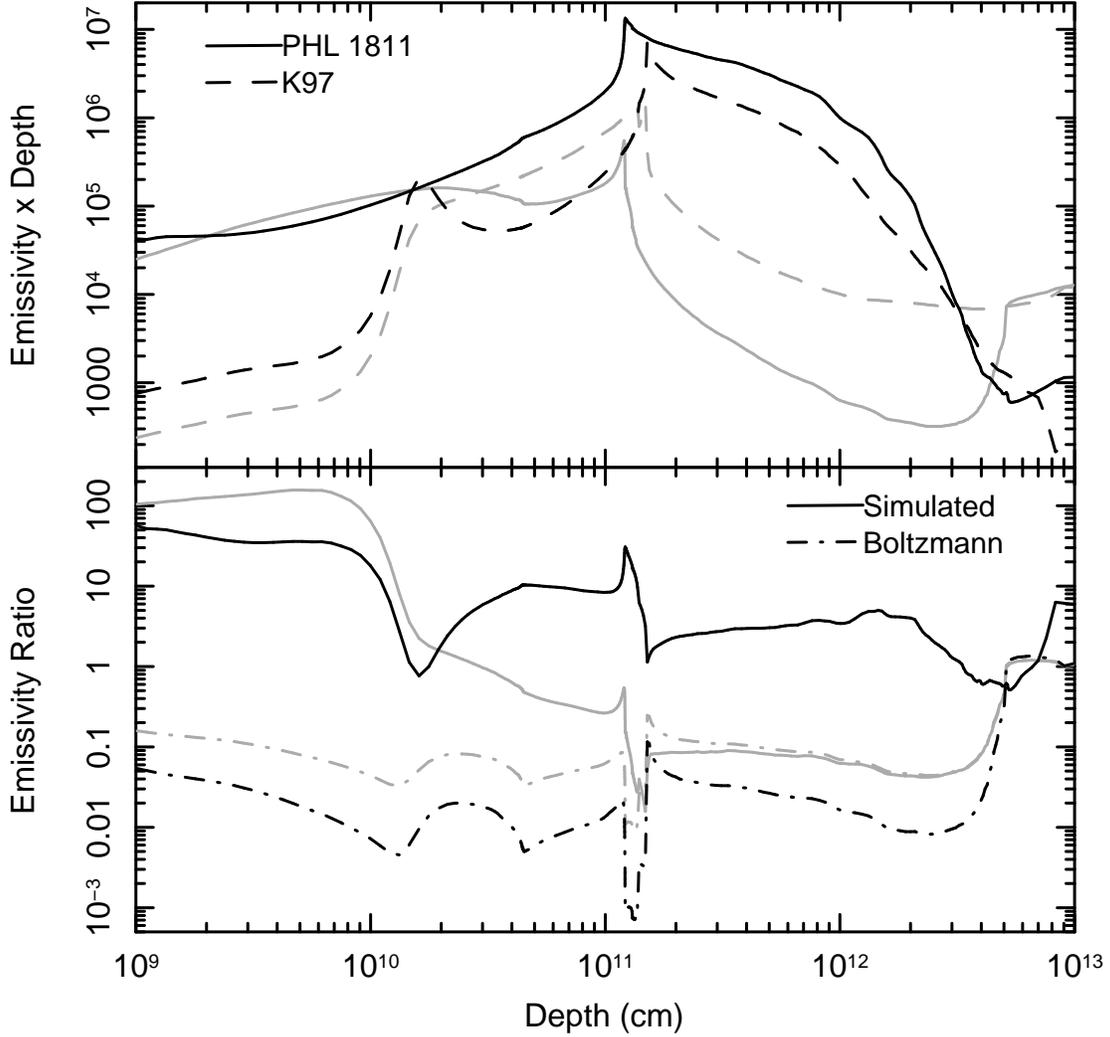}
\caption{A comparison of the line emissivity and ratios for the
  PHL~1811 continuum and the K97 continuum for \ion{Si}{2}~$\lambda
  1814$ (grey lines) and \ion{Si}{2}~$\lambda 1194$ (black lines).  The top
  panel shows the line emissivity multiplied by the depth as a
  function of the depth.  For the 1814\AA\/  line, the emissivity of
  PHL~1811 is smaller than that of K97 particularly in the
  partially-ionized zone where Si$^+$ dominates (depths $>\sim
  10^{11}\rm \,cm$).  In
  contrast, the emissivity of the 1194\AA\/ line is   much higher for
  PHL~1811.  The lower panel shows the ratios from the gas
  illuminated by the  PHL~1811 continuum to that of the gas
  illuminated by the K97 continuum.  The solid lines show the ratios
  computed from the {\it Cloudy} predictions of the line emissivity as
  a function of depth, while the dashed lines show the ratios computed
  using the electron temperatures and the Boltzmann excitation
  equation.   The two ratios match well for the
  collisionally-excited line at  1814\AA\/; deviations at low depths
  are a consequence of different fractional abundances of Si$^+$
  ions.  But the  observed   ratio is much larger than the Boltzmann
  equation ratio   for the   1194\AA\/ line.  This is because the
  1194\AA\/ line is  predominately excited by continuum fluorescence,
  and much brighter   in PHL~1811 due to its relatively brighter UV
  continuum for the same   ionization parameter and density. 
\label{fig21}} 
\end{figure}

\clearpage 

\begin{figure}
\epsscale{1.0} \plotone{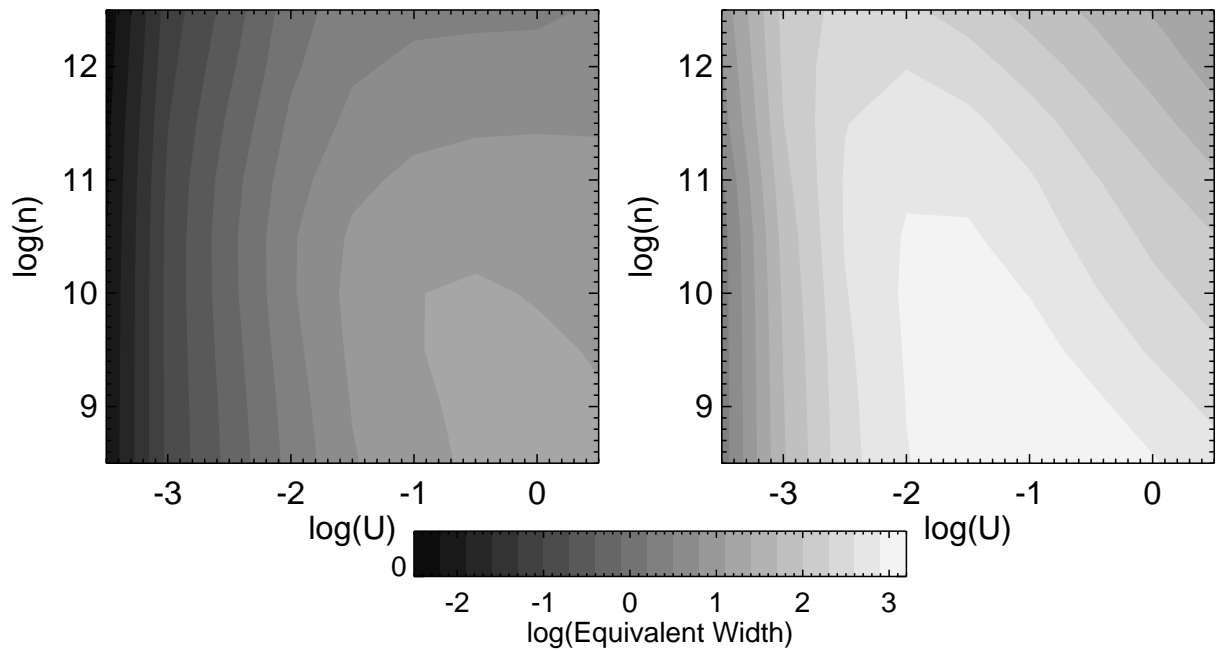}
\caption{A comparison of the line equivalent widths of \ion{C}{4} for
  the   PHL~1811 continuum (left) and the K97 continuum (right) as a
  function of ionization parameter and density.  \ion{C}{4} is
  predicted to be weak in PHL~1811 because of the low column of
  C$^{+3}$ ions and the lower temperature in the emitting gas of this
  collisionally excited line.  
\label{fig22}} 
\end{figure}

\clearpage 

\begin{figure}
\epsscale{1.0} \plotone{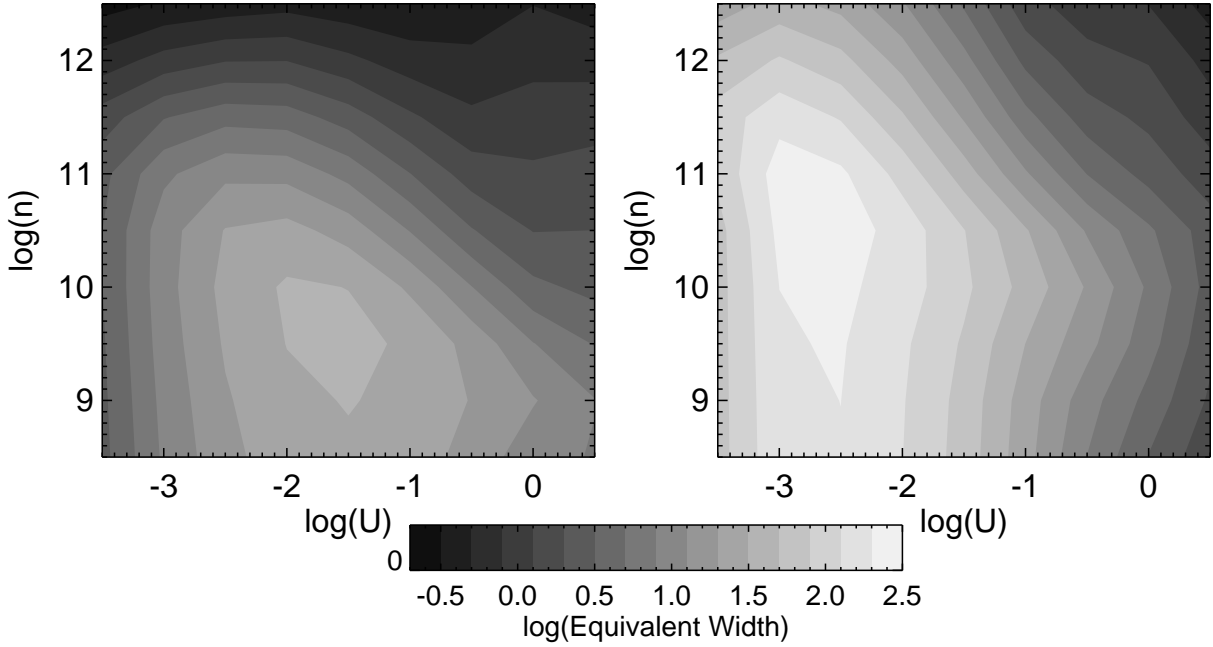}
\caption{A comparison of the line equivalent widths of \ion{Si}{3}] for
  the   PHL~1811 continuum (left) and the K97 continuum (right) as a
  function of ionization parameter and density. This semiforbidden
  line has a critical density of $3\times 10^{11}\rm \, cm^{-3}$, so a
  decline   toward higher densities in the K97 simulation is
  expected.  In contrast, in the PHL~1811 continuum simulations, the
  onset of the decline toward higher densities occurs at much lower
  densities, around $10^{10}\rm \,   cm^{-3}$.  This implies that for
gas illuminated by soft continua, the usual density  indicators are
  not reliable. 
\label{fig23}} 
\end{figure}

\clearpage

\begin{figure}
\epsscale{0.4} \plotone{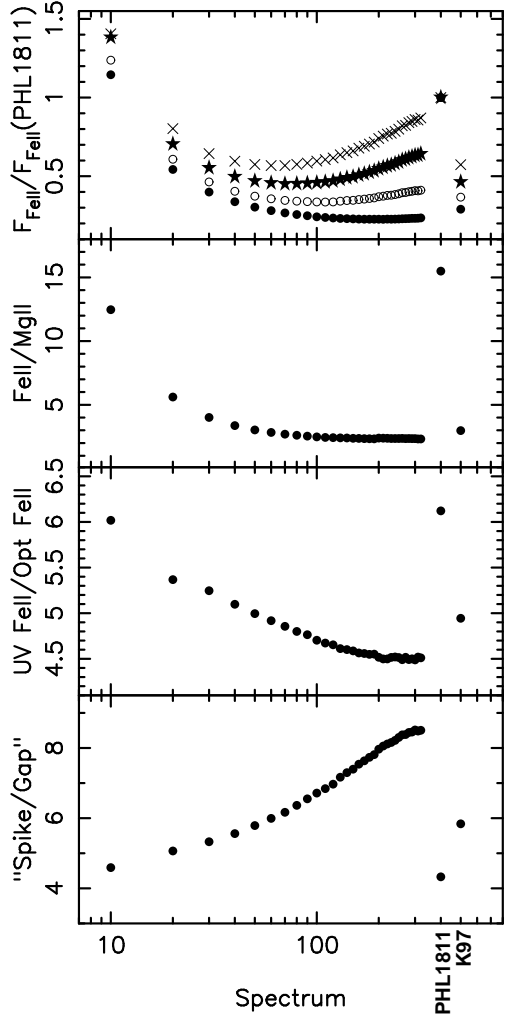}
\caption{Properties of \ion{Fe}{2} using the {\it Cloudy} 371-level
  atom for the  fiducial parameter ($\log U=-1.5$ and $\log n=11$).
{\it Top:} The ratio of the \ion{Fe}{2} emission in broad
  bands for the range of spectral energy distributions compared to that of
  PHL~1811.  Solid circles: 1000--2000\AA\/, solid stars:
  2000--3000\AA\/, $\times$'s: 
  4000--6000\AA\/, open circles: 7800\AA\/--3 microns.  {\it Second from top:}
  Ratio of \ion{Fe}{2} flux in the 2200--3050\AA\/ band to the
  \ion{Mg}{2} flux.  The large ratio for PHL~1811 is a combination of 
higher \ion{Fe}{2} flux, a consequence of stronger continuum pumping,
  and lower \ion{Mg}{2} flux, a consequence of lower electron
  temperature.  {\it Third from the top:}  The ratio of the
  \ion{Fe}{2} emission in the UV (2000-3000\AA\/) to the \ion{Fe}{2}
  emission in the optical (4000--6000\AA\/).  This ratio is high in
  PHL~1811 because the continuum pumping enhances the UV emission, and
  because of \ion{Fe}{2} emission in the H$^+$--neutral He region.
  {\it Bottom:} The ``spike/gap'' ratio, defined by
  \citet{baldwin04}.  This ratio is essentially the emission from the
  low-lying, collisionally-excited levels that produce the spikes, to
  the emission from high levels.  The high-level emission is enhanced
  in PHL~1811 due to continuum pumping.    
\label{fig24}} 
\end{figure}

\clearpage

\begin{figure}
\epsscale{1.0} \plotone{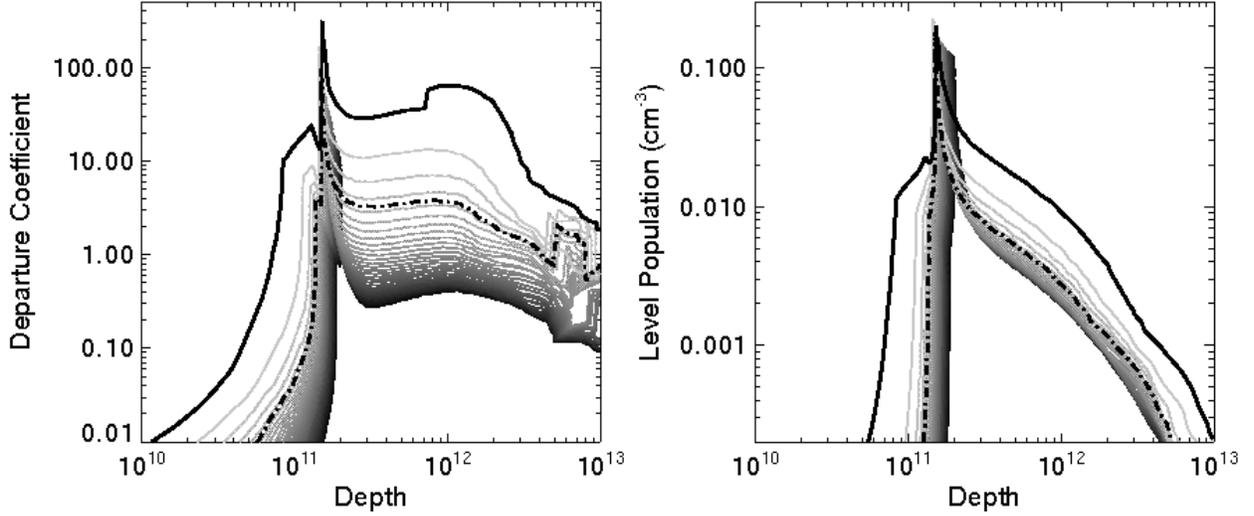}
\caption{Properties of level number 295 (10.85 eV) as a function of
  depth for a range of spectral energy distributions for the fiducial
  parameters ($\log U=-1.5$, $\log n=11$, and $\log N_H=24.5$).  The
  lines have the same meaning as in Fig.\ 14.  {\it Left:} The
  departure coefficient shows that level 295 is much more highly
  populated (by a factor of almost 100 in the partially-ionized zone)
  in the gas illuminated by the PHL~1811 spectral energy distribution
  than expected from the local electron temperature.  {\it Right:} The
  density of Fe$^+$ ions in level 295 shows that the departure
  coefficient is high for gas in PHL~1811 both because the temperature
  is low (Fig.\ 14) and because the number of atoms with electrons in
  this level is high.  
\label{fig25}} 
\end{figure}

\clearpage

\begin{figure}
\epsscale{1.0} \plotone{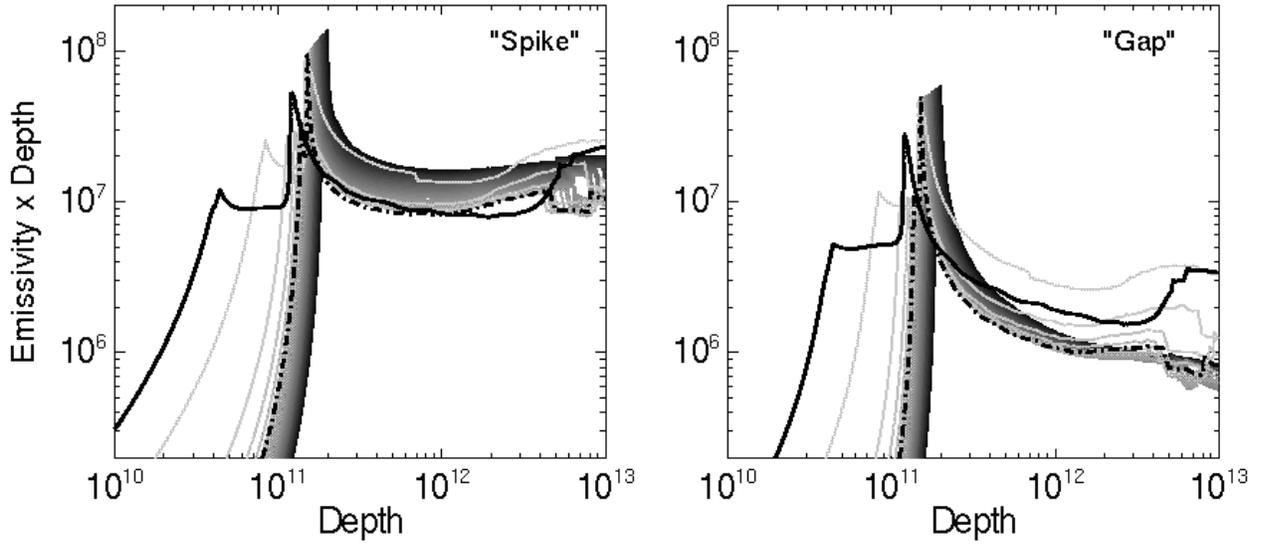}
\caption{The emissivity times depth as a function of depth for two
  regions of the \ion{Fe}{2} spectrum for the fiducial parameters
  ($\log U=-1.5$, $\log n=11$, and $\log N_H=24.5$). The lines have
  the same meaning as in Fig.\ 14.  Note that the emissivity has the
  opacity factored in; the value is the amount reaching the surface.
The left side shows the ``spike'' region of the spectrum (2312--2328
  and 2565--2665\AA\/) and the right side shows the ``gap'' region
  \citep[2462--2530\AA\/;][]{baldwin04}.  In both cases, the ``hook''
  between $\sim 3 \times 10^{10}$ and $\sim 10^{11}\rm \, cm$ for
  PHL~1811 originates in the H$^+$-neutral He region.  The spike is
  predominately resonance transitions, and the emission in the
  partially-ionized zone is approximately proportional to the
  temperature of the gas (although somewhat elevated for the softest
  continua).  The gap is predominately emission from higher levels;
  these levels are pumped by the continuum and Ly$\alpha$ when the SED
  is soft, and  they are higher in the partially-ionized zone.
\label{fig26}} 
\end{figure}

\clearpage

\begin{figure}
\epsscale{1.0} \plotone{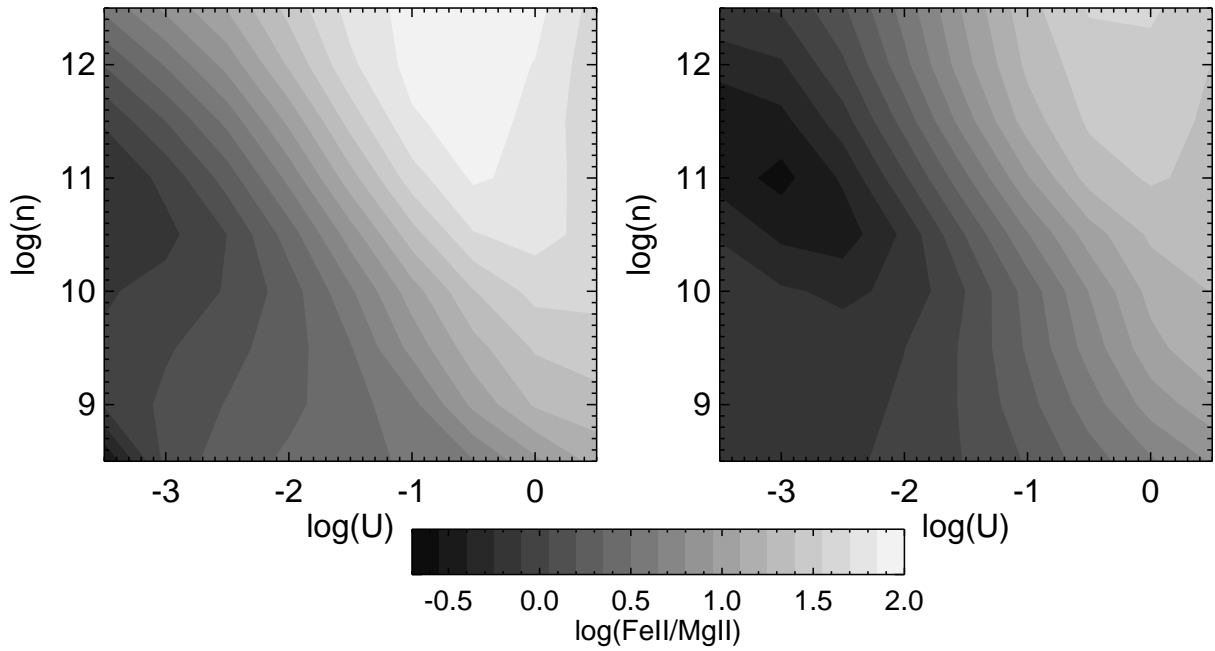}
\caption{The log of the ratio of the \ion{Fe}{2} emission in
  2200-3050\AA\/ band to the \ion{Mg}{2} emission as a function of
  ionization parameter and density.  In each case, the column density
  is proportional to the ionization parameter such that $\log N_H +
  \log U=26$.  The ratio is consistently higher for PHL~1811 (left)
  compared with the K97 continuum (right).
\label{fig27}} 
\end{figure}

\clearpage

\begin{figure}
\epsscale{1.0} \plotone{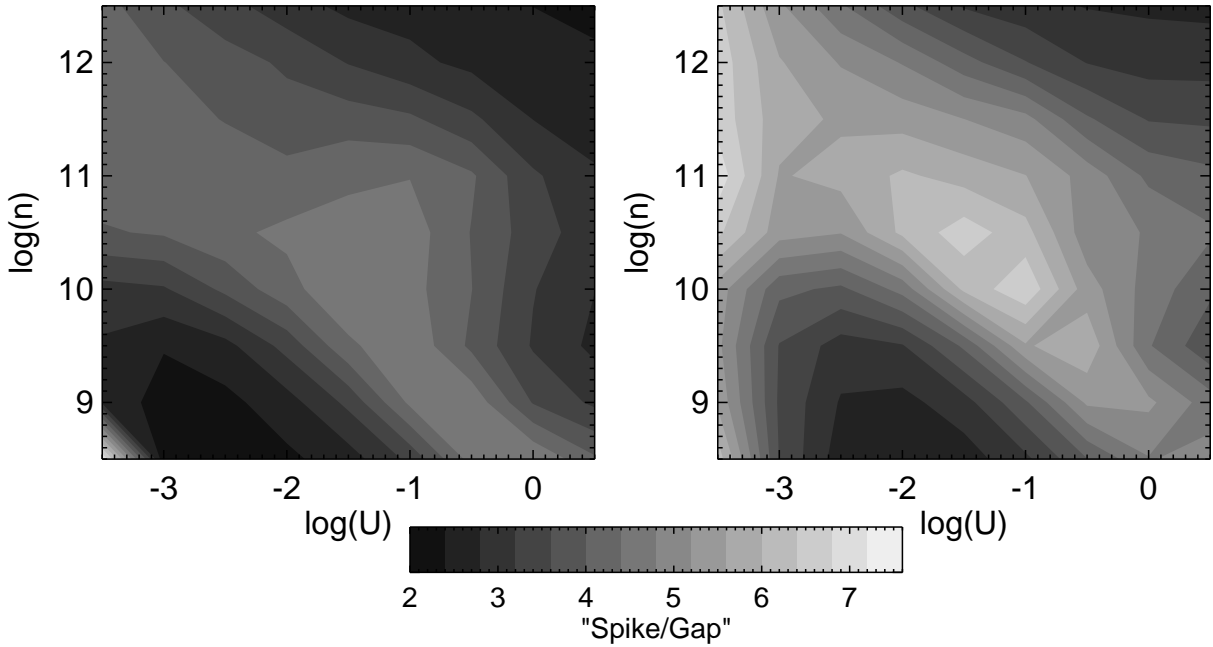}
\caption{The ``spike/gap'' ratio for the PHL~1811 (left) and K97
  (right) SEDs as a function of ionization parameter and density.  The
  ratio for the PHL~1811 continuum is consistently lower than for the
  K97 continuum, although still much larger than the observed value
  \citep[0.7;][]{baldwin04}.  As discussed in the text, the lower
  ratio originates from higher emission from high levels that comprise
  the ``gap'' emission in gas illuminated by the PHL~1811 continuum.
\label{fig28}} 
\end{figure}

\clearpage

\begin{figure}
\epsscale{0.9} \plotone{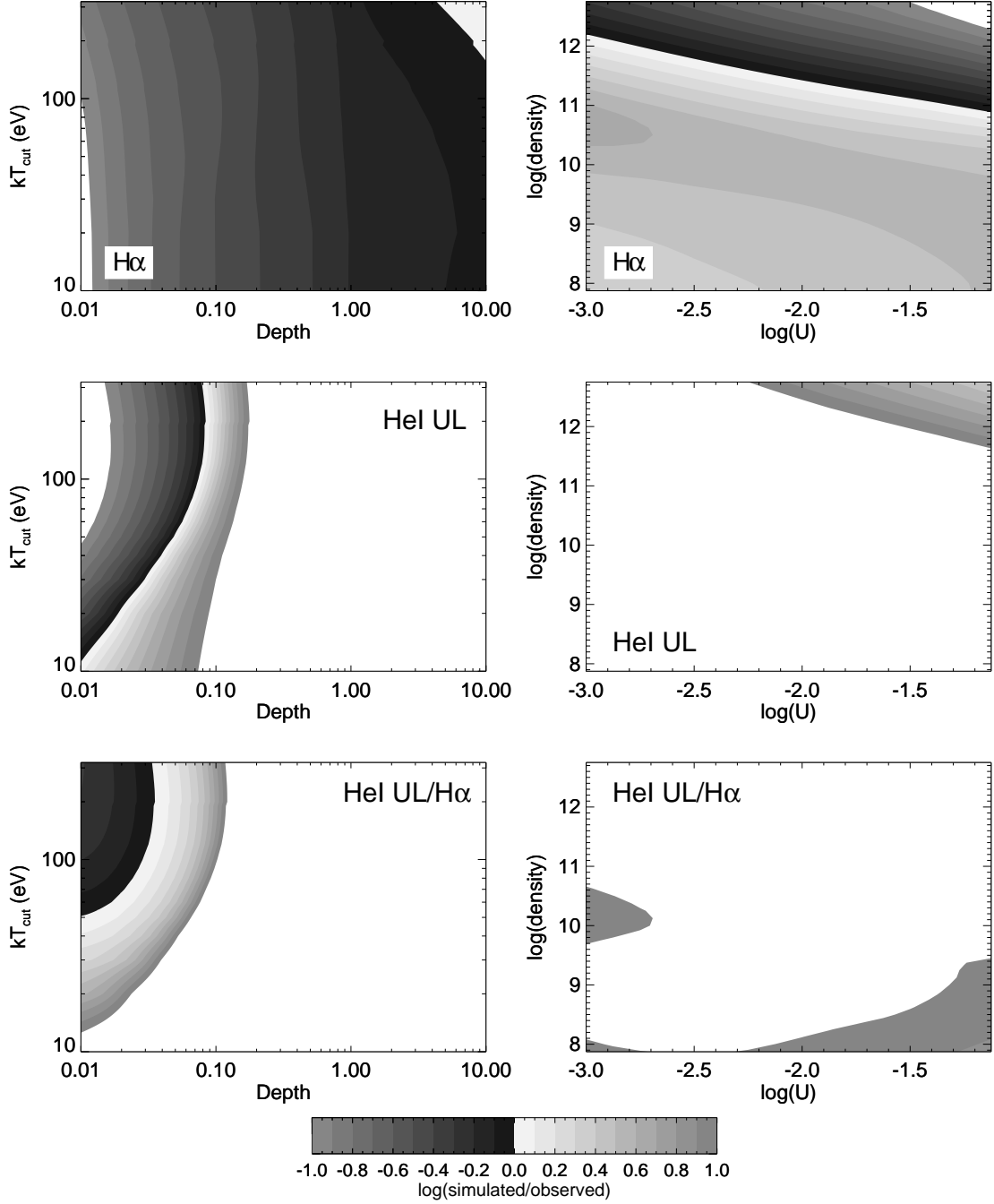}
\caption{Contours for {\it Cloudy} simulations of H$\alpha$, the upper
  limit for \ion{He}{1}~$\lambda$5875, and their ratio.  Only regions
  in which the log of the ratio of the simulated to observed is
  between $-1$ and 1 (i.e., 1 dex from a perfect match) are shaded.
  Four parameters ($kT_{cut}$, $D$, $\log(density)$ and $\log(U)$
  define the parameter space of the simulations, where $D$ is ratio of
  the depth to the depth of the hydrogen ionization front.  Shown are
  contours in planes defined by two of the four parameters: $kT_{cut}$
  and $D$ on the left side, and $\log(density)$ and $\log(U)$ on the
  right side.  These  planes intersect at   the point $kT_{cut}=10$,
  $\log(density)=11.5$, $\log(U)=-2.0$, and $D=10$.  
\label{fig29}}
\end{figure}

\clearpage

\begin{figure}
\epsscale{1.0} \plotone{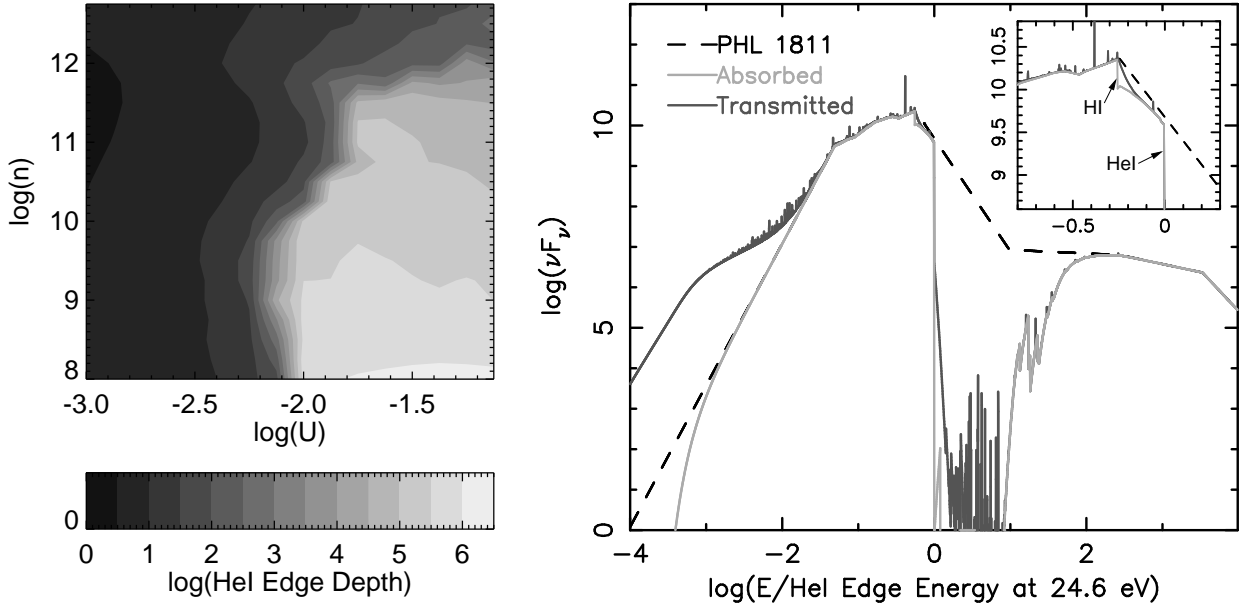}
\caption{Results from {\it Cloudy} simulations in which the column
  density is chosen to maximize the ratio of the He$^+$ and H$^+$
  columns.  {\it Left:} The log of the ratio of the transmitted
  continuum just below the \ion{He}{1} edge (23.9~eV) to the
  transmitted continuum just below the \ion{N}{2} edge (28.8~eV).  A
  plateau exists upon which the \ion{He}{1} edge is very prominent in
  the transmitted spectrum. {\it Right:} A representative example
  continuum for $\log(U)=-1.5$ and $\log(n)=11.5$.  The dashed black
  line shows the PHL~1811 continuum.  The light grey line shows the
  continuum after it is absorbed by the gas.  The dark grey line shows
  the absorbed plus diffuse continuum, or the transmitted continuum.
  The inset shows a magnified view of the region around the Lyman and
  \ion{He}{1} edges.  The diffuse continuum fills in the Lyman edge.
\label{fig30}} 
\end{figure}

\clearpage

\begin{figure}
\epsscale{0.9} \plotone{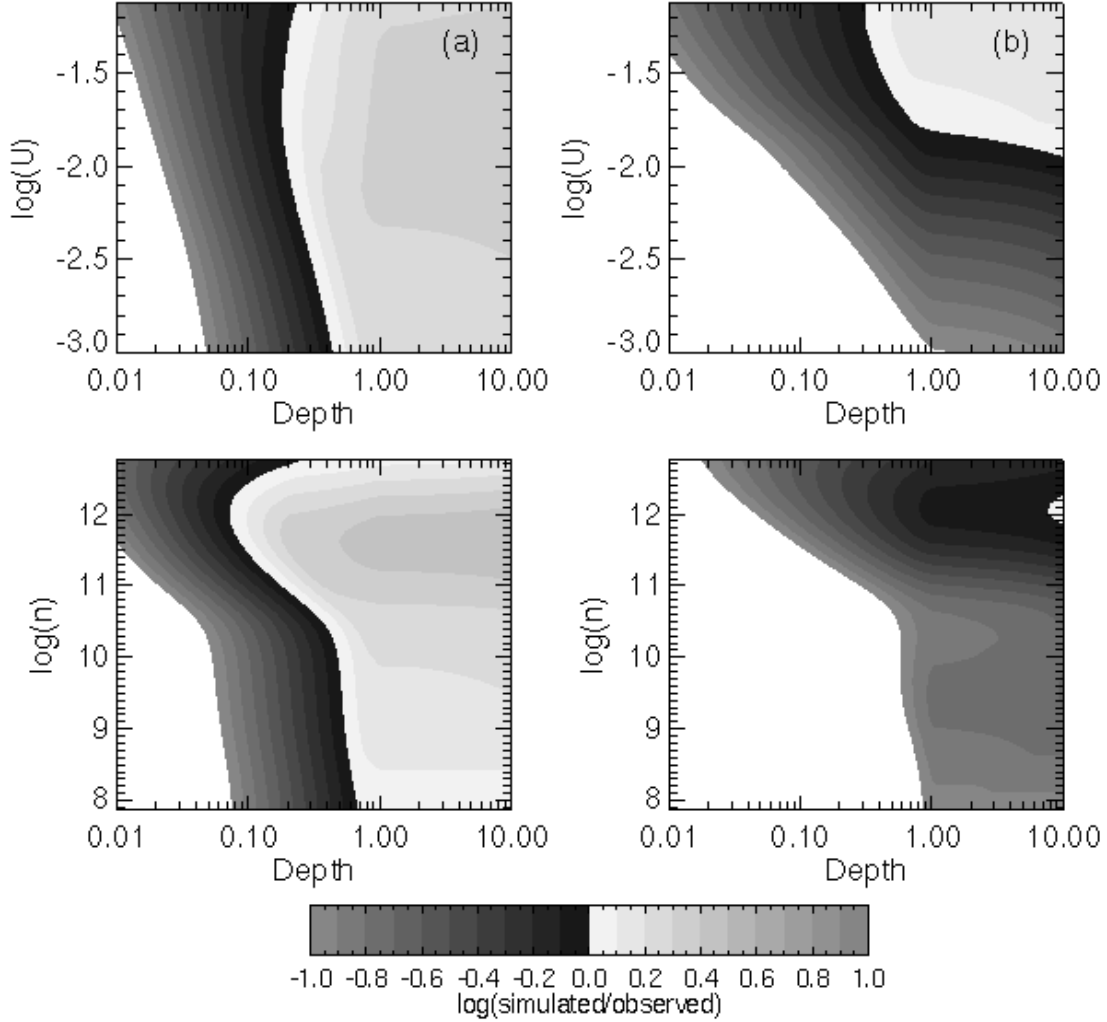}
\caption{The results of filtering the continuum on the
  \ion{He}{1}~$\lambda 5786$ emission for the case of the PHL~1811
  continuum.  The left column,
  labeled (a.), shows the \ion{He}{1} emission with respect to the
  upper limit when the emitting gas is illuminated directly by the
  continuum (no filtering).  A covering fraction of 10\% is assumed.
  Except for low column densities, the   predicted \ion{He}{1} exceeds
  the upper limit by a  large   factor.  The right column, labeled
  (b.), shows the \ion{He}{1}   emission with respect to the upper
  limit when the emitted gas is illuminated by the transmitted
  continuum shown in Fig.\ 30 (the filtered continuum).  The predicted
  \ion{He}{1} emission is lower than the  upper limit for low to
  moderate ionization parameters.  In both columns, the
  top panel shows the case where $\log(n)=10.5$, and the bottom panel
  shows the case where $\log(U)=-2.75$.  Note that the ionization
  parameter shown is the one appropriate for the unfiltered continuum
  in both cases. 
\label{fig31}} 
\end{figure}

\clearpage

\begin{figure}
\epsscale{0.9} \plotone{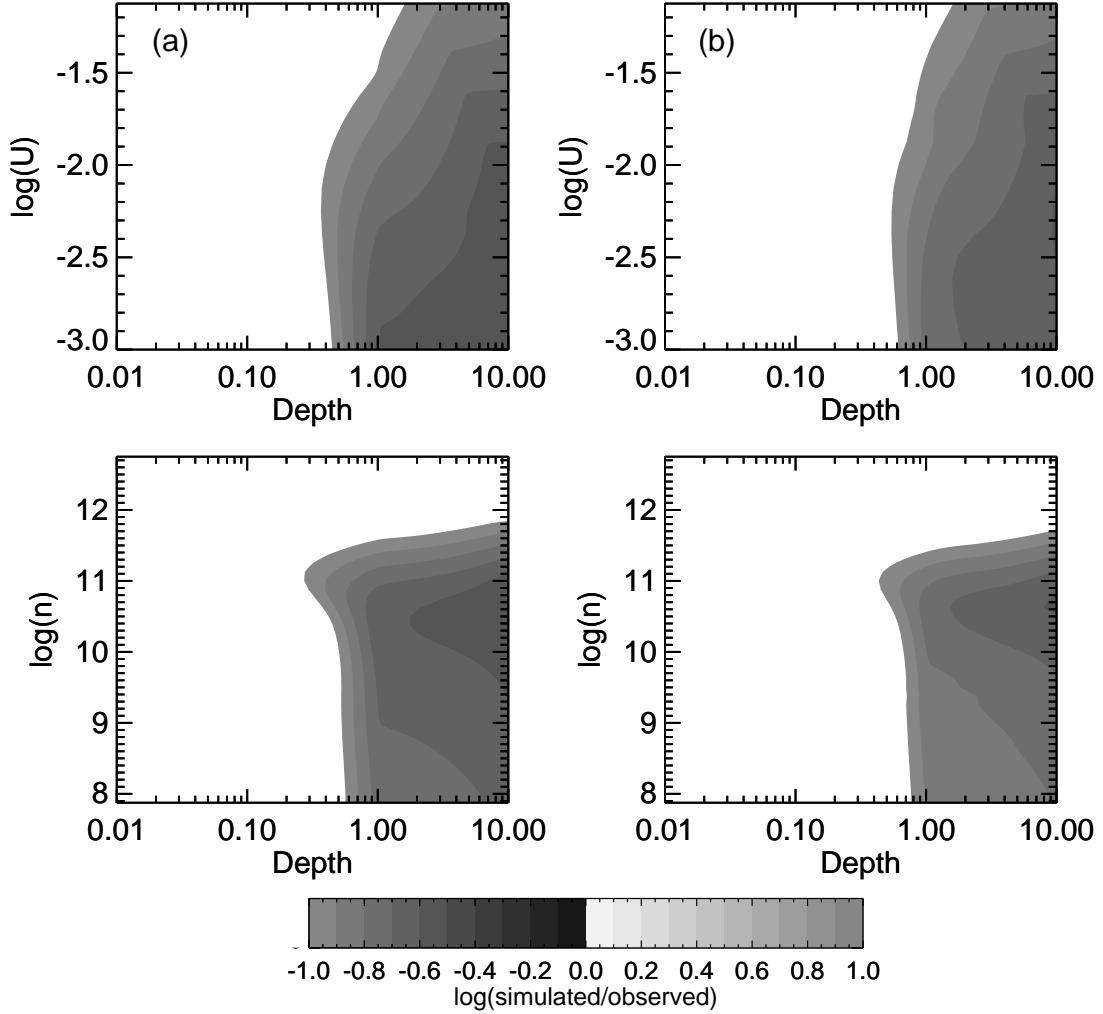}
\caption{The corresponding results of filtering the continuum on the
  H$\alpha$ emission for the same case as shown in Fig.\ 31. In both
  columns, the top panel shows the case where $\log(n)=10.5$, and the
  bottom panel shows the case where $\log(U)=-2.75$.  Again, a
  covering fraction of 10\% is assumed. Note that the ionization
  parameter shown is the one appropriate for the unfiltered continuum
  in both cases.  There is little difference between the
  non-filtered case (left) and the filtered case (right).  This is
  expected because the Lyman continuum is not much altered by the
  transmission, as shown in Fig.\ 30.  H$\alpha$ is in the vicinity of
  the observed value (although lower by a factor of two) at high
  column densities and moderate densities.   
\label{fig32}} 
\end{figure}

\clearpage

\begin{figure}
\epsscale{1.0} \plotone{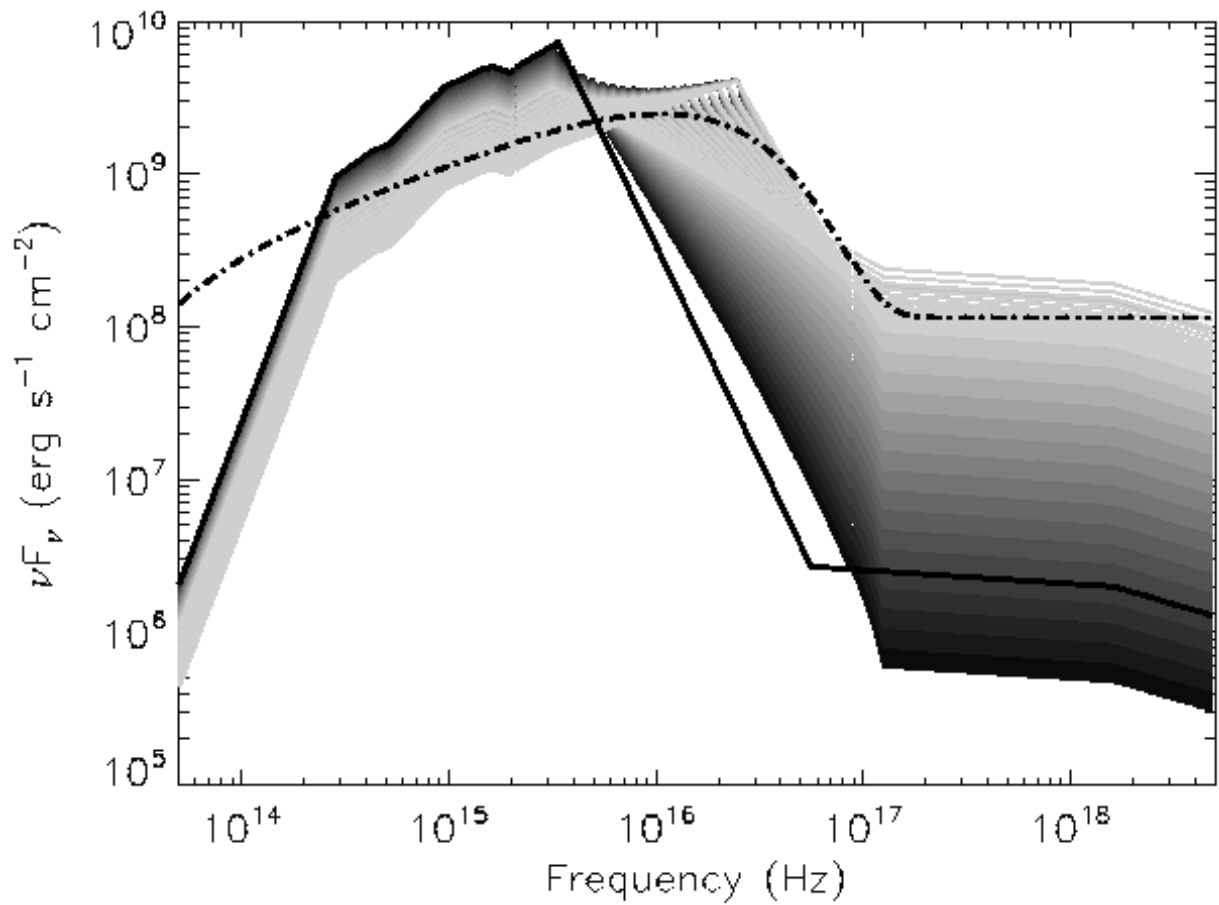}
\caption{The spectral energy distributions used in Appendix A.1.  The
  SEDs are   based on the PHL~1811 and are described by two parameters:
  $\alpha_{ox}$, and the break between the UV and the X-ray.  All SEDs
  are shown for the same value of the photoionizing flux.  These
  SEDS span parameter space described by the strength of
  the extreme UV and the strength of the X-ray between the PHL~1811
  (thick solid line) and the K97 (thick dash-dot line).
\label{fig33}} 
\end{figure}

\clearpage

\begin{figure}
\epsscale{0.9} \plotone{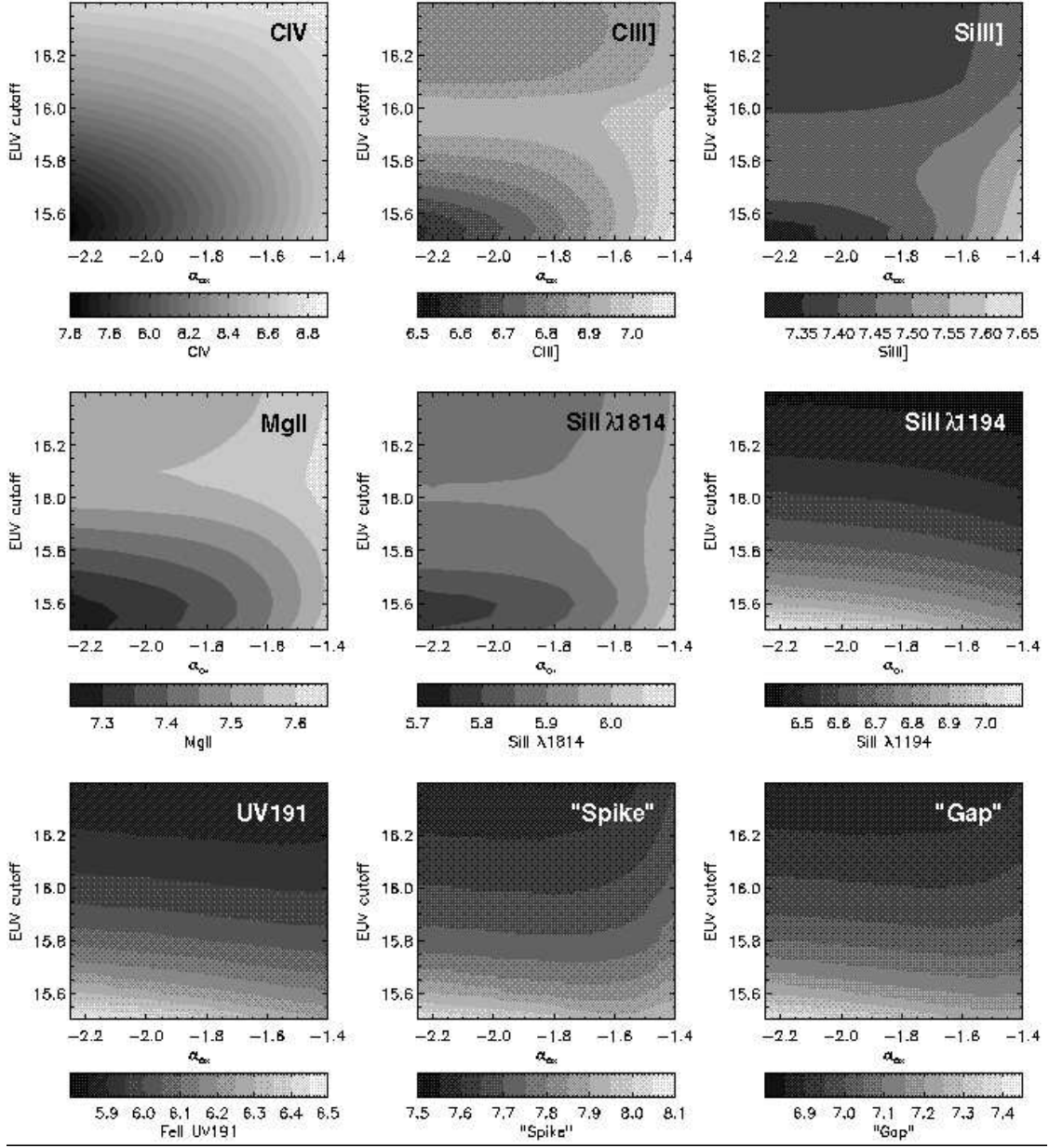}
\caption{Contours of the log of the line flux as a function of the
  parameters describing the SEDs shown in Fig.\ 33, $\alpha_{ox}$
  and the EUV cutoff, evaluated for the fiducial parameters ($\log
  U=-1.5$, $\log n=11.0$, and $\log N_H=24.5$).  A variety of lines are
  shown, including lines emitted in the \ion{H}{2} region (top row),
  the partially-ionized zone (middle row), and UV \ion{Fe}{2} lines
  (bottom row).  In each graph the shaded contour interval is 0.05.
 The behavior of high-ionization lines is governed by the ionization
  potential of the lines.  Collisionally excited intermediate- and
  low-ionization lines such as \ion{C}{3}], \ion{Si}{3}], \ion{Mg}{2},
  and \ion{Si}{2}~$\lambda 1814$ depend primarily on the temperature.
  Pumped lines including \ion{Si}{2}~$\lambda 1194$, and \ion{Fe}{2}
  lines UV~191, and the \citet{baldwin04} ``spike'' and ``gap'' lines
  depend on the UV flux, which is stronger for a given photoionizing
  flux for lower values of the EUV cutoff, and the temperature.  
\label{fig34}} 
\end{figure}

\clearpage

\begin{figure}
\epsscale{1.0} \plotone{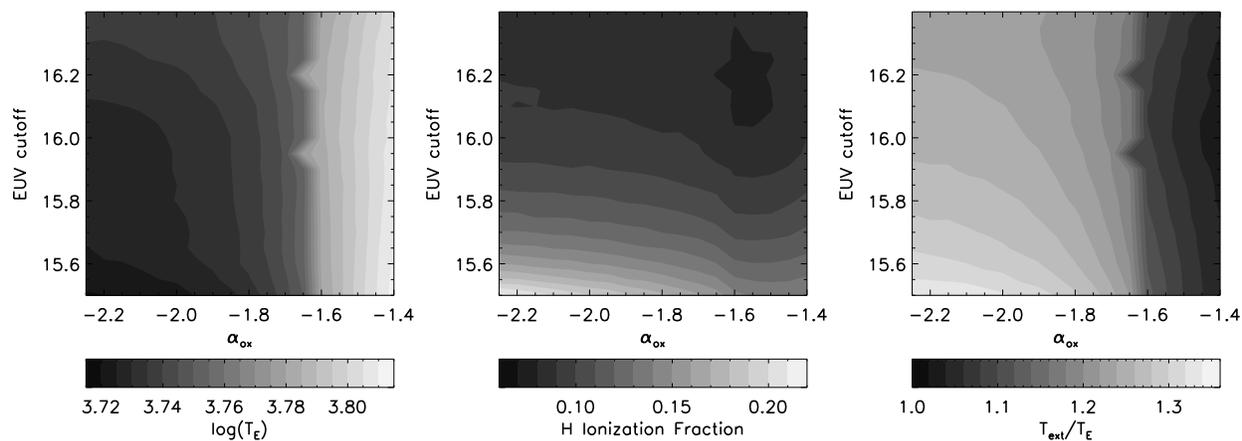}
\caption{Properties of gas illuminated by SEDs shown in Fig.\ 33,
  evaluated at a depth of ten times the depth of the hydrogen
  ionization front.  The electron temperature (left), the hydrogen
  ionization fraction (middle) and the ratio of the excitation
  temperature to the electron temperature for $n=2$ are shown as
  functions of the parameters describing the SEDs, $\alpha_{ox}$
  and the EUV cutoff. 
\label{fig35}} 
\end{figure}

\clearpage

\begin{figure}
\epsscale{1.0} \plotone{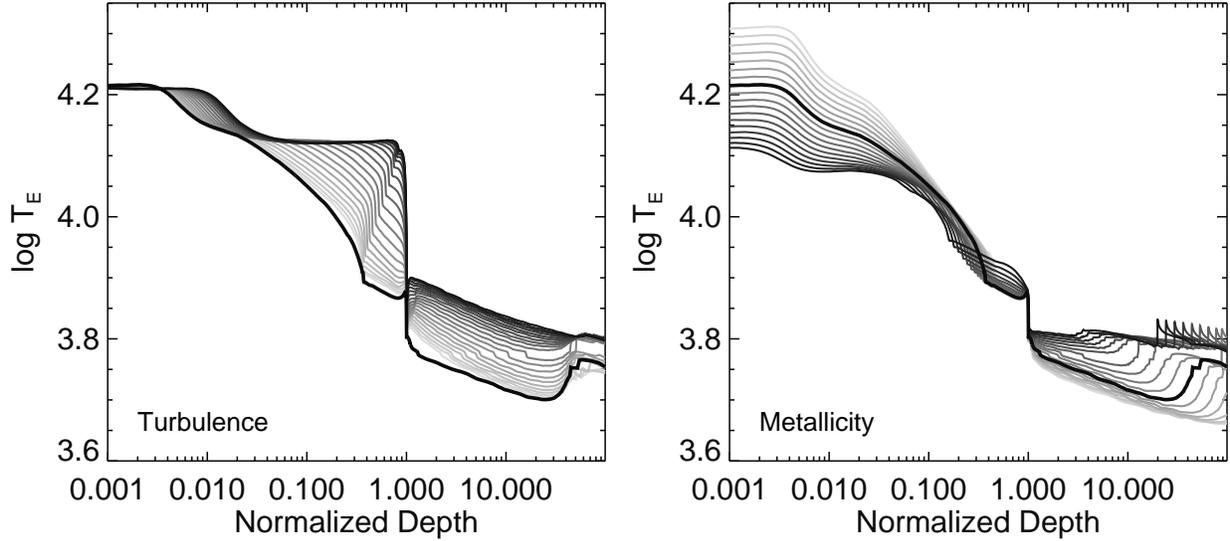}
\caption{The electron temperature as a function of the ratio of the
  depth to the depth of the hydrogen ionization front for the fiducial gas
  parameters ($\log U=-1.5$, $\log n=11$, and $\log N_H=24.5$) as
  microturbulence (left) and metallicity (right) are varied.  The
  solid line shows the result for the PHL~1811 continuum with
  $v_{turb}=0$ and $Z/Z_\odot=1$.  Light to dark shows the results as
  the parameters are increased.  Increase of either turbulence or
  metallicity results in an increase in temperature in the
  partially-ionized zone to values comparable to those obtained for
  typical AGN continua (Fig.\ 14).   
\label{fig36}} 
\end{figure}

\clearpage

\begin{figure}
\epsscale{1.0} \plotone{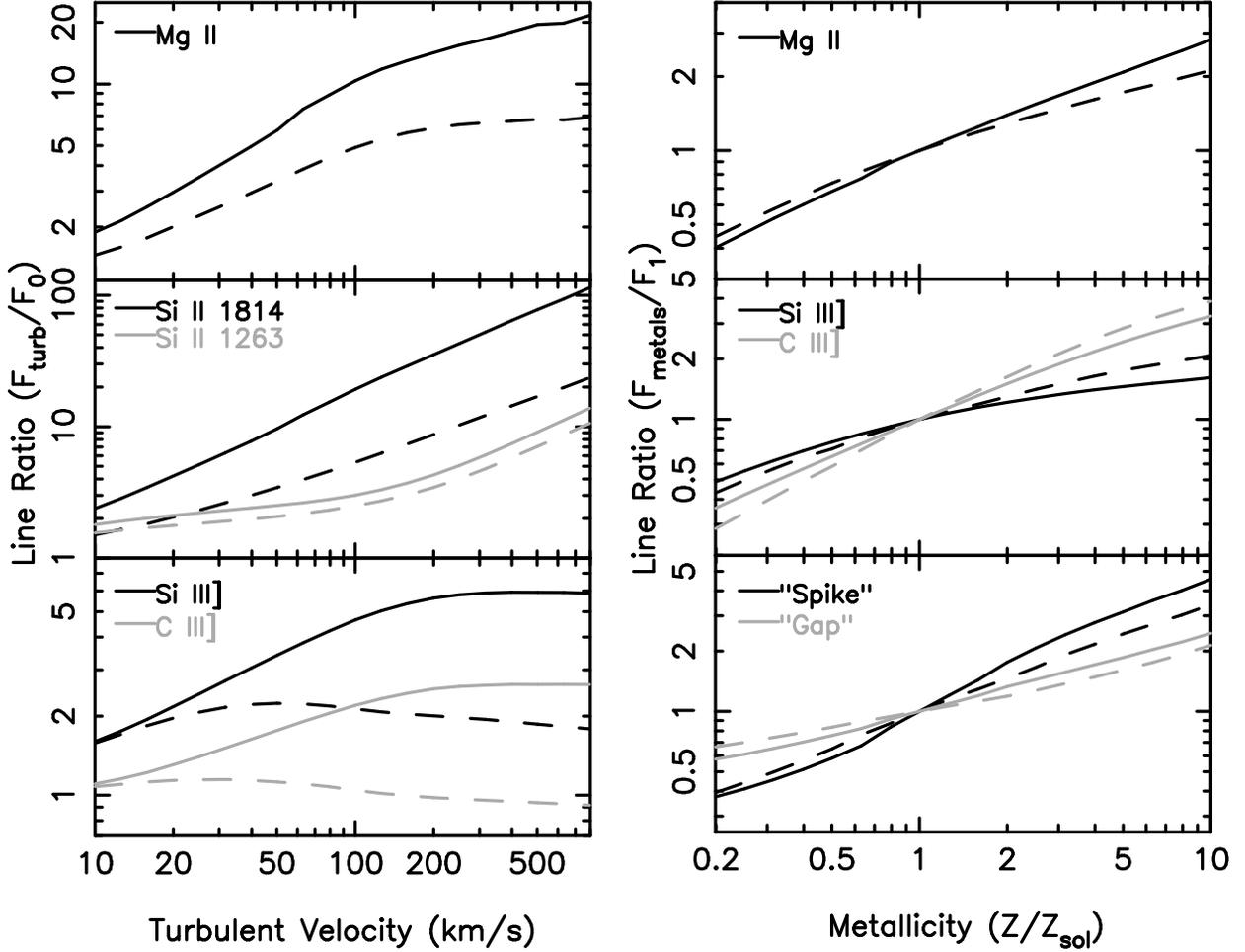}
\caption{The ratio of lines as a function of turbulence (left) and
  metallicity (right) to the values obtained for models in which
  $v_{turb}=0$ and $Z/Z_\odot=1$, respectively, evaluated for the
  fiducial parameters ($\log U=-1.5$, $\log n=11.0$, and $\log
  N_H=24.5$).  The solid and he dashed lines show the case for the
  PHL~1811 and the K97 SEDs.  More rapid increases in the line ratios
  for the PHL~1811 SED compared with the K97 SED are a consequence of
  temperature increases with turbulence and metallicity. 
\label{fig37}} 
\end{figure}

\clearpage



\begin{thebibliography}{}
\bibitem[Anderson et al.\ (2001)]{anderson01} Anderson, S.\ F.,  et al.\ 2001,
  \aj, 122, 503

\bibitem[Baldwin (1977)]{baldwin77} Baldwin, J.\ A.\ 1977, \mnras,
  178, 67

\bibitem[Baldwin et al.\ (1996)]{baldwin96} Baldwin, J.\ A., Ferland,
  G.\ J., Korista, K.\ T., Carswell, R.\ F., Hamann, F., Phillips, M.\
  M.\, Verner, D., Wilkes, B.\ J., \& Williams, R.\ E.\ 1996, \apj,
  461, 664

\bibitem[Baldwin et al.\ (2004)]{baldwin04} Baldwin, J.\ A., Ferland,
  G.\ J., Korista, K.\ A., Hamann, F., \& LaCluyz\'e, A.\ 2004, \apj,
  615, 610

\bibitem[Baldwin et al.\ (1995)]{baldwin95} Baldwin, J., Ferland, G.,
  Korista, K., \& Verner, D.\ 1995, \apjl, 455, 119

\bibitem[Baskin \& Laor (2004)]{bl04} Baskin, A., \& Laor, A.\ 2004,
  \mnras, 350, 31

\bibitem[Becker, White, \& Helfand (1995)]{bwh95} Becker, R.\ H.,
  White, R.\ L., \& Helfand, D.\ J.\ 1995, \apj, 450, 599

\bibitem[Begelman (2002)]{begelman02} Begelman, M.~C.\ 2002, \apj,
  568, L97

\bibitem[Blundell, Beasley, \& Bicknell (2003)]{blundell03} Blundell,
  K.~M., Beasley, A.~J., \& Bicknell, G.~V.\ 2003, \apj, 59, L103

\bibitem[Boroson \& Green (1992)]{bg92} Boroson, T.~A., \& Green,
  R.~F.\ 1992, \apjs, 80, 109

\bibitem[Bottorff et al.\ (2000)]{bottorff00} Bottorff, M., Ferland,
  G., Baldwin, J., \& Korista, K.\ 2000, \apj, 542, 644

\bibitem[Brandt et al.\ (1999)]{bbfr99} Brandt, W.\ N., Boller, T.,
  Fabian, A.\ C., \& Ruszkowski, M.\ 1999, \mnras, 303, 53

\bibitem[Brandt \& Hasinger (2005)]{bh05} Brandt, W.~N., \& Hasinger,
  G.\ 2005, ARAA, 43, in press

\bibitem[Brandt, Laor, \& Wills (2000)]{blw00} Brandt, W.~N., Laor,
  A., \& Wills, B.~J.\ 2000, \apj, 528, 637

\bibitem[Brotherton et al.\ (2001)]{brotherton01} Brotherton, M.\ S., Tran,
  H.\ D., Becker, R.\ H., Gregg, M.\ D., Laurent-Muehleisen, S.\ A.,
  \& White, R.\ L.\ 2001, ApJ, 546, 775

\bibitem[Cardelli, Clayton \& Mathis (1989)]{ccm89} Cardelli, J.\ A.,
  Clayton, G.\ C., \& Mathis, J.\ S.\ 1992, \apj, 345, 245

\bibitem[Casebeer, Leighly \& Baron (2006)]{clb05} Casebeer, D.~A.,
  Leighly, K.~M., \& Baron, E.\ 2006, \apj, 637, 157

\bibitem[Collinge et al.\ (2005)]{col05} Collinge, M.\ J., et al.\ 2005, \aj,
  129, 2542

\bibitem[Collin-Souffrin \& Dumont (1989)]{cd89} Collin-Souffrin, S.,
  \& Dumont, A.~M.\ 1989, \aap, 213, 29

\bibitem[Corbin \& Boroson (1996)]{cb96} Corbin, M.\ R., \& Boroson, T.\
  A.\ 1996, \apjs, 107, 69

\bibitem[Crenshaw (1986)]{crenshaw86} Crenshaw, D.\ M.\ 1986, \apjs, 62, 821

\bibitem[de Grijp, Lub \&  Miley (1987)]{degrijp87} de Grijp,
  M.~H.~K, Lub, J.,   \& Miley, G.~K.\ 1987, A\&AS, 70, 95

\bibitem[Dhanda et al.\ (2007)]{dhanda07} Dhanda, N., Baldwin, J.~A.,
  Bentz, M.~C., \& Osmer, P.~S., 2007, \apj, in press, astro-ph/0612610

\bibitem[Dietrich, Crenshaw, \& Kraemer (2005)]{dck05} Dietrich, M.,
  Crenshaw, D.\ M., \& Kraemer, S.~B.\ 2005, \apj, 623, 700

\bibitem[Dietrich et al.\ (2003)]{dietrich03} Dietrich, M., Hamann, F.,
  Appenzeller, I., \& Vestergaard, M.\ 2003, \apj, 596, 817

\bibitem[Dietrich et al.\ (2002)]{dietrich02} Dietrich, M., Hamann,
  F., Shields, J.\ C., Constantin, A., Vestergaard, M., Chaffee, F.,
  Foltz, C.\ B., \& Junkkarinen, V.\ T.\ 2002, \apj, 581, 912

\bibitem[Fan et al.\ (1999)]{fan99} Fan, X., et al.\ 1999, \apj, 526,
  57 

\bibitem[Ferland (1999)]{ferland99} Ferland, G.\ J.\ 1999, in Proc.\
  ``Quasars and Cosmology'', Eds.\ G.\ Ferland \& J.\ Baldwin (ASP:
  San Francisco), p.\ 147

\bibitem[Ferland et al.\ (1996)]{ferland96} Ferland, G.~J., Baldwin,
  J.~A., Korista, K.~T., Hamann, F., Carswell, R.~F., Phillips, M.,
  Wilkes, B., \& Williams, R.~E.\ 1996, \apj, 461, 683

\bibitem[Ferland \& Persson (1989)]{fp89} Ferland, G.~J., \& Persson,
  S.~E.\ 1989, \apj, 347, 656

\bibitem[Forster et al.\ (2001)]{forster01}Forster, K., Green, P.\ J.,
  Aldcroft, T.\ L., Vestergaard, M., Foltz, C.\ B., \& Hewitt, P.\ C.\
  2001, \apjs, 134, 35

\bibitem[Forster \& Halpern (1996)]{fh96} Forster, K., \& Halpern, J.\
  P.\ 1996, \apj, 468, 565

\bibitem[Francis et al.\ (1991)]{francis91} Francis, P.\ J., Hewett, P.\
C., Foltz, C.\ B., Chaffee, F.\ H., Weymann, R.\ J., \& Morris, S.\
L.\ 1991, ApJ, 373, 465

\bibitem[Gallagher et al.\ (2002)]{gall02} Gallagher, S.~C., Brandt,
  W.~N., Chartas, G., \& Garmire, G.~P.\ 2002, \apj, 567, 37

\bibitem[Gallo et al.\ (2004)]{gallo04} Gallo, L.\ C., Boller, T.,
  Brandt, W.\ N., Fabian, A.\ C., \& Grupe, D.\ 2004, \mnras, 352, 744

\bibitem[Goodrich (2000)]{goodrich00} Goodrich, R.~W., 2000, NewAR,
  44, 519

\bibitem[Grupe et al.\ (2004)]{grupe04} Grupe, D., Leighly, K.\ M.,
  Burwitz, V., Predehl, P., \& Mathur, S.\ 2004, \aj, 128, 1524

\bibitem[Grupe et al.\ (2000)]{grupe00} Grupe, D., Leighly, K.\ M.,
  Thomas, H.-C., \& Laurent-Muehleisen, S.\ A.\ 2000, \aap, 356, 11

\bibitem[Grupe et al.\ (2001)]{grupe01} Grupe, D., Thomas, H.-C., \&
  Leighly, K.\ M.\ 2001, \aap, 369, 450

\bibitem[Hall et al.\ (2004)]{hall04} Hall, P.\ B., et al.\ 2004, Proc.\
  ``Multiwavelength AGN Surveys," eds.\ R.\ Mujica \& R.\ Maiolino,
  astro-ph/0403347 

\bibitem[Hamann \& Ferland (1993)]{hf93} Hamann, F., \& Ferland, G.\
  1993, \apj, 418, 11

\bibitem[Hamann et al.\ (2002)]{hkfwb02} Hamann, F., Korista, K.\ T.,
  Ferland, G.\ J., Warner, C., \& Baldwin, J.\ 2002, \apj, 564, 592

\bibitem[Haro \& Luyten (1962)]{hl62} Haro, G., \& Luyten, W.\ J.\
  1962, Bol.\ Inst.\ Tonantzintla, 3, 37

\bibitem[Jenkins et al.\ (2005)]{jenkins05} Jenkins, E.\ B., Bowen,
  D.~V., Tripp, T.~M., \& Sembach, K.~R.\ 2005, \apj, 623, 767

\bibitem[Jenkins et al.\ (2003)]{jenkins03} Jenkins, E.\ B., Bowen,
  D.~V.\, Tripp, T.~M., Sembach, K.~R., Leighly, K.~M., Halpern,
  J.~P., \& Lauroesch, J.~T.\ 2003, \aj, 125, 2824

\bibitem[Johansson et al.\ (1995)]{johansson95} Johansson, S., Brage,
  T., Leckrone, D.\ S., Nave, G., \& Wahlgren, G.~M.\ 1995, \apj, 446,
  361 

\bibitem[Johansson \& Hansen (1988)]{jh88} Johansson, S., \& Hansen,
  J.~E.\ 1988, in ``Physics of Formation of \ion{Fe}{2} lines outside
  LTE'', eds R.\ Viotti, A.\ Vittone, \& M.\ Friedjung (Dordrecht:
  Reidel), 13

\bibitem[Johansson et al.\ (2000)]{johansson00} Johansson, S., Zethson,
  T., Hartman, H., Ekberg, J.\ O., Ishibashi, K., Davidson, K., \&
  Gull, T.\ 2000, \aap, 361, 977

\bibitem[Joly (1989)]{joly89} Joly, M.\ 1989, \aap, 208, 47

\bibitem[Kim Quijano et al.\ (2003)]{quijano03} Kim Quijano, J., et
  al., 2003, ``STIS Instrument Handbook'', Version 7.0, (Baltimore:
  STScI) 

\bibitem[Korista et al.\ (1995)]{korista95} Korista, K.~T., et al.\
  1995, \apjs, 97, 285

\bibitem[Korista et al. \ (1997)]{k97} Korista, K., Baldwin, J.,
  Ferland, G., \& Verner, D.\ 1997, ApJS, 108, 401 (K97)

\bibitem[Krolik \& Kallman (1988)]{kk88} Krolik, J.~H., \& Kallman,
  T.~R.\ 1988, \apj, 324, 714

\bibitem[Kwan \& Krolik (1981)]{kk81} Kwan, J., \& Krolik, J.~H.\
  1981, \apj, 250, 478

\bibitem[Laor (2000)]{laor00} Laor, A.\ 2000, NewA Rev., 44, 503

\bibitem[Laor et al.\ (1997b)]{laor97b} Laor, A., Jannuzi, B.\ T., Green,
  R.\ F., \& Boroson, T.\ A.\ 1997, ApJ, 489, 656

\bibitem[Laor \& Netzer (1989)]{ln89} Laor, A., \& Netzer, H.\ 1989,
  \mnras, 238, 897

\bibitem[Leighly (1999a)]{leighly99a} Leighly, K.~M.\ 1999, \apjs, 125,
  297 
 
\bibitem[Leighly (1999b)]{leighly99b} Leighly, K.~M.\ 1999, \apjs, 125,
  317

\bibitem[Leighly (2001)]{leighly01} Leighly, K.~M.\ 2001, in ``X-ray
  Emission from Accretion onto Black Holes'', eds.\ T.\ Yaqoob \& J.\
  H.\ Krolik, published electronically
  (http://www.pha.jhu.edu/groups/astro/workshop2001/)

\bibitem[Leighly (2004)]{leighly04} Leighly, K.\ M.\ 2004, \apj, 611, 125

\bibitem[Leighly et al.\ (2001)]{lhhbi01} Leighly, K.\ M., Halpern,
  J.\ P., Helfand, D.\ J., Becker, R.\ H., \& Impey, C.\ D.\ 2001,
  \aj, 121, 2889

\bibitem[Leighly, Halpern \& Jenkins (2004)]{lhj04} Leighly, K.\ M.,
  Halpern, J.\ P., \& Jenkins, E.\ B.\ 2004, in Proc. ``AGN Physics
  with the Sloan Digital Sky Survey'', eds.\ G.\ T.\ Richards \& P.\
  B.\ Hall (ASP: San Francisco), 277

\bibitem[Leighly et al.\ (2007)]{leighly06} Leighly, K.\ M., Halpern,
  J.\ P., Jenkins, E.\ B., Grupe, D., Choi, J., \& Prescott, K.\ B.,
  2007, accepted for publication in ApJ, astro-ph/0611349

\bibitem[Leighly \& Moore (2004)]{lm04} Leighly, K.\ M., \& Moore, J.\
  R.\ 2004, \apj, 611, 107

\bibitem[Leighly \& Moore (2006)]{lm06} Leighly, K.\ M., \& Moore, J.\
  R.\ 2006, \apj, 644, 748

\bibitem[Londish et al.\ (2004)]{londish04} Londish, D., Heidt, J.,
  Boyle, B.\ J., Croom, S.\ M.,  \& Kedziora-Chudczer, L.\ 2004,
  \mnras, 352, 903

\bibitem[Matsumoto, Leighly \& Kawaguchi (2004)]{mlk04} Matsumoto, C.,
  Leighly, K.\ M., \& Kawaguchi, T.\ 2004, Progress of Theoretical
  Physics Supplement, 155, 377

\bibitem[McDowell et al.\ (1995)]{mcdowell95} McDowell, J.\ C.,
  Canizares, C., Elvis, M., Lawrence, A., Markoff, S., Mathur, S., \&
  Wilkes, B.\ J., 1995, \apj, 450, 585

\bibitem[Netzer (1987)]{netzer87} Netzer, H.\ 1987, \mnras, 225, 55

\bibitem[Nomoto, Nakamura, Kobayashi (1992)]{nnk92} Nomoto, K., Nakamura,
  T., \& Kobayashi, C.\ 1999, \apjs, 265, 37

\bibitem[Persson (1988)]{persson88} Persson, S.~E., 1988, \apj, 330, 751

\bibitem[Proga (2005)]{proga05} Proga, D., 2005, \apj, 630, L9

\bibitem[Ptak et al.\ (2003)]{ptak03} Ptak, A., Heckman, T., Levenson,
  N.~A., Weaver, K., \& Strickland, D.\ 2003, \apj, 592, 782

\bibitem[Reimers et al.\ (2005)]{reimers05} Reimers, D., Janknecht,
  E., Fechner, C., Agafonova, I.\ I., Levshakov, S.\ A., \& Lopez, S.\
  2005, \aap, 435, 17

\bibitem[Rees, Netzer \& Ferland (1989)]{rnf89} Rees, M.~J., Netzer,
  H., \& Ferland, G.~J.\ 1989, \apj, 347, 640

\bibitem[Rudy et al.\ (2000)]{rudy00} Rudy, R.\ J., Mazuk, S., Puetter,
  R.\ C., \& Hamann, F.\ 2000, \apj, 539, 166

\bibitem[Schlegel, Finkbeiner, \& Davis (1998)]{sfd98} Schlegel, D.\ J.,
  Finkbeiner, D.\ P., \& Davis, M.\ 1998, \apj, 500, 525

\bibitem[Schneider et al.\ (2005)]{schneider05} Schneider, D.\ P., et
  al.\ (2005), \aj, 130, 367

\bibitem[Shang et al.\ (2005)]{shang05} Shang, Z., et al., 2005, \apj,
  619, 41

\bibitem[Snedden \& Gaskell (1999)]{sg99} Snedden, S.~A., \& Gaskell,
  C.~M.\ 1999, \apj, 521, L91

\bibitem[Stalin \& Srianand (2005)]{ss05} Stalin, C.\ S., \& Srianand,
  R.\ 2005, \mnras, 359, 1022

\bibitem[Steffen et al.\ (2006)]{steffen06} Steffen, A.\ T., Strateva,
  I., Brandt, W.\ N., Alexander, D.\ M., Koekemoer, A.\ M., Lehmer,
  B.\ D., Schneider, D.\ P., \& Vignali, C., 2006, \aj, 131, 2826

\bibitem[Telfer et al.\ (2002)]{telfer02} Telfer, R.~C., Zheng, W.,
  Kriss, G.~A., \& Davidsen, A.~F.\ 2002, \apj, 565, 773

\bibitem[Thompson (1991)]{thompson91} Thompson, K.\ L.\ 1991, \apj,
  374, 496

\bibitem[Vanden Berk et al.\ (2001)]{vandenberk01} Vanden Berk,
  D.~E., et al.\ 2001, \aj, 122, 549

\bibitem[V\'eron-Cetty, Joly, \& V\'eron (2004)]{vjv04} V\'eron-Cetty,
  M.-P., Joly, M., \& V\'eron, P.\ 2004, \aap, 417, 515

\bibitem[V\'eron-Cetty et al.\ (2006)]{veroncetty06} V\'eron-Cetty,
  M.-P., Joly, M., V\'eron, P., Boroson, T., Lipari, S., \& Ogle, P.\
  2006, \aap, 451, 851

\bibitem[Verner et al. (2004)]{verner04} Verner, E., Bruhweiler, F.,
  Verner, D., Johansson, S., Kallman, T., \& Gull, T.\ 2004, \apj,
  611, 780

\bibitem[Vestergaard \& Wilkes (2001)]{vw01} Vestergaard, M., \& Wilkes,
  B.\ J.\ 2001, \apjs, 134, 1

\bibitem[Vignali et al.\ (2001)]{vignali01} Vignali, C., Brandt, W.\
  N., Fan, X., Gunn, J.\ E., Kaspi, S., Schneider, D.\ P., \& Strauss,
  M.\ A.\ 2001, \aj, 122, 2143

\bibitem[Walter et al.\ (1994)]{walter94} Walter, R., Orr, A.,
  Courvoisier, T.\ J.-L., Fink, H.\ H., Makino, F., Otani, C., \&
  Wamsteker, W.\ 1994, \aap, 285, 119

\bibitem[White et al.\ (1997)]{white97} White, R.\ L.\, Becker, R.\
  H., Helfand, D.\ J., \& Gregg,   M.\ D.\ 1997, \apj, 475, 479

\bibitem[Wilkes et al.\ (1994)]{wilkes94} Wilkes, B.~J., Tananbaum,
  H., Worrall, D.~M., Avni, Y., Oey, M.~S., \& Flanagan, J.\ 1994,
  \apjs, 92, 53

\bibitem[Wilkes et al.\ (1999)]{wilkes99} Wilkes, B.~J.,
  Kuraszkiewicz, J., Green, P.~J., Mathur, S., \& McDowell, J.~C.\
  1999, \apj, 513, 76

\bibitem[Zheng et al.\ (1997)]{zheng97} Zheng, W., Kriss, G.~A.,
  Telfer, R.~C., Grimes, J.~P., \& Davidsen, A.~F.\ 1997, \apj, 475,
  469 

\end{thebibliography}
\end{document}